\documentclass[aps, prb, reprint]{revtex4-2}

\def\ket#1{\mathinner{|{#1}\rangle}}

\def\braket#1{\mathinner{\langle{#1}\rangle}}
\usepackage{graphicx}
\usepackage{amsmath}
\usepackage{amsfonts}
\usepackage{dsfont}
\usepackage{bm}

\begin{document}
	\title{Tower of quantum scars in a partially many-body localized system}
	
	\author{Michael Iversen}
	\affiliation{Department of Physics and Astronomy, Aarhus University, DK-8000 Aarhus C, Denmark}
		
	\author{Anne E. B. Nielsen}
	\affiliation{Department of Physics and Astronomy, Aarhus University, DK-8000 Aarhus C, Denmark}
	
	\begin{abstract}
		Isolated quantum many-body systems are often well-described by the eigenstate thermalization hypothesis.
		There are, however, mechanisms that cause different behavior: many-body localization and quantum many-body scars.
		Here, we show how one can find disordered Hamiltonians hosting a tower of scars by adapting a known method for finding parent Hamiltonians.
		Using this method, we construct a spin-1/2 model which is both partially localized and contains scars.
		We demonstrate that the model is partially localized by studying numerically the level spacing statistics and bipartite entanglement entropy. As disorder is introduced, the adjacent gap ratio transitions from the Gaussian orthogonal ensemble to the Poisson distribution and the entropy shifts from volume-law to area-law scaling.
		We investigate the properties of scars in a partially localized background and compare with a thermal background.
		At strong disorder, states initialized inside or outside the scar subspace display different dynamical behavior but have similar entanglement entropy and Schmidt gap.
		We demonstrate that localization stabilizes scar revivals of initial states with support both inside and outside the scar subspace.
		Finally, we show how strong disorder introduces additional approximate towers of eigenstates.
	\end{abstract}

	\maketitle
	\section{Introduction}
	The eigenstate thermalization hypothesis (ETH) describes how isolated quantum systems reach thermal equilibrium \cite{Deutsch1991, Srednicki1994, Rigol2008}.
	The hypothesis is a statement about generic quantum many-body systems and has been verified for a wide variety of physical models \cite{Rigol2008, Rigol2009Sep, Rigol2009Nov, Rigol2010, Heidrick-Meisner2014, Marquardt2012, Gemmer2014, Srednicki2015, Prelovsek2013, Huse2014NovE, Rigol2016}.
	Despite the effectiveness of ETH, several phenomena are known to cause non-thermal behavior.
	
	One such mechanism is many-body localization (MBL) \cite{Basko2006, Polyakov2005, Huse2007, Huse2010}.
	MBL appears in many-body interacting systems and may originate from different sources such as disordered potentials \cite{Huse2007}, disordered magnetic fields \cite{Znidaric2008, Huse2010}, quasi-periodic potentials \cite{Iyer2013, Setaiwan2017, Zhang2018, Singh2021}, disordered interactions \cite{Sierant2017, Pollmann2014}, bond disorder \cite{Potter2016}, gradient fields \cite{Schulz2019, Evert2019, Zhang2021}, periodic driving \cite{Bairey2017, Choi2018, Bhakuni2020, Yousefjani2023}, etc.
	In the case of quench disorder, all the energy eigenstates become localized at strong disorder and an extensive set of quasi-local integrals of motion (LIOM) emerges \cite{Abanin2013, Huse2014NovB}.
	Consequently, all energy eigenstates behave non-thermally and MBL represents a strong violation of ETH.	
	Signatures of MBL have been observed in experimental setups with ultra-cold fermions \cite{Schreiber2015}, ultra-cold bosons \cite{Choi2016}, ultra-cold ions representing an effective spin-$1/2$ chain \cite{Smidt2016}, superconducting qubits \cite{Xu2018}, etc.
	While MBL is well-established for finite systems, the stability of MBL in the thermodynamic limit is still an open question \cite{Imbrie2016, Suntajs2020, Luitz2020, Suntajs2020Dec, Kiefer-Emmanouilidis2021, Abanin2021}.

	Another mechanism leading to non-thermal behavior was found in the Affleck-Kennedy-Lieb-Tasaki model \cite{Moudgalya_I,Moudgalya_II} and in experiments with kinetically constrained Rydberg atoms \cite{Bernien2017}.
	The atoms were arranged with strong nearest neighbor interactions so the simultaneous excitation of neighboring atoms was prohibited.
	When initializing the system in the Néel state, observables displayed abnormal persistent oscillations -- contrary to the predictions by ETH.
	Subsequent theoretical works uncovered that the revivals were caused by a small number of non-thermal eigenstates dubbed quantum many-body scars (QMBS) \cite{Papic2018Jul, Papic2018Oct, Olexei2019, Iadecola2019}.		
	The scar states are uncommon and represent a vanishingly small part of an otherwise thermalizing spectrum.
	Therefore, QMBS represent a weak violation of ETH.
	After their initial discovery, QMBS were uncovered in numerous different models \cite{Schecter2019, Iadecola_Schecter, MarkDaniel2020, MoudgalyaSanjay2020, Shibata2020}.
	Furthermore, the scarred models have been categorized under several unifying formalisms, e.g.\ a spectrum generating algebra \cite{Mark2020, Moudgalya2020}, the Shiraishi-Mori formalism \cite{Shiraishi2017}, quasi-symmetry based formalisms \cite{Ren2021, Ren2022}, scar states constructed from the Einstein-Podolsky-Rosen state in bilayer systems \cite{Wildeboer2022}, etc.
	These formalisms are generally overlapping and each formalism only describes a subset of the known scarred models.
	In addition to being widely investigated theoretically, scarred models have also been realized experimentally in different setups \cite{Chen2022, Bluvstein2021, ZhangPengfei2021}.
	
	In this work, we realize both ETH-breaking mechanisms simultaneously.
	We study a one-dimensional disordered spin-$1/2$ chain hosting a tower of QMBS.
	As the disorder strength is increased, the model transitions from the thermal phase to being partially localized while preserving the scar states. In earlier works, a single scar state was embedded in an otherwise MBL spectrum \cite{Anne2020, Iversen2022, Srivatsa2022}.
	Our work adds to these studies by considering a full tower of QMBS in an MBL spectrum.
	The presence of multiple scar states, enables us to study the effect of localization on the dynamical revivals characteristic of scar states.
	Using this model, we demonstrate how scar states can be distinguished from a localized background.
	We also find two phenomena originating from the interplay between QMBS and localization: disorder stabilization of scar revivals and disorder induced revivals.

Our results show that the phenomenon of quantum many-body scars can be robust to disorder, and in some cases scar revivals can even be stabilized by disorder. MBL systems have properties that are interesting for quantum memories \cite{nandkishore2015}, while quantum many-body scars can be utilized for metrology and sensing \cite{Dooley2021,Dooley2023}. Quantum many-body scars in an MBL background provides a device, in which part of the Hilbert space can be utilized for quantum storage (the MBL states), while other parts of the Hilbert space can be utilized for processing (the scar states).

	The paper is structured as follows.
	In Sec.\ \ref{subsec:initial_model}, we summarize the model by Iadecola and Schecter which is the starting point of our analysis.
	In Sec.\ \ref{subsec:discover}, we explain how we find Hamiltonians having a set of scar states with equal energy spacing.
	In Sec.\ \ref{subsec:generalized_models}, we use this method to determine all local $1$- and $2$-body Hamiltonians for the tower of scar states in the Iadecola and Schecter model.
	In Sec.\ \ref{sec:partial MBL}, we show that a subset of these Hamiltonians partially localize as disorder is introduced.
	We quantify the partial localization as a special structure in the energy eigenstates and compare with results from exact diagonalization.
	We verify the localization by studying the level spacing statistics in Sec.\ \ref{subsec:spectral statistic} and the entanglement entropy in Sec.\ \ref{subsec:entropy}.
	In Sec.\ \ref{sec:features}, we show that the fidelity between initial states and the corresponding time evolved states can be utilized to distinguish the scar states from the partially localized background.
	We further show that the bipartite entanglement entropy and Schmidt gap are ineffective tools for distinguishing scar states from a partially localized background.
	In Sec.\ \ref{sec:stabilization}, we demonstrate how scar revivals are stabilized by strong disorder.
	In Sec.\ \ref{sec:disorder induced scars}, we uncover additional approximate towers of eigenstates which emerge as disorder is introduced.
	Finally, we summarize our results in Sec.\ \ref{sec:conclusion}.
	
	\section{Model}\label{sec:model}
	\subsection{Model by Iadecola and Schecter}\label{subsec:initial_model}
	We take the model by Iadecola and Schecter as our starting point \cite{Iadecola_Schecter}.
	Consider a one-dimensional spin-$\frac{1}{2}$ chain of even length $L$ with periodic boundary conditions.
	The local Hilbert space on each site is described by the eigenkets $\ket \uparrow$ and $\ket \downarrow $ of the Pauli $z$-matrix, i.e.\ $\hat \sigma^z \ket \uparrow = \ket \uparrow$ and $\hat \sigma^z \ket \downarrow = - \ket \downarrow$.
	The model by Iadecola and Schecter is given by
	\begin{align}\label{eq:Iadecola_Scheter}
		\hat H_0 = \sum_{i=1}^L \Big[ \lambda(\hat \sigma_i^x - \hat \sigma_{i-1}^z \hat \sigma_i^x \hat \sigma_{i+1}^z) + \Delta \hat \sigma_i^z + J \hat \sigma_i^z \hat \sigma_{i+1}^z \Big],
	\end{align}
	with $\lambda, \Delta, J \in \mathbb R$.
	All indices are understood as modulo $L$, i.e.\ the index $i + L$ is identified as $i$.
	The operators $\hat \sigma_i^x$, $\hat \sigma_i^y$ and $\hat \sigma_i^z$ are the Pauli matrices acting on site $i$.
	The first term in Eq.\ \eqref{eq:Iadecola_Scheter} flips the spin $s_i$ at site $i$ if its nearest neighbors are in different states, i.e.\ $s_{i - 1} \neq s_{i + 1}$.
	The second term is a magnetic field along the $z$-direction with strength $\Delta$.
	The third term represents nearest neighbor interactions with strength $J$.
	
	Two adjacent spins in different states represent a domain wall, i.e.\ $\uparrow\downarrow$ or $\downarrow\uparrow$.
	The Hamiltonian conserves the number of domain walls $N_\text{dw}$ because only spins with different neighbors are allowed to change their state.
	Furthermore, the Hamiltonian is invariant under spatial inversion and translation, but these symmetries are broken when disorder is introduced in section \ref{sec:MBL} and we will not consider them any further.
	
	For nonzero values of $\lambda$, $\Delta$ and $J$, the energy eigenstates are thermal except for a small number of ETH-violating scar states grouped into two towers.
	Throughout this work, we only focus on one of these towers.
	This tower contains $L/2+1$ eigenstates and the $n$-th state $\ket{\mathcal S_n}$ is constructed by acting $n$ times with the operator $\hat Q^\dagger$ on the ``all-spin-down'' state
	\begin{align}\label{eq:scars}
		\ket{\mathcal S_n} \propto \big(\hat Q^\dagger\big)^n \ket{\downarrow \downarrow \ldots \downarrow}.
	\end{align}
	The operator $\hat Q^\dagger$ is given by
	\begin{align}\label{eq:Q}
		\hat Q^\dagger = \sum_{i=1}^L (-1)^i \hat P_{i-1}^\downarrow \hat \sigma_i^+ \hat P_{i+1}^\downarrow,
	\end{align}
	where $\hat \sigma_i^+ = (\hat \sigma_i^x + i \hat \sigma_i^y) / 2$ is the raising operator and $\hat P_i^\downarrow = (\hat{\mathds 1} - \hat \sigma_i^z)/2$ is the local projection onto spin down.
	The $n$-th scar state has energy $E_n = 2(\Delta - 2J)n + (J-\Delta)L$, number of domain walls $N_\text{dw} = 2n$ and generally appears central in the spectrum after resolving all symmetries.
	Since the scar states are equally spaced in energy, any initial state in the scar subspace displays the dynamical revivals characteristic of QMBS.
	Furthermore, it was shown in Ref.\ \cite{Iadecola_Schecter} that the bipartite entanglement entropy of the scar states displays logarithmic scaling with system size.
	
	\subsection{Determining Hamiltonians}\label{subsec:discover}
	All eigenstates of $\hat H_0$ located near the middle of the spectrum are thermal except the scar states.
	We wish to extend the model so the scar states are embedded in a MBL background instead of a thermal background.
	MBL is possible in disordered systems.
	Unfortunately, disorder cannot be introduced naively to the Hamiltonian $\hat H_0$.
	When promoting any parameter to being site-dependent $\lambda \to \lambda_i$, $\Delta \to \Delta_i$ or $J \to J_i$, the scar states are no longer eigenstates.
	Therefore, disorder must be introduced through new terms.
	In this section, we uncover all local few-body Hamiltonians which share the scar states as eigenstates and maintain equal energy spacing.
	In the next section, we show that a subset of these Hamiltonians are partially localized.
	
	We search for local Hamiltonians following Refs.\ \cite{Chertkov, Greiter}.
	The set of $2^L\times 2^L$ Hermitian operators form a vector space.
	Most of these operators are long-ranged, contain many-body interactions and are difficult to realize in experiments.
	Therefore, we restrict ourselves to Hamiltonians containing local 1- and 2-body Hermitian operators.
	This subspace is spanned by the operator basis
	\begin{align}\label{eq:H_basis}
	\begin{split}
		\mathcal B_2 = & \Big\{\hat \sigma_i^a \Big| a \in \{ x, y, z\}, \enspace i \in \mathbb Z_L \Big\} \\
		&\cup \Big\{\hat \sigma_i^a \hat \sigma_{i+1}^b \Big|  a, b \in \{ x, y, z\}, \enspace i \in \mathbb Z_L \Big\},
	\end{split}
	\end{align}
	where $\mathbb Z_L = \{1, 2, \ldots, L\}$ are the first $L$ integers.
	This subspace is considerably smaller than the full operator vector space and has dimension $|\mathcal B_2| = 12L$ where $|\cdot|$ denotes the number of elements in a set. Any local 1- or 2-body interacting Hamiltonian can be expressed as a linear combination of the basis elements
	\begin{align}\label{eq:general_linearcombination}
		\hat H = \sum_{i=1}^{|\mathcal B_2|} \alpha_i \hat h_i, \qquad \hat h_i \in \mathcal B_2,
	\end{align}
	where $\alpha_i \in \mathbb R$ are free coefficients.
	To simplify notation, we collect the coefficients in a vector $\bm \alpha = (\alpha_1, \alpha_2, \ldots, \alpha_{|\mathcal B_2|})^T$ where $T$ is the transpose.
	
	We search for the vector of parameters $\bm \alpha$ so the resulting Hamiltonian has $\ket{\mathcal S_n}$ as eigenstates for $n = 0, 1, \ldots, L/2$.
	The scar state $\ket{\mathcal S_n}$ is an eigenstate of $\hat H$ if and only if the energy variance of $\ket{\mathcal S_n}$ is exactly zero
	\begin{align}\label{eq:variance}
		\braket{\mathcal S_n | \hat H^2 | \mathcal S_n} - \braket{\mathcal S_n | \hat H | \mathcal S_n}^2 = 0.
	\end{align}
	Inserting Eq.\ \eqref{eq:general_linearcombination}, the expression becomes
	\begin{align}\label{eq:aCa}
		\bm \alpha^T C_n \bm \alpha = 0,
	\end{align}
	where $C_n$ is the quantum covariance matrix
	\begin{align}
		[C_n]_{ij} = \braket{\mathcal S_n| \hat h_i \hat h_j |\mathcal S_n} - \braket{S_n|\hat h_i|S_n} \braket{S_n|\hat h_j|S_n}.
	\end{align}
	Equation \eqref{eq:aCa} is satisfied when the vector of coefficients lies in the null space of the quantum covariance matrix $\bm \alpha \in \mathrm{Null}(C_n)$, i.e.\ $C_n \bm \alpha = \bm 0$.
	We ensure all scar states $\ket{\mathcal S_n}$ are simultaneously eigenstates of $\hat H$ by demanding the vector of coefficients $\bm \alpha$ lies in the null space of every covariance matrix $\bm \alpha \in \mathrm{Null}(C_0) \cap \mathrm{Null}(C_1) \cap \ldots \cap \mathrm{Null}(C_{L/2})$.
	While this condition ensures all scar states are eigenstates of $\hat H$, they are not necessarily equally spaced in energy.
	Equal energy spacing is established by imposing another set of requirements
	\begin{align}
		\begin{split}
		\braket{\mathcal S_{n+2}|\hat H|\mathcal S_{n+2}} - \braket{\mathcal S_{n+1}|\hat H|\mathcal S_{n+1}} \\
		= \braket{\mathcal S_{n+1}|\hat H|\mathcal S_{n+1}} - \braket{\mathcal S_n|\hat H|\mathcal S_n},
		\end{split}
	\end{align}
	for all $n=0, 1, \ldots, L/2-2$.
	Inserting Eq.\ \eqref{eq:general_linearcombination}, we find
	\begin{align}
		G \bm \alpha = 0,
	\end{align}
	where we introduce the rectangular matrix of energy gap differences
	\begin{align}
		[G]_{ij} = &\braket{\mathcal S_{i+2}|\hat h_j|\mathcal S_{i+2}} - 2\braket{\mathcal S_{i+1}|\hat h_j|\mathcal S_{i+1}} + \braket{\mathcal S_i|\hat h_j|\mathcal S_i}.
	\end{align}
	We observe that the scar states are equally spaced in energy when the coefficient vector resides in the null space of the gap matrix.
	In summary, the scar states appear as eigenstates of the Hamiltonian with equal energy spacing when the vector of coefficients lies in the intersection
	\begin{align}\label{eq:null_space}
		\bm \alpha \in \bigcap\limits_{n=0}^{L/2} \mathrm{Null}(C_n) \cap \mathrm{Null}(G).
	\end{align}
	It is straightforward to determine this subspace numerically since the scar states are known analytically.
	Note however, that while the matrices $C_n$ and $G$ are complex, we only search for real vectors $\bm \alpha \in \mathbb R^{|\mathcal B_2|}$ (for complex vectors $\bm \alpha \in \mathbb C^{|\mathcal B_2|}$, the linear combination in Eq.\ \eqref{eq:general_linearcombination} is not necessarily Hermitian).
	We find real coefficient vectors by stacking the real and imaginary parts of the matrices $(\mathrm{Re}(C_0), \allowbreak \mathrm{Im}(C_0),\allowbreak \ldots, \allowbreak \mathrm{Re}(C_{L/2}), \allowbreak \mathrm{Im}(C_{L/2}), \allowbreak \mathrm{Re}(G), \allowbreak \mathrm{Im}(G))^T$ and determining the null space of the resulting rectangular matrix by e.g.\ singular value decomposition.
	
	The vectors $\bm \alpha_i$ produced by this numerical method are typically dense, i.e.\ have few nonzero entries.
	As a consequence, the corresponding operator $\sum_i \alpha_i \hat h_i$ is difficult to interpret.
	We overcome this difficulty by noting that if $\{\bm \alpha_i|i = 1, 2, \ldots\}$ lies in the null space Eq.\ \eqref{eq:null_space}, then any linear combination of these vectors also lies in the null space.
	We apply a heuristic algorithm to determine sparse vectors in the subspace \cite{QU}.
	
	\subsection{Generalized models}\label{subsec:generalized_models}
	\renewcommand{\arraystretch}{2}
	\begin{table}
		\centering
		\begin{tabular}{ll}
			\toprule
			(\textit{i}) & $\hat H_z = \sum_{i=1}^L \hat \sigma_i^z$ \\
			(\textit{ii}) & $ \hat D_i = \hat \sigma_i^z + \hat \sigma_{i+1}^z + \hat \sigma_i^z \hat \sigma_{i+1}^z$, \quad for $i \in \mathbb Z_L$ \\
			(\textit{iii}) & $ \hat H_{zz}^\text{odd} =  \sum_{i=1}^{L/2} \hat \sigma_{2i-1}^z \hat \sigma_{2i}^z$ \\
			(\textit{iv}) & $ \hat H_{xz}^\text{alt} = \sum_{i=1}^{L} (-1)^i(\hat \sigma_i^x  \hat \sigma_{i+1}^z + \hat \sigma_i^z \hat \sigma_{i+1}^x)$ \\
			(\textit{v}) & $ \hat H_{yz}^\text{alt} = \sum_{i=1}^{L} (-1)^i(\hat \sigma_i^y \hat \sigma_{i+1}^z + \hat \sigma_i^z \hat \sigma_{i+1}^y)$ \\
			\toprule
		\end{tabular}
		\caption{Local 1- and 2-body operators which have $\ket{\mathcal S_n}$ for $n=0,1,\ldots,L/2$ as energy eigenstates with equal energy spacing. The operators are determined by applying the numerical method presented in Sec.\ \ref{subsec:discover} and Appendix \ref{apx:new_operators} proves the statement rigorously.}
		\label{tab:operators}
	\end{table}
	\renewcommand{\arraystretch}{1}
	We apply the numerical method for system sizes $L=8$, $10$, $12$, $14$ and for all sizes find $L+4$ linearly independent vectors $\bm \alpha_i$ satisfying Eq.\ \eqref{eq:null_space}.
	The corresponding operators are summarized in Tab.\ \ref{tab:operators}.
	The first operator $\hat H_z$ was already present in the initial model Eq.\ \eqref{eq:Iadecola_Scheter} and adds nothing new.
	The $L$ operators $\hat D_i$ act locally on sites $i$ and $i+1$ and represent good candidates for adding quench disorder into the model in Eq.\ \eqref{eq:Iadecola_Scheter}.
	Indeed, in Sec.\ \ref{sec:MBL}, we demonstrate the system partially localizes when introducing sufficiently strong disorder via these operators.
	The third operator $\hat H_{zz}^\text{odd}$ represents an interaction between every odd site and its right neighbor with equal interaction strength.
	The fourth and fifth operators $\hat H_{xz}^\text{alt}$ and $\hat H_{yz}^\text{alt}$ flip spins with the sign of the term determined by the nearest neighbors.
	
	Using the numerical method, we rediscover the 1- and 2-body terms of the model in Eq.\ \eqref{eq:Iadecola_Scheter} by starting from the scar states.
	As noted above, the operator $\hat H_z$ was already present in the original model. Furthermore, the third term in Eq.\ \eqref{eq:Iadecola_Scheter} is a linear combination of the operators in Tab.\ \ref{tab:operators}: $\sum_{i=1}^L \hat \sigma_i^z \hat \sigma_{i+1}^z = \sum_{i=1}^L \hat D_i - 2 \hat H_z$.
	Hence, the operators in Tab.\ \ref{tab:operators} only represent $L+2$ non-trivial extensions to the initial model.
	
	The numerical method presented in Sec.\ \ref{subsec:discover} finds all operators in the operator subspace $\operatorname{span}(\mathcal B_2)$ hosting the tower of scars for finite $L$ (up to length $L=14$ in our case).
	However, in principle, the scar states may not be eigenstates of these operators at larger $L$.
	Therefore, in Appendix \ref{apx:new_operators} we prove analytically for all even $L$ that the scar states remain eigenstates with equal energy spacing for all operators in Tab.\ \ref{tab:operators}.
	
	The method from Sec.\ \ref{subsec:discover} can be extended by including all 3-body terms to the basis $\mathcal B_3 = \mathcal B_2 \cup \{\hat \sigma_i^a \hat \sigma_{i+1}^b \hat \sigma_{i+2}^c \Big| a,b,c \in \{x,y,z\}, \enspace i\in \mathbb Z_L\}$.
	This results in a myriad of new operators -- including the first term from Eq.\ \eqref{eq:Iadecola_Scheter}.
	Hence, with a large enough operator basis, the numerical method fully recovers the original model.
	Since long-ranged many-body interactions are less relevant experimentally, we will not explore this possibility any further.
	
	In addition to hosting the tower of scar states $\{\ket{\mathcal{S}_n}\}$, the model from Eq.\ \eqref{eq:Iadecola_Scheter} also hosts another tower of scar states $\{\ket{\mathcal{S}_n'}\}$ \cite{Iadecola_Schecter}.
	However, by construction, the numerical method from Sec.\ \ref{subsec:discover} is only guaranteed to preserve $\{\ket{\mathcal{S}_n}\}$.
	Therefore, the second tower of scar states may be destroyed when extending the model with operators from Tab.\ \ref{tab:operators}.
	All scar states in the second tower $\{\ket{\mathcal{S}_n'}\}$ are, e.g., not eigenstates of the Hamiltonian $\hat H_0 + \sum_i d_i\hat D_i$ for general choices of the coefficients $d_i$.
	
	Finally, we remark that the effectiveness of this approach is highly non-trivial.
	For an eigenstate of a generic local Hamiltonian, it is unlikely for another local Hamiltonian to exist that shares the same eigenstate \cite{Qi2019}.
	Contrary to this, we find a large subspace of local Hamiltonians sharing a full tower of scar states.
	We attribute the effectiveness of our study to the analytical structure of the scar states, i.e.\ Eq.\ \eqref{eq:scars} and \eqref{eq:Q}.
	Our methods are not expected to be valuable starting from generic eigenstates but may be equally effective in other scarred models with similar amount of structure.

	\section{Many-body localization}\label{sec:MBL}
	In the last section, we determined a subspace of Hamiltonians with the scar states $\ket{\mathcal S_n}$ as eigenstates equally spaced in energy.
	Now, we study a concrete Hamiltonian from this subspace
	\begin{align}
		\hat H = \hat H_0 + \sum_{i=1}^L d_i \hat D_i,
		\label{eq:Hamiltonian}
	\end{align}
	 with $d_i$ chosen randomly from the uniform probability distribution $d_i \in [-W, W]$ where $W>0$ is the disorder strength.
	 The action of $\hat D_i$ is given by
	 \begin{align}
	 	\begin{split}
	 		\hat D_i &\ket{s_1\ldots s_i s_{i+1}\ldots s_L} \\
	 		&=
	 		\begin{cases}
	 			3 \ket{s_1\ldots s_i s_{i+1}\ldots s_L}, & \text{if } s_i= s_{i+1} = \, \uparrow \\
	 			- \ket{s_1\ldots s_i s_{i+1}\ldots s_L}, & \text{otherwise}
	 		\end{cases}
	 	\end{split}
	 	\label{eq:Hi}
	 \end{align}
	The operator $\hat D_i$ is related to the projection operators through $\hat D_i = 4 \hat P_i^\uparrow \hat P_{i+1}^\uparrow - \hat {\mathds 1}$ with $\hat P_i^\uparrow =  (\hat{\mathds 1} + \hat \sigma_i^z)/2$.	
	 We remark that Ref.\ \cite{Iadecola_Schecter} also observes that the operator $\hat P_i ^\uparrow \hat P_{i + 1}^\uparrow$ preserves the scar states.
	
	 This Hamiltonian is described by $L + 3$ parameters: $\lambda$, $\Delta$, $J$ and $\{d_i\}_{i=1}^L$.
	 The results presented in the following sections rely on numerical simulations for concrete values of $\lambda$, $\Delta$ and $J$.
	 While the results are calculated for specific values of these parameters, e.g.\ $\lambda = \Delta = J = 1$, one obtains qualitatively similar results for other values, e.g. $\lambda \neq \Delta \neq J$.
	
	 The model conserves the number of domain walls.
	 The dimension of the symmetry sector containing $N_\text{dw}$ domain walls is given by the binomial coefficient $2(\begin{smallmatrix}L \\ N_\text{dw} \end{smallmatrix})$.
	 We generally consider the largest symmetry sector with $N_\text{dw} = 2\lfloor L/4 \rfloor$ domain walls where $\lfloor \cdot \rfloor$ is the function rounding down to the nearest integer.
	 \subsection{Partial many-body localization}
	 \label{sec:partial MBL}
	 A physical system may transition to the MBL phase when disorder is introduced.
	 MBL is usually realized with the disorder term in the Hamiltonian acting uniquely on each basis state.
	 Consequently, a complete set of LIOMs emerge and all energy eigenstates are fully described by their eigenvalues of the LIOMs.
	
	 The situation is slightly different in our model because the disorder term $\sum_i d_i \hat D_i$ treats some basis states the same.
	The operator $\hat D_i$ is only sensitive to whether spins $i$ and $i+1$ are both up (it acts identically on states where spins $i$ and $i+1$ are $\downarrow \downarrow$, $\downarrow \uparrow$ or $\uparrow \downarrow$).
	 Therefore, the operator $\sum_i d_i \hat D_i$ has the same action on product states with all consecutive spin-ups placed identically.
	 We do not expect these to localize in the usual sense.
	 Instead, we anticipate the spectrum to separate into fully MBL eigenstates and partially localized eigenstates.
	
	This structure is most easily described when the product states $\ket{s_1 s_2 \ldots s_L}$ are relabeled to reflect the action of $\sum_i d_i \hat D_i$.
	 In this spirit, we define $\ket{N_\mathrm{dw}, \bm D, n}$ as a simultaneous eigenstate of the $\hat D_i$'s with eigenvalues $\bm D = (D_1, D_2, \ldots D_L)$ where $D_i \in \{-1, 3\}$.
	 We will refer to $\bm D$ as the disorder indices.
	 As discussed above, the state $\ket{s_1 s_2 \ldots s_L}$ is not fully described by $\bm D$ since multiple states can have the same eigenvalues.
	 Therefore, we further label the states by their number of domain walls $N_\mathrm{dw}$ and introduce a dummy index $n = 1, 2, \ldots, N_{\bm D}^{(N_\mathrm{dw})}$ to distinguish states with identical $N_\mathrm{dw}$ and $\bm D$.
	 For instance, if two states $\ket{s_1 s_2 \ldots s_L}$ and $\ket{s_1' s_2' \ldots s_L'}$ have the same number of domain walls $N_\mathrm{dw}$ and disorder indices $\bm D$, then they are relabeled as $\ket{N_\mathrm{dw}, \bm D, n}$ for $n=1, 2$.
	 Note that some labelings are invalid.
	 Consider the vector of eigenvalues $\bm D = (3, -1, 3, 3)$ for a small system $L=4$.
	 The ``$3$''s imply all spins are up, while the ``$-1$'' entail at least one spin is down.
	 In the following, we study a single symmetry sector and hence omit the $N_\mathrm{dw}$ index for clarity but reintroduce it in Secs.\ \ref{sec:stabilization} and \ref{sec:disorder induced scars} when studying multiple symmetry sectors at once.
	 	
	 Upon introducing strong disorder, we expect LIOMs to emerge which are localized on the operators $\hat D_i$ and energy eigenstates are characterized by their eigenvalues of the LIOMs.
	 Therefore, we expect the energy eigenstates to be close to linear combinations of product states with the same disorder indices
	 \begin{align}\label{eq:eigenstates}
	 	\ket{E_{\bm D, m}} \approx \sum_{n=1}^{N_{\bm D}} \alpha_{mn} \ket{\bm D, n}.
	 \end{align}
     with $\alpha_{mn} \in \mathbb R$ and $m = 1, 2, \ldots, N_{\bm D}$.
     This expression is an approximation rather than an equality due to an exponentially small overlap with states $\ket{\bm D', n}$ with different disorder indices $\bm D' \neq \bm D$.
     The special case $N_{\bm D} = 1$ corresponds to the disorder term acting uniquely on the basis state $\ket{\bm D, 1}$.
     We expect the corresponding energy eigenstate $\ket{E_{\bm D, 1}} \approx \ket{\bm D, 1}$ to be MBL.
	 For $N_{\bm D} > 1$, the states $\{\ket{E_{\bm D, m}}| m = 1, 2, \ldots, N_{\bm D} \}$ are only partially MBL since the LIOMs do not fully describe each state and all additional structure is captured by the extra index $m$.
	
\begin{figure}
		\includegraphics{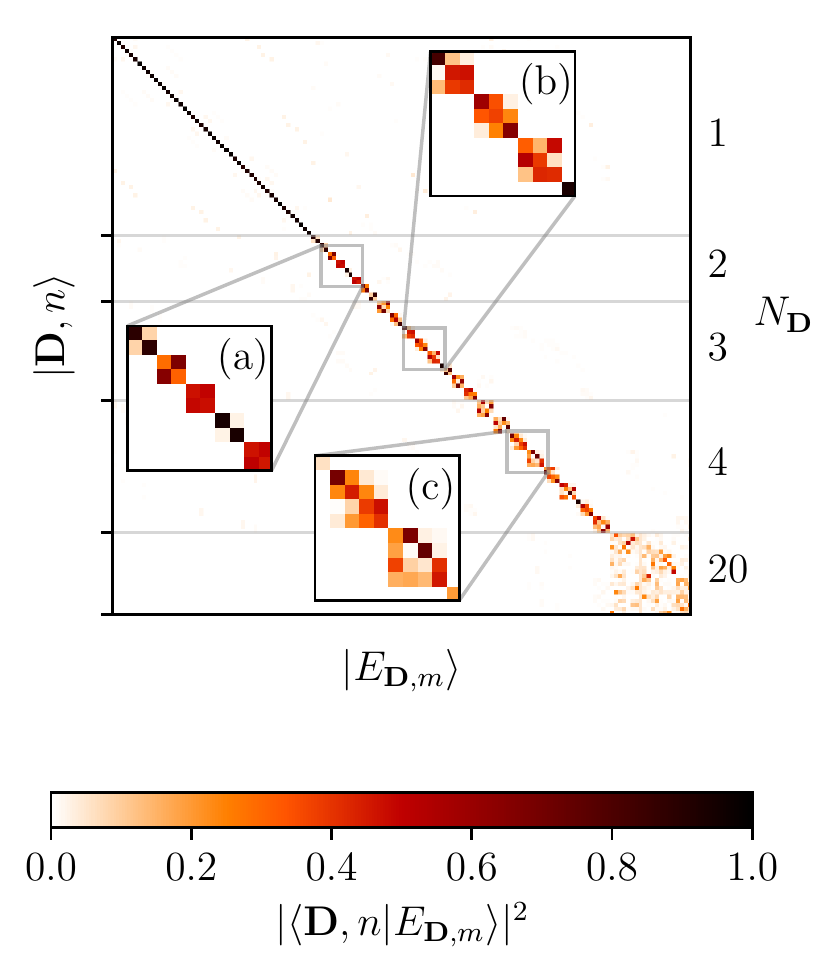}
		\caption{
			The norm squared overlap of the energy eigenstates with the product states $|\braket{\bm D, n| E_{\bm D, m}}|^2$ for system size $L=8$, disorder strength $W=10$ and parameters $\lambda = \Delta = J = 1$.
			The color of pixel $(i,j)$ displays the overlap between the $i$'th product state and the $j$'th eigenstate.
			The product states are sorted into ascending order according to $N_{\bm D}$.
			The second axis on the right hand side groups the product states according to $N_{\bm D}$.
			The insets show eigenstates with significant weight on (a) two, (b) three and (c) four product states.
			The figure verifies that all energy eigenstates are approximately linear combinations of product states with the same disorder indices. }
		\label{fig:subspaces}
	\end{figure}

	The above considerations are verified in numerical simulations by considering a system of size $L=8$ at strong disorder $W = 10$.
	Figure \ref{fig:subspaces} illustrates the norm squared overlap of all energy eigenstates $\ket{E_{\bm D, m}}$ with the product states $\ket{\bm D, n}$.
	The $(i, j)$-th pixel displays the norm squared overlap between the $i$-th product state and $j$-th energy eigenstate.
	The product states on the second axis are sorted according to $N_{\bm D}$.
	The energy eigenstates are reordered to allow the diagonal shape in Fig.\ \ref{fig:subspaces}.
	In the upper left corner of Fig.\ \ref{fig:subspaces}, each eigenstate has high overlap with a single product state.
	Numerical analysis reveals that these product states exactly coincide with those being fully described by their disorder indices, i.e.\ $N_{\bm D} = 1$.
	These results support the claim that such eigenstates fully localize.
	The next eigenstates shown in Fig.\ \ref{fig:subspaces}(a) each has significant overlap with exactly two product states of the same disorder indices.
	The pattern continues: we find eigenstates that are linear combinations of Fig.\ \ref{fig:subspaces}(b) three, Fig.\ \ref{fig:subspaces}(c) four, and (bottom right corner) twenty product states.
	In each case, the product states have the same disorder indices and hence correspond to $\{\ket{\bm D, n}|n=1, 2, \ldots, N_{\bm D}\}$ for $N_{\bm D} = 3$, $4$, $20$.
	These observations are not restricted to $L=8$, but seem universal at all system sizes.
	For larger system sizes, the number and sizes of the blocks increase.
	Finally, we note that the scar state within the considered symmetry sector is located in the block $N_{\bm D} = 20$ in Fig.\ \ref{fig:subspaces}.
	The scar state is generally an equal weight linear combination of product states with the maximum $N_{\bm D}$.
	This fact will play an important role when we explore the system dynamics in Sec.\ \ref{sec:stabilization}.
	
\begin{figure}
		\includegraphics{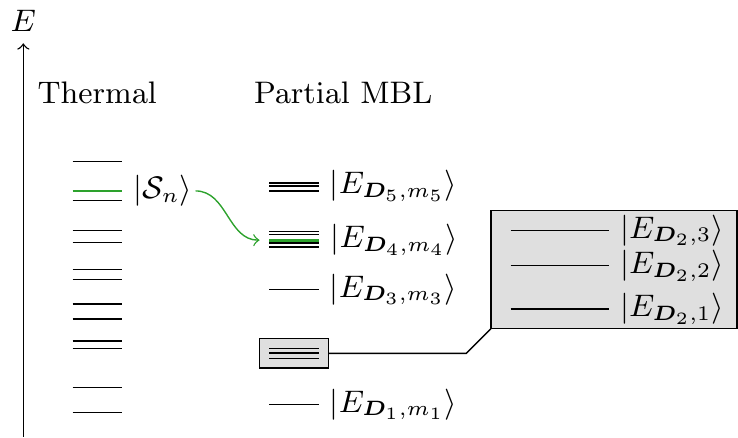}
		\caption{
			Sketch of the spectrum in the thermal phase (left) and in the partially localized phase (right).
			In the thermal phase, the energy levels follow the Wigner-Dyson surmise.
			As disorder is introduced, the spectrum experiences partial localization.
			Eigenstates with similar indices $\bm D$ are near degenerate and the spectrum forms clusters of such eigenstates.
			The scar state lies in the largest of these clusters.
		}
		\label{fig:energy_scales}
	\end{figure}

	Next, we discuss how the eigenstates are distributed in energy.
	The magnetization $M_{\bm D} = \sum_i \sigma_i^z$ of a product state $\ket{\bm D, n}$ is fixed by the symmetry sector $N_\mathrm{dw}$ and disorder indices $\bm D$.
	Likewise, the number $\mathcal N^{(\uparrow\uparrow,\downarrow\downarrow)}_{\bm D}$ of adjacent spins pointing in the same direction ($\uparrow\uparrow$ or $\downarrow\downarrow$) and the number $\mathcal N^{(\uparrow\downarrow, \downarrow\uparrow)}_{\bm D}$ of adjacent spins pointing in opposite directions  ($\uparrow\downarrow$ or $\downarrow\uparrow$) are also fully determined.
	Therefore, the terms $\Delta \sum_i \hat \sigma_i^z$, $J \sum_i \hat \sigma_i^z \hat \sigma_{i+1}^z$ and $\sum_id_i \hat D_i$ have the same action on all product states with the same number of domain walls and disorder indices: $\{\ket{\bm D, n}| n = 1, 2, \ldots, N_{\bm D}\}$.
	At strong disorder, the energy of an eigenstate is approximately $E_{\bm D, m} \approx \Delta M_{\bm D} + J(\mathcal N_{\bm D}^{(\uparrow\uparrow, \downarrow\downarrow)} - \mathcal N_{\bm D}^{(\uparrow\downarrow, \downarrow\uparrow)}) + \sum_i d_i D_i$ with a small correction that depends on the value of the $m$ index.
	The slight additional contribution originates from the term $\sum_i \lambda(\hat \sigma_i^x - \hat \sigma_{i-1}^z \hat \sigma_i^x \hat \sigma_{i+1}^z)$ and scales with $\lambda$.
	Consequently, at large disorder, the set of eigenstates $\{\ket{E_{\bm D, m}}|m = 1, 2, \ldots, N_{\bm D} \}$ are near degenerate and form clusters.
	A scar state resides in the largest of these clusters in all symmetry sectors.
	Figure \ref{fig:energy_scales} illustrates the spectral structure.
	Note that Fig.\ \ref{fig:energy_scales} is highly idealized to highlight the structure described above.
	In practice, it is highly likely for two or more clusters to overlap making the structure less apparent.
	
	\subsection{Spectral statistics}\label{subsec:spectral statistic}
	The distribution of energy gaps distinguishes the thermal and MBL phases.
	Let $E_i$ be the energies of the Hamiltonian in ascending order  and $\delta_i = E_{i+1}-E_i \geq 0$ the $i$-th energy gap.
	In the thermal phase, the number of energy levels in an interval $[E, E + \Delta E]$ is known to follow the Wigner-surmise \cite{Luca, Guhr}.
	In particular, it follows the Gaussian orthogonal ensemble (GOE) since the model in Eq.\ \eqref{eq:Hamiltonian} is time-reversal invariant.
	On the other hand, the number of energy levels in an interval follows the Poisson distribution in the MBL phase.
	Since our model only partially localizes, we review how the Poisson distribution accurately describes the MBL phase and investigate the validity of these arguments in our model.
	Consider two adjacent eigenstates with energies $E_i$ and $E_{i+1}$.
	At large disorder, the energy of these states are dominated by the disorder term $\sum_i d_i \hat D_i$.
	If the states have different disorder indices $\ket{E_{\bm D, m}}$ and $\ket{E_{\bm D', m'}}$, then their energies originate from different linear combinations of the random numbers $d_i$: $\sum_i d_i D_i \approx \sum_i d_i D'_i$ with $D_i \neq D_i'$ for some $i$'s.
	Consequently, the eigenstates ``arrive'' at this energy independently of each other and hence follow the Poisson distribution.
	These arguments are no longer valid when two adjacent eigenstates have the same disorder indices and different $m$ indices. In this case, we expect the level spacing distribution to follow GOE.
	Thus, the distribution of energy levels still identifies the transition to partial localization if we only consider level spacings between eigenstates of different disorder indices.
	
	Instead of working directly with the level spacing distribution, it is convenient to analyze the adjacent gap ratio since it removes the need for unfolding the spectrum \cite{Guhr, Abdulmagd}.
	The adjacent gap ratio is defined by \cite{Huse2007}
	\begin{align}
		r_i =  \frac{\operatorname{min}(\delta_i, \delta_{i+1})}{\operatorname{max}(\delta_i, \delta_{i+1})}.
	\end{align}
	This quantity is bounded by the interval $r_i \in [0, 1]$ and follows the distributions below in the thermal and MBL phases respectively \cite{Atas}
	\begin{subequations}\label{eq:adjacent_gap_ratio_distributions}
	\begin{align}
	P_\mathrm{GOE}(r) &= \frac{27}{4} \frac{r(1 + r)}{(1 + r + r^2)^{5/2}}, \label{eq:adjacent_gap_ratio_GOE}\\
	P_\mathrm{Poisson}(r) &= \frac{2}{(1 + r)^2}. \label{eq:adjacent_gap_ratio_Poisson}
	\end{align}
	\end{subequations}
	The mean values of the distributions in Eq.\ \eqref{eq:adjacent_gap_ratio_distributions} are given by $\braket{r}_\text{GOE} = 2(2 - \sqrt 3) \approx 0.536$ and $\braket{r}_\text{Poisson} = 2\ln 2 - 1 \approx 0.386$.
	\begin{figure}
		\centering
		\includegraphics[width=\linewidth]{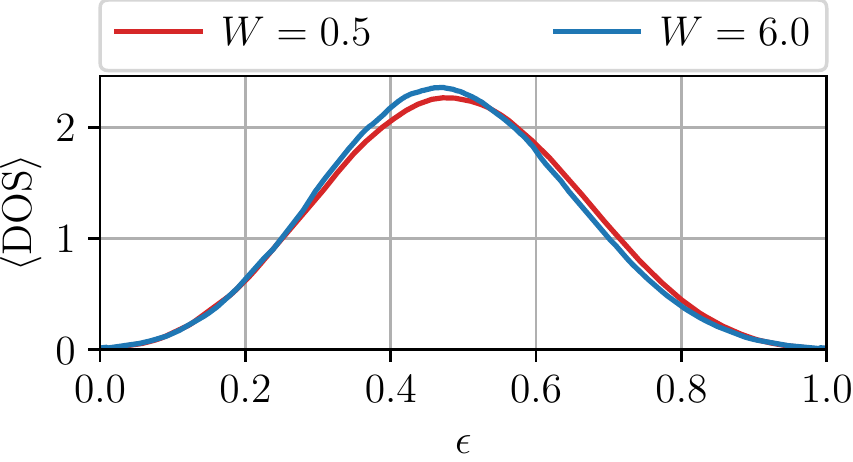}
		\caption{
			Disorder averaged density of states $\braket{\mathrm{DOS}}$ as a function of normalized energy $\epsilon$ for system size $L = 14$ at weak disorder $W = 0.5$ and strong disorder $W = 6.0$.
			We average the density of states over $10^3$ disorder realizations.
		}
		\label{fig:DOS}
	\end{figure}
	\begin{figure*}
		\centering
		\includegraphics{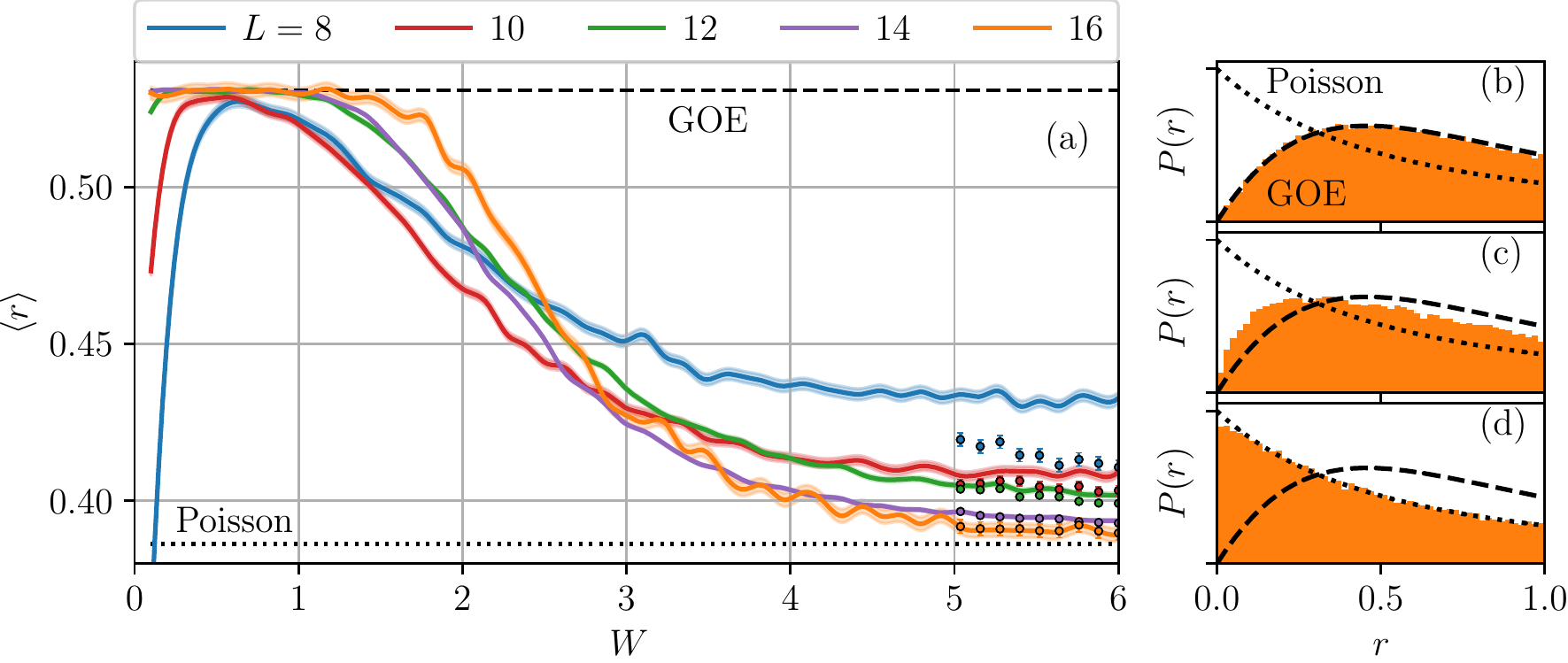}
		\caption{
			(a) Mean adjacent gap ratio $\braket r$ (solid line) as a function of disorder strength $W$ for different system sizes $L$ with parameters $\lambda = \Delta = J = 1$.
			The shaded areas display two standard deviations on the estimate of $\braket r$ when assuming a Gaussian distribution of data.
			For $L = 8$, the adjacent gap ratio is averaged over $2\times 10^3$ disorder realizations, for $L = 10, 12, 14$ we use $10^3$ disorder realizations and for $L = 16$ we use $500$ disorder realizations.
			For system sizes $L=8, 10, 12, 14$, we average over all energies $E_i \in [E^{(q=1/3)}, E^{(q=2/3)}]$ where $E^{(q)}$ is the $q$-th quantile.
			For system size $L = 16$, we average over the $10^3$ energies closest to $(E_\text{min} +E_\text{max}) / 2$ where $E_\text{min}$ and $E_\text{max}$ are the smallest and largest energies in the spectrum.
			At low disorder $0 \lesssim W \lesssim 1$, the system is thermal and $\braket{r}$ coincides with the Gaussian orthogonal ensemble $\braket{r}_\mathrm{GOE} \approx 0.536$ (upper dashed line). At strong disorder $5 \lesssim W$, the mean adjacent gap ratio agrees with the Poisson distribution $\braket{r}_\mathrm{Poisson} \approx 0.386$ (lower dotted line). The agreement between data and the GOE and Poisson values improves with system size.
			Additionally, the transition from the thermal phase to partial localization happens more rapidly as a function of disorder strength for larger system sizes.
			The figure also illustrates the mean adjacent gap ratio when only averaging over neighboring energy eigenstates with different disorder indices (dots).
			The errorbars show two standard deviations on the estimate of the mean.
			This average coincides with the naive calculation at large system sizes.
			The figure also shows the adjacent gap ratio distribution for $L = 16$ at (b) weak disorder $W = 0.46$, (c) intermediate disorder strength $W = 2.27$ and (d) strong disorder $W = 6$.
			These plots include the distributions Eq.\ \eqref{eq:adjacent_gap_ratio_GOE} (dashed curve) and Eq.\ \eqref{eq:adjacent_gap_ratio_Poisson} (dotted curve).
			The data agrees with Eq.\ \eqref{eq:adjacent_gap_ratio_GOE} at weak disorder and transitions to the distribution \eqref{eq:adjacent_gap_ratio_Poisson} at strong disorder.
		}
		\label{fig:level_spacing_statistic}
	\end{figure*}
	Figure \ref{fig:level_spacing_statistic}(a) illustrates the mean adjacent gap ratio as a function of disorder strength for different system sizes.
	We average the adjacent gap ratio over $2 \times 10^3$ disorder realizations for $L=8$, $10^3$ disorder realizations for $L=10, 12, 14$ and $500$ disorder realizations for $L = 16$.
	For each disorder realization, we average over all energies in the interval $E_i \in [E^{(q=1/3)}, E^{(q=2/3)}]$ where $E^{(q)}$ is the $q$-th quantile of the energy distribution for the current disorder realization.
	For system size $L = 16$, we average over the $10^3$ energies closest to $(E_\text{min} +E_\text{max}) / 2$ where $E_\text{min}$ and $E_\text{max}$ are the smallest and largest energies in the spectrum.
	The errorbars indicate two standard deviations of the average when assuming a Gaussian distribution.
	The disorder averaged density of states (DOS) is illustrated in Fig.\ \ref{fig:DOS} as a function of normalized energy $\epsilon$ for weak disorder $W = 0.5$ and strong disorder $W = 6.0$.
	This figure illustrates that the energies in the interval $[E^{(q=1/3)}, E^{(q=2/3)}]$  and closest to $(E_\text{min} +E_\text{max}) / 2$ generally correspond to high density of states.
	
	As discussed above, the distribution of adjacent gap ratios only converges to Eq.\ \eqref{eq:adjacent_gap_ratio_Poisson} if the analysis is restricted to adjacent energy levels with different disorder indices.
	In practice, however, it is unlikely for two neighboring eigenstates to have the same disorder indices.
	Furthermore, the likelihood of neighboring eigenstates having the same disorder indices decreases rapidly with system size.
	With this in mind, we study the mean adjacent gap ratio using all eigenstates in the central third of the spectrum.
	We verify the considerations above by also computing the mean adjacent gap ratio using only adjacent eigenstates with different disorder indices at large disorder.
	For each energy gap $\delta_i = E_{i + 1} - E_i$, we inspect the eigenstates $\ket{E_{\bm D, m}}$ and $\ket{E_{\bm D', m'}}$ corresponding to the energies $E_i$ and $E_{i + 1}$.
	At large disorder, the disorder indices $\bm D$ are accurately determined by computing which $\bm D$ yields $\sum_{n=1}^{N_{\bm{D}}} |\braket{\bm D, n | E_{\bm D, m}}|^2 \approx 1$.
	The mean of the adjacent gap ratio is then restricted to energy gaps with $\bm D \neq \bm D'$.
	For small system sizes, there is a large difference between the two methods, but the difference is seen to be small for large systems.
	
	The mean adjacent gap ratio agrees well with the GOE value at weak disorder $0 \lesssim W \lesssim 1$.
	The discrepancy between the GOE value and data for small system sizes $L$ and small non-zero $W$ is caused by the model possessing additional symmetries at $W = 0$, i.e.\ translational and inversion symmetry.
	The proximity to a model with further symmetries causes the adjacent gap ratio to differ from $\braket{r}_\text{GOE}$.
	This deviation decreases with increasing system size.
	
	As the disorder strength is increased, the mean adjacent gap ratio decreases and ultimately approaches the Poisson value at $5 \lesssim W$.
	The agreement of data with the GOE and Poisson values improves with increasing system size and the transition between the thermal and localized phase becomes steeper for larger systems.
	
	Figures\ \ref{fig:level_spacing_statistic}(b)-(d) illustrate the adjacent gap ratio distribution at (b) weak disorder $W = 0.46$, (c) intermediate disorder strength $W = 2.27$ and (d) strong disorder $W = 6$.
	The figures display the distributions in Eq.\ \eqref{eq:adjacent_gap_ratio_distributions} for comparison.
	As expected, the data agrees with Eq.\ \eqref{eq:adjacent_gap_ratio_GOE} at weak disorder and \eqref{eq:adjacent_gap_ratio_Poisson} at strong disorder.
	Figure \ref{fig:level_spacing_statistic} indicates the system transitions from the thermal phase to being partially localized as disorder is introduced.
	
	\subsection{Bipartite entanglement entropy}\label{subsec:entropy}
	In this section, we further verify the transition from the thermal phase to partial localization by studying the bipartite entanglement entropy.
	We separate the system into a left part $\mathcal L$ containing the first $L/2$ sites and a right part $\mathcal R$ containing the remaining sites.
	The reduced density matrix of the left part is obtained by tracing out the right part
	\begin{align}
		\rho_{\mathcal L} = \operatorname{Tr}_{\mathcal R}(\rho)
	\end{align}
	where $\rho$ is the density matrix of the full system and $\operatorname{Tr}_\mathcal{R}(\cdot )$ is the partial trace over $\mathcal R$.
	The entanglement entropy between the left and right halves is given by
	\begin{align}
		S = - \operatorname{Tr}_{\mathcal L}\big[\rho_{\mathcal L} \ln(\rho_{\mathcal L})\big].
	\end{align}

	In the thermal phase, we expect eigenstates near the center of the spectrum to display volume-law scaling with system size.
	Specifically, the entropy is approximately described by the Page value $S_\text{Page} =[L\ln(2) - 1]/2$  \cite{Page1993}.
	On the other hand, the entanglement entropy displays area-law scaling for MBL eigenstates \cite{Bauer2013}.
	While some eigenstates in our model are fully MBL, others are only partially localized.
	Hence, the precise scaling behavior of the entanglement entropy is not clear.
	Nonetheless, we expect the entropy of partially localized eigenstates to grow slower with system size than thermal eigenstates and we use the entropy to identify the onset of partial localization.
	
\begin{figure}
		\centering
		\vspace{5mm}
		\includegraphics{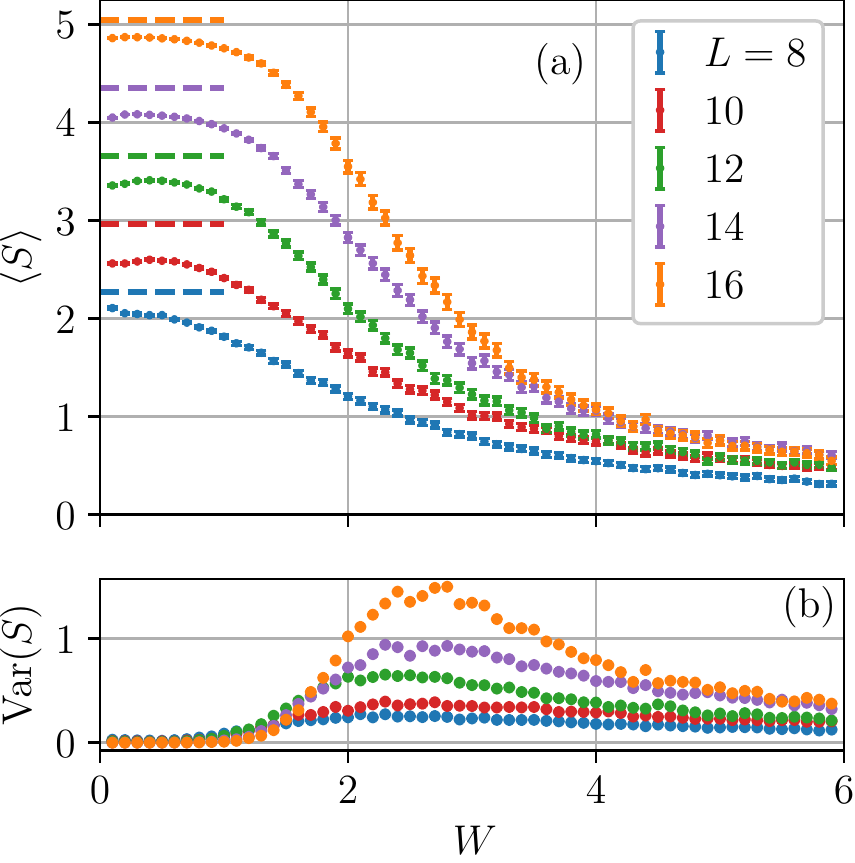}
		\caption{
			(a) Average bipartite entanglement entropy of the eigenstate closest to the center of the spectrum $\braket S$ as a function of disorder strength $W$ for different system sizes $L$.
			The entropy is averaged over $10^3$ disorder realizations with system parameters $\lambda = \Delta = J = 1$.
			Errorbars display two standard deviations on the estimate of average entropy assuming a Gaussian distribution.
			At low disorder, the entropy displays volume-law scaling with system size and approaches the Page value (dashed lines) as expected in the thermal phase.
			At large disorder, the entropy follows area-law scaling with system size.
			(b) Variance of bipartite entanglement entropy of the eigenstate closest to the center of the spectrum.
			The variance is computed from $10^3$ disorder realizations.
			As the disorder strength is increased, the variance displays a sudden peak.
			This indicates a transition from the thermal phase to partial localization.
			The peak becomes higher at larger system sizes.
		}
		\label{fig:entropy}
	\end{figure}

	Figure \ref{fig:entropy}(a) shows the entropy of the eigenstate with energy closest to $(E_\text{min} + E_\text{max}) / 2$ as a function of disorder strength $W$ for different system sizes $L$.
	Each data point represents the average entropy over $10^3$ disorder realizations with errorbars displaying two standard deviations of the mean when assuming a Gaussian distribution.
	For low disorder, the entanglement entropy scales linearly with the system size and hence agrees with the expected volume-law scaling in the thermal phase.
	Additionally, the entropy approaches the Page value with increasing system size.
	At large disorder, the entropy seems to be roughly independent of system size.
	Thus, the scaling of entropy is consistent with area-law for partially localized eigenstates.
		
	The sudden shift in scaling behavior of the entropy verifies the transition from the thermal phase to partial localization at strong disorder.
	The transition point is identified by analyzing the variance of entanglement entropy.
	Figure \ref{fig:entropy}(b) illustrates the sample variance of the entropy over $10^3$ disorder realizations.
	The variance displays a peak when the system transitions from volume-law to area-law scaling.
	
	\section{distinguishable features of scar states in a partially localized background}\label{sec:features}
	Scar states are commonly distinguished from a thermal background by their low entanglement and oscillatory dynamics.
	In this section, we show that oscillatory dynamics can also be utilized to distinguish scar states from a partially localized background, while entanglement entropy turns out not to be an effective tool to identify the scar states. We also find that although the Schmidt gap can distinguish the scar states from fully MBL states, it does not distinguish the scar states from partial MBL states.
		
	\subsection{Entanglement entropy}\label{subsec:features_S}
	The entanglement entropy of the scar states scales logarithmically with system size \cite{Iadecola_Schecter}, while thermal states display volume-law scaling.
	Therefore, the entanglement entropy provides a way to identify the scar states in a thermal background.
	\begin{figure}
		\centering
		\vspace{5mm}
		\includegraphics{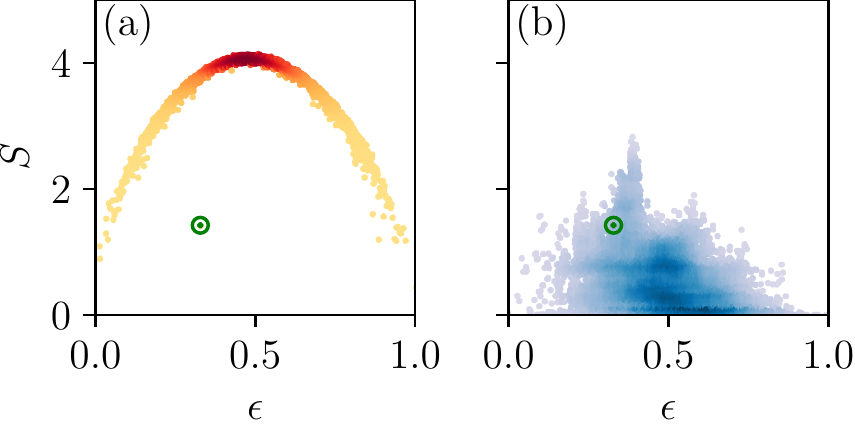}
		\caption{
			The entanglement entropy $S$ as a function of normalized energy $\epsilon = (E - E_\text{min})/(E_\text{max} - E_\text{min})$ where $E_\text{min}$ and $E_\text{max}$ are the smallest and largest energies in the spectrum.
			Lighter (darker) colors indicate lower (higher) density of points.
			(a) We consider a thermal system of size $L=14$, disorder strength $W=0.5$ and system parameters $\lambda = \Delta = J = 1$.
			In the thermal phase, the energy eigenstates form a narrow band with maximum at the center of the spectrum.
			The scar state (inside the green ring) is easily identified since it appears isolated below the curve.
			(b) We consider a partially localized system at strong disorder $W = 6$. The energy eigenstates are spread out at low entropy with the scar state embedded among them.
			The entanglement entropy is hence not an effective tool to distinguish the scar state from a partially localized background.
		}
		\label{fig:SE}
	\end{figure}
	Figure \ref{fig:SE}(a) illustrates the entropy as a function of energy of a thermal system with size $L=14$ and disorder strength $W = 0.5$.
	The thermal states form a narrow arc with maximum in the middle of the spectrum while the scar state appears as an outlier at much lower entropy.
	The situation is different in a partially localized background.
	Figure \ref{fig:SE}(b) illustrates the entropy as a function of energy at strong disorder $W = 6$.
	As discussed above, partially localized eigenstates are weakly entangled making it difficult to identify the scar state.
	We conclude that entanglement entropy is an ineffective tool for distinguishing scar states from a partially localized background.
	
	\subsection{Schmidt gap}\label{subsec:schmidt_gap}
	The Schmidt gap effectively distinguishes thermal eigenstates from MBL eigenstates \cite{Gray2018}.
	Here we find that the Schmidt gab distinguishes the scar states from MBL states, but not from thermal or partial MBL states.
	Similar to Sec.\ \ref{subsec:entropy}, we consider the reduced density matrix of the first $L/2$ sites $\rho_\mathcal{L}$.
	Let $\{\Lambda_i\}$ be the eigenvalues of $\rho_{\mathcal L}$ in descending order.
	The Schmidt gap is given by
	\begin{align}
		\Delta_{\mathrm{SG}} = \Lambda_1 - \Lambda_2
	\end{align}
	and is bounded by the interval $0 \leq 	\Delta_{\mathrm{SG}} \leq 1$.
	
\begin{figure}
		\centering
		\vspace{5mm}
		\includegraphics[width=\linewidth]{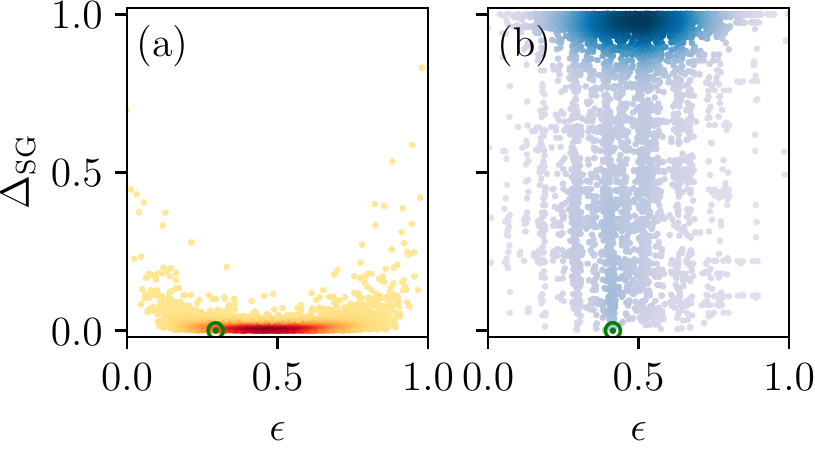}
		\caption{
			The Schmidt gap $\Delta_{\mathrm{SG}}$ for a single disorder realization with system size $L=14$ as a function of normalized energy $\epsilon$ at (a) weak disorder $W = 0.5$ and (b) strong disorder $W = 6$.
			The color illustrates the  density of points with darker (lighter) colors signifying higher (lower) density of points.
			(a) Thermal eigenstates close to the middle of the spectrum are highly entangled and the Schmidt gap vanishes.
			The Schmidt gap of the scar state (inside green circle) is also close to zero, and the scar state is hence indistinguishable from the thermal background.
			(b) The Schmidt gap of fully MBL eigenstates are close to one and the Schmidt gap of partial MBL eigenstates can take any value between zero and one.
			The scar state is hence distinguishable from fully MBL eigenstates but indistinguishable from partial MBL eigenstates.
			}
		\label{fig:schmidt-gap}
	\end{figure}

	Figure \ref{fig:schmidt-gap} illustrates the Schmidt gap for each energy eigenstate in a single disorder realization at (a) weak disorder $W = 0.5$ and (b) strong disorder $W = 6$.
	In the thermal phase, an eigenstate in the middle of the spectrum is highly entangled and the eigenvalues $\{\Lambda_i\}$ have similar magnitude.
	Consequently, the Schmidt gap is expected to vanish in accordance with Fig.\ \ref{fig:schmidt-gap}(a).
	The Schmidt gap of the scar state is close to zero and hence cannot be distinguished from the thermal background.
	Fully MBL eigenstates are localized on a single product state and the Schmidt gap is close to one.
	These eigenstates are visible in Fig.\ \ref{fig:schmidt-gap}(b) as the high density of points close to one.
	Partial MBL eigenstates are localized on multiple product states and no predictions can be made about the value of the Schmidt gap.
	In Fig.\ \ref{fig:schmidt-gap}(b), these eigenstates are scattered across all values $\Delta_\mathrm{SG} \in \, [0, 1]$.
	Since the Schmidt gap of the scar state is close to zero, it is easily distinguished from fully MBL eigenstates.
	However, it is not possible to distinguish the scar state from partial MBL eigenstates since they may have a Schmidt gap close to zero.
	
	\subsection{Fidelity}\label{subsec:fidelity}
	States initialized in the scar subspace distinguish themselves from a thermal background by displaying persistent dynamic revivals.
	We now show that this behavior also enables the identification of scar states from a partially localized background.
	We quantify the dynamics of quantum systems by the fidelity $F(t)$.
	Let $\ket{\psi(0)}$ be the initial state and $\ket{\psi(t)} = e^{-i\hat Ht}\ket{\psi(0)}$ the time evolved state.
	The fidelity is given by
	\begin{align}\label{eq:F1}
		F(t) = |\braket{\psi(0)|\psi(t)}|^2.
	\end{align}
	The time evolution of fidelity is most clearly understood by considering the overlap of the initial state with all energy eigenstates.
	Let $\ket{\phi_i}$ be the $i$-th energy eigenstate with corresponding energy $E_i$ and let $c_i$ be the inner product between the $i$-th energy eigenstate and the initial state $c_i = \braket{\phi_i|\psi(0)}$.
	The relation between fidelity and the expansion coefficients $c_i$ is highlighted by rewriting the fidelity according to
	\begin{align}\label{eq:F2}
		F(t) &= \sum_i |c_i|^4 + \sum_{i\neq j}|c_i|^2 |c_j|^2 e^{i(E_i - E_j)t}
	\end{align}
	It is clear from this expression that the dynamics of fidelity is sensitive to the distribution of $|c_i|^2$.
	We generally display this distribution along with the fidelity for clarity.
	
\begin{figure}
		\centering
		\includegraphics{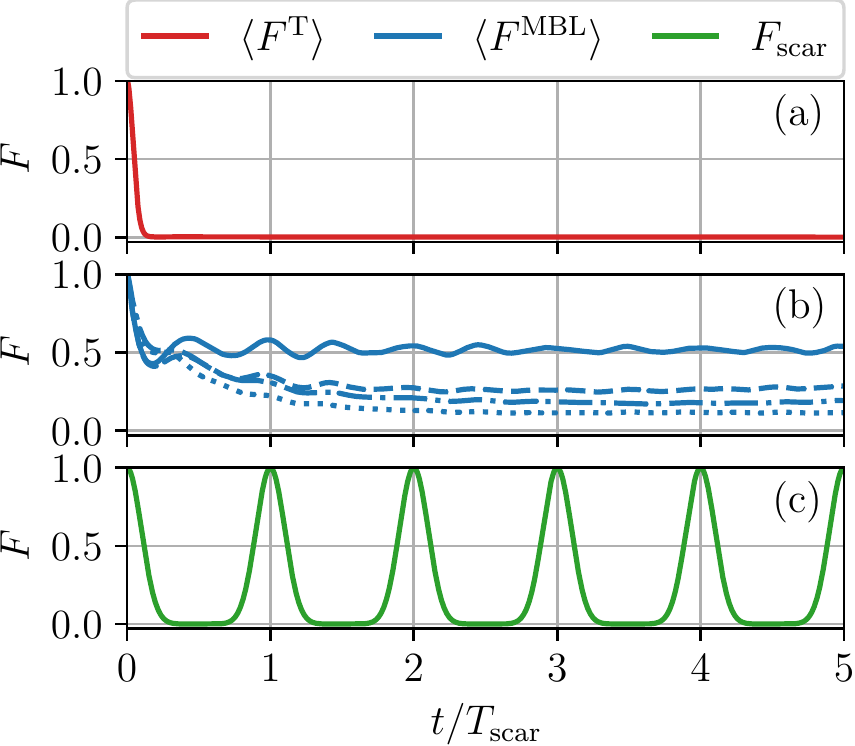}\\
		\vspace{3mm}
		\includegraphics{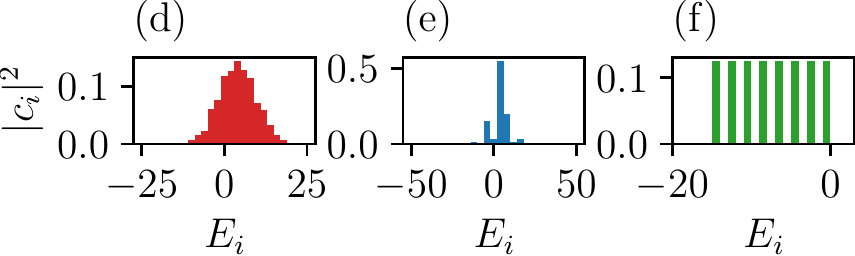}
		\caption{
			(a) The average fidelity of a random product state in a thermal system at disorder strength $W = 0.5$.
			(b) The average fidelity in a partially localized system at disorder strength $W = 10$.
			The system is initialized in a product state which fully localizes (solid line).
			For comparison, the system is initialized in a random product state which only partially localizes $\ket{\psi(0))} = \ket{\bm D, n}$ with $N_{\bm D} = 5$ (dashed line), $10$ (dashed dotted line) and $35$ (dotted line).
			(c) The system is initialized in the scar subspace at any disorder strength.
			The average fidelity is in all cases calculated over $10^3$ disorder realizations.
			The bottom panel displays the distribution of expansion coefficients $|c_i|^2$ across energy in a single disorder realization.
			(d) For the thermal phase $W = 0.5$.
			(e) For partial MBL $W = 10$ with initial state $\ket{\psi(0)} = \ket{\bm D, n}$ for $N_{\bm D} = 5$.
			(f) For the initial state $\ket{\psi(0)} = \ket{\psi_\text{scar}}$ residing in the scar subspace.
		}
		\label{fig:fidelity}
	\end{figure}

	We demonstrate the different dynamical behavior of the thermal and partial MBL phases by initializing a system of size $L = 14$ in a product state.
	First, we consider a thermal system at disorder strength $W = 0.5$.
	The initial state is chosen as a random product state with all product states having the same probability of being drawn.
	We ensure the initial state resides outside the scar subspace by drawing a new product state if the first has non-zero overlap with a scar state.
	We consider $10^3$ disorder realizations and draw a  random product state in each realization.
	In the $i$-th realization, the fidelity is computed as a function of time $F_i(t)$ and Fig.\ \ref{fig:fidelity}(a) shows the average fidelity $\braket{F(t)} = 10^{-3} \sum_{i=1}^{10^{3}} F_i(t)$ over all realizations.	
	Figure \ref{fig:fidelity}(d) shows the expansion coefficients $|c_i|^2$ of a single disorder realization following the Gaussian distribution as expected \cite{Santos2012a, Santos2012b}.
	Since the initial state has large overlap with many different eigenstates, the second sum in Eq.\ \eqref{eq:F2} rapidly vanishes due to cancellation between terms with different phase factors.
	As a consequence, the fidelity quickly decreases and saturates at $F_i(t) \approx \sum_i|c_i|^4 \approx 0$ at long times $T_\text{scar} \ll t $ for all disorder realizations.
	These considerations agree with the observed time evolution of the average fidelity in Fig.\ \ref{fig:fidelity}(a) which rapidly decreases to a value near zero.
	
	Next, we consider the same setup when the system is partially localized at large disorder $W = 10$.
	As discussed in Sec.\ \ref{sec:partial MBL}, the spectrum separates into fully MBL eigenstates and partially localized eigenstates.
	Consequently, the dynamics depend greatly on the initial state.
	The solid blue line in Fig.\ \ref{fig:fidelity}(b) is the average fidelity over $10^3$ disorder realizations when initialing the system in a random product state which fully localizes, i.e.\ $\ket{\psi(0)} = \ket{\bm D, n}$ with $N_{\bm D} = 1$.
	Fully MBL eigenstates have significant overlap with only one product state, and the average fidelity remains far from zero at all times as observed in Fig.\ \ref{fig:fidelity}(b).
	We note that a stronger disorder strength is needed to achieve MBL in larger systems.
	Therefore, the average fidelity saturates significantly below unity in Fig.\ \ref{fig:fidelity}(b) even though all product states with $N_{\bm D} = 1$ in Fig.\ \ref{fig:subspaces} are near identical to an energy eigenstate.
	The average fidelity saturates closer to unity at larger disorder strengths.
	
	When the initial state is chosen as a product state that only partially localizes, it has significant overlap with multiple eigenstates.
	Consequently, the average fidelity drops closer to zero as illustrated by the dashed and dotted curves in Fig.\ \ref{fig:fidelity}(b).
	For these curves, we choose the initial state randomly as $\ket{\psi(0)} = \ket{\bm D, n}$ with $N_{\bm D} = 5, 10$ and $35$.
	These initial states have significant support on up to $N_{\bm D}$ eigenstates causing the average fidelity to decrease with increasing $N_{\bm D}$.
	Figure \ref{fig:fidelity}(e) illustrates the distribution of $|c_i|^2$ for a single disorder realization for a random initial state $\ket{\psi(0)} = \ket{\bm D, n}$ with $N_{\bm D} = 5$.
	The distribution is more sparse than the thermal case.
	
	Finally, we consider the initial state being a linear combination of scar states.
	For a complex number $\xi \in \mathbb C$, we consider the state
	\begin{align}\label{eq:init_scar}
		\ket{\xi} = \frac{1}{{\mathcal N}_\xi} \prod_{i=1}^{L} \Big[1 + (-1)^i \xi {\hat P}^\downarrow_{i - 1} {\hat \sigma}_i^+ {\hat P}^\downarrow_{i + 1} \Big] \ket{\downarrow \downarrow \ldots \downarrow}
	\end{align}
	where $\mathcal N_\xi$ is a normalization constant.
	This special state is area-law entangled and the ground state of a simple Hamiltonian \cite{Iadecola_Schecter}.
	We choose the initial state $\ket{\psi_\text{scar}} = \ket{\xi = 1}$ which fully resides in the scar subspace.
	When the initial state is chosen within the scar subspace, the equal energy spacing causes the fidelity to display persistent periodic revivals.
	Revivals occur at times $t_\ell = T_\text{scar}\ell = 2\pi \ell/\Delta E_\mathrm{scar}$ where $\ell\in \mathbb N$ and $\Delta E_\mathrm{scar}$ is the energy spacing between consecutive scar states.
	Figure \ref{fig:fidelity}(c) illustrates the fidelity of this initial state and Fig.\ \ref{fig:fidelity}(f) shows the distribution of the expansion coefficients.
	
	In the thermal phase, states initialized respectively inside and outside the scar subspace behave differently.
	The fidelity of states outside the scar subspace quickly drops to zero, while any linear combination of scar states display persistent revivals.
	In our analysis, we specifically initialized the system as a product state, but the same conclusions hold for generic linear combinations of product states.
	In a partially localized background, the average fidelity distinguishes between states with support inside and outside the scar subspace.
	The average fidelity of partially localized states saturates while scar states display revivals.
	Again, our analysis concerns the special case of initializing the system as a random product state.
	If instead the initial state is a generic linear combination of a large number of product states, the second term of Eq.\ \eqref{eq:F2} will generally vanish due to phase cancellation, and the average fidelity saturates near zero.
	While this is true for generic linear combinations, there exists particular states where the phase cancellation happens exceptionally slowly.
	We discuss these special initial states in section \ref{sec:disorder induced scars} and how to distinguish them from the scar states.
	Summing up, the average fidelity represents an effective tool for identifying scar states in both a thermal and localized background.
	
	Finally, we remark that the fidelity of individual disorder realizations are enough to distinguish initial states with support inside and outside the scar subspace.
	This statement is simple in the thermal phase where initial states outside the scar subspace rapidly converges to zero.
	At large disorder, the fidelity of individual disorder realizations may oscillate rapidly contrary to the average fidelity.
	However, these oscillations are generally composed of frequencies different from the scar revivals.
	The amplitude of the oscillations are also typically different from the scar revivals.
	Thus, the scar states can be distinguished from a partially localized background.
	
	\section{Disorder stabilization of scar revivals}\label{sec:stabilization}
\begin{figure*}
		\centering
		\includegraphics[width=\linewidth]{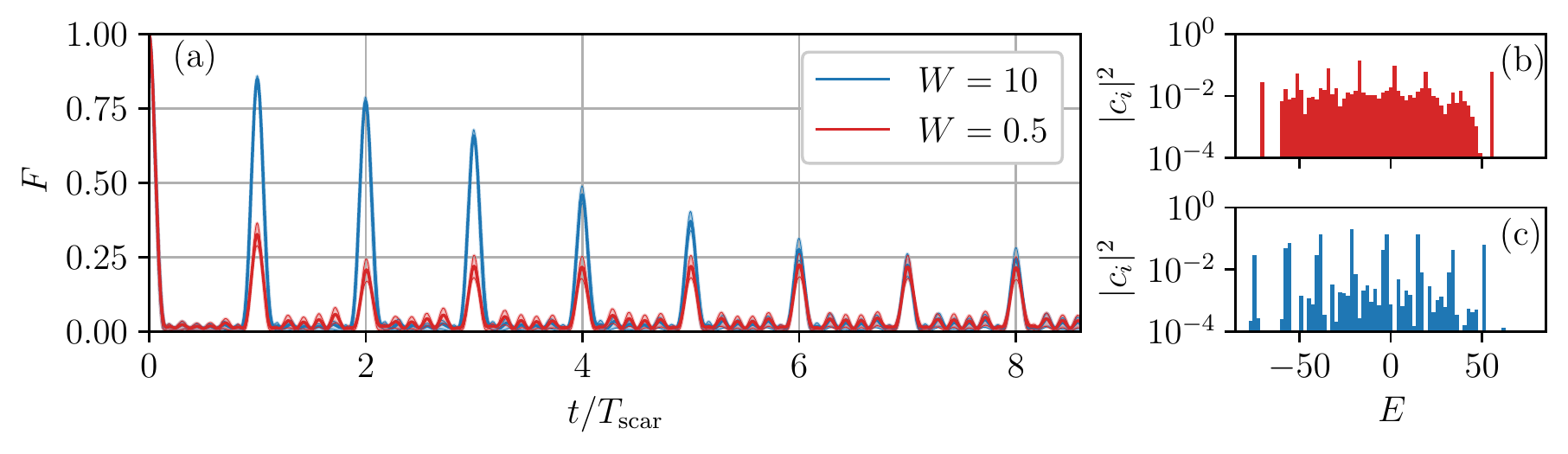}
		\caption{
			A system of size $L = 14$ with parameters $\Delta = 1$, $J = 5$, $\lambda = 1$ is initialized according to Eq.\ \eqref{eq:psi_mix} in the thermal phase at disorder strength $W = 0.5$ and the partial MBL phase at disorder strength $W = 10$.
			(a) The average fidelity over $10^3$ disorder realizations when the system is thermal and partially MBL.
			The interquartile range (middle $50\%$) of the disorder realizations are shown by the shaded areas.
			The disorder protects the scar revivals and the fidelity amplitude decays much slower compared to the thermal case.
			The right panels illustrate the distribution of expansion coefficients $|c_i|^2$ over energy $E_i$ for a single disorder realization at disorder strength (b) $W = 0.5$ and (c) $W = 10$.
			The distribution of the expansion coefficients is wide in the thermal phase and consists of narrow peaks near the scar states in the localized phase.
		}
		\label{fig:fidelity_mixed}
	\end{figure*}

	We study the dynamics of initial states with support both inside and outside the scar subspace across all symmetry sectors.
	In this case, we generally expect the scar revivals to diminish.
	The scar revivals are stabilized when the initial state only has support on product states with the same disorder indices as the scar states $\bm D_0 = (-1, -1, \ldots, -1)$.
	We demonstrate this behavior by initializing the system in a generic state only having support on product states with disorder indices $\bm D_0$
	\begin{align}\label{eq:psi_mix}
			\begin{split}
			\ket{\psi_\text{stable}} =& \frac{1}{\mathcal N_\text{stable}} \Big( \ket{\psi_\text{scar}} + \sum_{N_\text{dw}, n} \beta_n^{(N_\text{dw})} \ket{N_\text{dw}, \bm D_0, n} \Big),
		\end{split}
	\end{align}
	where $\mathcal N_\mathrm{stable}$ is a normalization constant and $\beta_n^{(N_\mathrm{dw})}$ are drawn randomly from the interval $\beta_n^{(N_\mathrm{dw})} \in [0, 1/\sqrt{N_{\bm D_0}^{(N_\mathrm{dw})}}]$.
	We reintroduce the index $N_\text{dw}$ to describe product states with the same disorder indices in different symmetry sectors.
	The time evolution of fidelity is investigated at weak and strong disorder in $10^3$ realizations.
	The coefficients $\beta_n^{(N_\mathrm{dw})}$ are redrawn in each disorder realization.
	Figure \ref{fig:fidelity_mixed}(a) displays the disorder averaged fidelity for a thermal system and a partially localized system.
	In both cases, the average fidelity displays persistent revivals with the revival amplitude decaying and eventually saturating at a value around $0.5$.
	
	\begin{figure}
	\centering
	\includegraphics{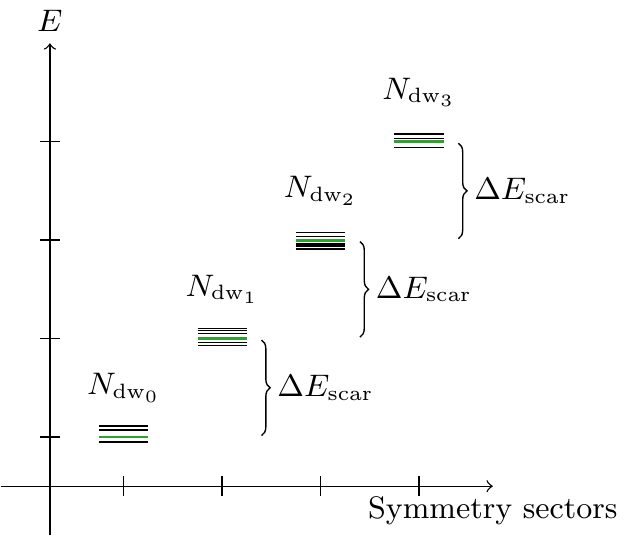}
	\caption{
		At large disorder, the initial state Eq.\ \eqref{eq:psi_mix} has significant overlap with a small number of energy eigenstates (black lines) as sketched in the figure.
		These eigenstates appear in clusters around the energy of the scar states (green lines).
		A single cluster exists in every symmetry sector and the energy gap between two adjacent clusters equals the energy gap between scar states $\Delta E_\mathrm{scar}$.
	}
	\label{fig:clustering}
	\end{figure}
	The fidelity amplitude quickly decays for a thermal system.
	The explanation can be found by studying the expansion coefficients $|c_i|^2$ as illustrated in Fig.\ \ref{fig:fidelity_mixed}(b).
	Because	the system is thermal, the initial state has support on many energy eigenstates.
	Consequently, terms with different phases quickly cancel causing the fidelity amplitude to saturate almost immediately.
		
	At large disorder, the fidelity amplitude decays at a much slower rate and only saturates alongside the thermal graph after many revivals $t \sim 7T_\text{scar}$.
	We understand this behavior by recalling the spectral structure at large disorder.
	First, recall that the energy eigenstates $\{\ket{E_{\bm D_0, m}}|m=1,2, \ldots, N_{\bm D_0}\}$ are near degenerate and only have significant overlap with product states of the same disorder indices as described in Eq.\ \eqref{eq:eigenstates}.
	Therefore, the second term in Eq.\ \eqref{eq:psi_mix} can be rewritten as a sum of near degenerate eigenstates,
	\begin{align}
		\sum_{n=1}^{N_{\bm D_0}} \beta_n^{(N_\text{dw})} \ket{N_\text{dw}, \bm D_0, n} \approx \sum_{m=1}^{N_{\bm D_0}} \gamma_m^{(N_\mathrm{dw})}\ket{E_{N_\mathrm{dw}, \bm D_0, m}},
		\label{eq:degenerate linear combination}
	\end{align}
	with $\gamma_m^{(N_\mathrm{dw})} = \sum_{n} \beta_n^{(N_\mathrm{dw})} \braket{E_{N_\mathrm{dw},\bm D_0, m}|N_\mathrm{dw}, \bm D_0, n}$.
	Furthermore, the scar states themselves are described by the disorder indices $\bm D_0$, so the eigenstates $\ket{E_{N_\mathrm{dw}, \bm D_0, m}}$ are close in energy to a scar state.
	Consequently, the eigenstates outside the scar subspace having large overlap with $\ket{\psi_\text{stab}}$ are always close in energy to a scar state.
	We sketch this structure in Fig.\ \ref{fig:clustering} where the eigenstates $\ket{E_{N_\mathrm{dw}, \bm D_0, m}}$ have similar energy to the scar states for all $N_\mathrm{dw}$.
	These considerations agree with the observed distribution of $|c_i|^2$ for a single disorder realization illustrated in Fig.\ \ref{fig:fidelity_mixed}(c).
	The expansion coefficients are sharply peaked around the scar states and consequently the cancellation of terms with different phases takes place at much larger times.
	
	In this way, the partially localized background stabilizes the scar revivals by rearranging the support outside the scar subspace.
	The stabilization takes place whenever the initial state is predominantly a linear combination of product states with the same disorder indices as the scar states $\bm D_0$.
	If product states with other disorder indices $\bm D' \neq \bm D_0$ are included, the stabilization will be less pronounced.
	
\section{Disorder induced revivals}\label{sec:disorder induced scars}
	Revivals appear when the system is initialized in the scar subspace.
	However, revivals can also be observed from initial states with no support in the scar subspace.
	This dynamical behavior is caused by different symmetry sectors containing energy eigenstates with the same disorder indices. For instance, the eigenstates $\ket{E_{2, \bm D, 1}} \approx \ket{\uparrow\uparrow\downarrow\downarrow\downarrow\downarrow}$ and $\ket{E_{4, \bm D, m}} \approx \alpha_{m1} \ket{\uparrow\uparrow\downarrow\uparrow\downarrow\downarrow} + \alpha_{m2} \ket{\uparrow\uparrow\downarrow\downarrow\uparrow\downarrow}$ for $m=1, 2$ have the same disorder indices $\bm D = (3, -1, -1, -1, -1, -1)$ but different number of domain walls $N_\mathrm{dw}$.
	Recall from Sec.\ \ref{sec:partial MBL} that the energy of an eigenstate at large disorder is approximately given by,
	\begin{align}
		\begin{split}
			E_{N_\mathrm{dw},\bm D, m} \approx& \enspace \Delta M_{N_\mathrm{dw}, \bm D} + J \Big( \mathcal N_{N_\mathrm{dw}, \bm D}^{(\uparrow\uparrow, \downarrow\downarrow)} - \mathcal N_{N_\mathrm{dw}, \bm D}^{(\uparrow\downarrow, \downarrow\uparrow)} \Big) \\
			 &+ \sum_i d_i D_i,
		\end{split}
		\label{eq:approximate energy}
	\end{align}
	If an eigenstate $\ket{E_{N_\mathrm{dw}, \bm D, m}}$ is described by the values $M_{N_\mathrm{dw}, \bm D}$, $\mathcal N_{N_\mathrm{dw}, \bm D}^{(\uparrow\uparrow, \downarrow\downarrow)}$ and $\mathcal N_{N_\mathrm{dw}, \bm D}^{(\uparrow\downarrow, \downarrow\uparrow)}$, then another eigenstate $\ket{E_{N_\mathrm{dw}+2, \bm D, m}}$ with $N_\mathrm{dw}+2$ domain walls and identical disorder indices $\bm D$ is described by
	\begin{subequations}
	\label{eq:adding two domain walls}
	\begin{align}
        M_{N_\mathrm{dw}+2, \bm D} &= M_{N_\mathrm{dw}, \bm D} + 2,\\
		\mathcal N_{N_\mathrm{dw}+2, \bm D}^{(\uparrow\uparrow, \downarrow\downarrow)} &= \mathcal N_{N_\mathrm{dw}, \bm D}^{(\uparrow\uparrow, \downarrow\downarrow)} - 2, \\
		\mathcal N_{N_\mathrm{dw}+2, \bm D}^{(\uparrow\downarrow, \downarrow\uparrow)} &= \mathcal N_{N_\mathrm{dw}, \bm D}^{(\uparrow\downarrow, \downarrow\uparrow)} + 2.
	\end{align}
	\end{subequations}
	Using Eq.\ \eqref{eq:approximate energy} and \eqref{eq:adding two domain walls}, one can show the energy difference between two eigenstates with the same disorder indices $\bm D$ and number of domain walls $N_{\bm D}$ and $N_{\bm D} + 2$ is approximately
	\begin{align}
		E_{N_\mathrm{dw} + 2, \bm D, m} - E_{N_\mathrm{dw}, \bm D, m} \approx \Delta E_\mathrm{scar},
	\end{align}
	where $\Delta E_\mathrm{scar} = 2(\Delta - 2J)$ is the energy gap between the scar states.
	This calculation demonstrates that the spectrum contains eigenstates outside the scar subspace with an approximate energy separation $\Delta E_\text{scar}$ at large disorder.
	Hence, approximate towers of eigenstates appear as disorder is introduced.
	\begin{figure*}
		\includegraphics{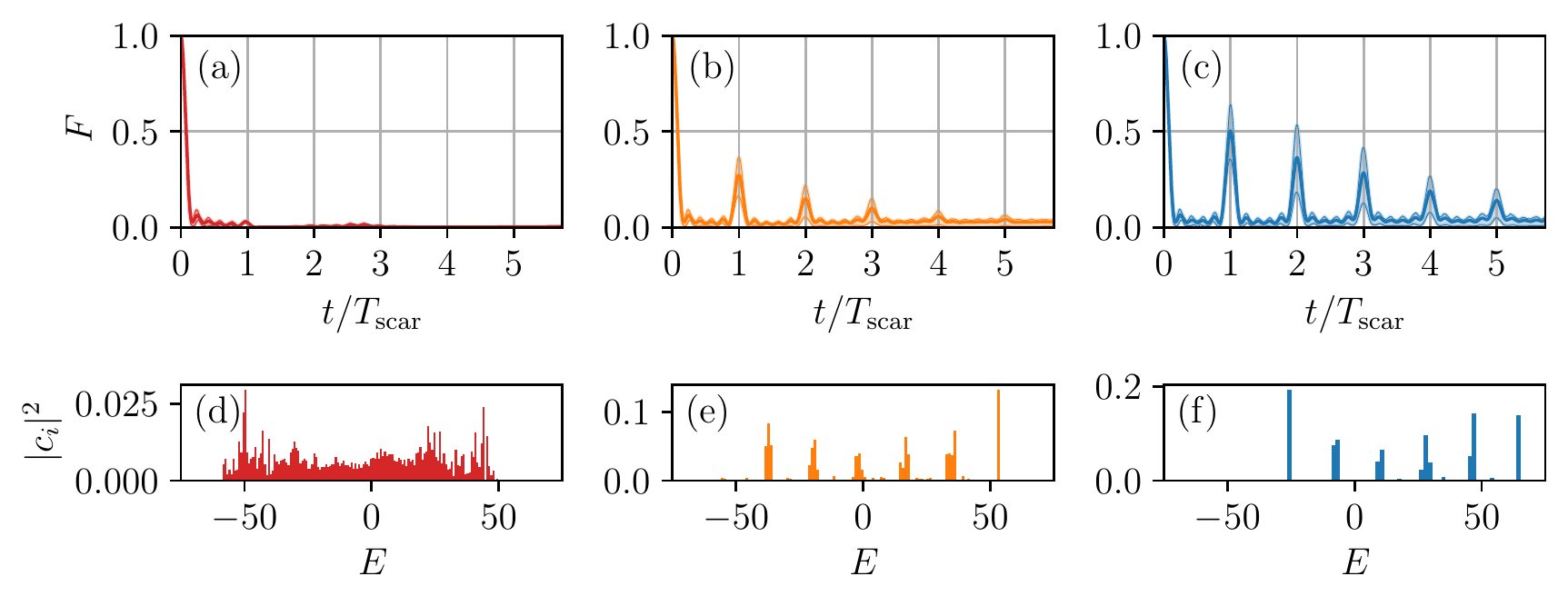}
		\caption{
			The average fidelity of the initial state Eq.\ \eqref{eq:psi induced} over $10^3$ disorder realizations for system size $L=14$ with parameters $\lambda = \Delta = 1$, $J = 5$ at disorder strength (a) $W = 0.5$, (b) $W = 5$ and (c) $W = 10$.
			The shaded areas show the interquartile range (middle $50\%$) of the disorder realizations.
			The corresponding distribution of expansion coefficients $|c_i|^2$ of a single disorder realization at disorder strength (d) $W = 0.5$, (e) $W = 5$ and (f) $W = 10$.
			At weak disorder, the initial state has significant overlap with many energy eigenstates and the average fidelity quickly decays to zero.
			As the disorder strength is increased, the initial state has significant overlap with a small number of energy eigenstates with equal energy spacing. Consequently, the average fidelity shows persistent revivals.
		}
		\label{fig:fidelity induced}
	\end{figure*}

	We demonstrate how the appearance of approximate towers of eigenstates generates non-trivial dynamics.
	The system is initialized in a generic linear combination of product states with disorder indices $\bm D_1 = (3, -1, -1, \ldots, -1)$
		\begin{align}
		\ket{\psi_{\bm D_1}^\text{induced}} &= \frac{1}{\mathcal N_\text{induced}} \sum_{N_\mathrm{dw}, n} \zeta^{(N_\mathrm{dw})}_n\ket{N_\mathrm{dw},\bm D_1, n}.
		\label{eq:psi induced}
	\end{align}
	The coefficients are chosen randomly from the interval $\zeta_n^{(N_{\mathrm{dw}})} \in [0, 1]$ and $\mathcal N_\text{induced}$ is a normalization constant.
	We study this initial state because, at large disorder, it is a linear combination of eigenstates in an approximate tower.
	We consider $10^3$ disorder realizations at different disorder strengths and the fidelity is computed for each realization.
	Figure \ref{fig:fidelity induced}(a) displays the average fidelity of a thermal system at weak disorder $W = 0.5$.
	In this case, there is nothing special about the initial state in Eq.\ \eqref{eq:psi induced} and it quickly decays to zero similar to Fig.\ \ref{fig:fidelity}(a).
	The dynamical behavior changes remarkably as the disorder strength is increased as illustrated in Fig.\ \ref{fig:fidelity induced}(b)-(c).
	At stronger disorder, the initial state Eq.\ \eqref{eq:psi induced} has large overlap with eigenstates that are approximately equidistant in energy.
	Consequently, the average fidelity oscillates with a period given by the energy gap $T_\mathrm{scar} = 2\pi/\Delta E_\mathrm{scar}$.
	The revival amplitude increases with disorder strength.
	The shaded area in Fig.\ \ref{fig:fidelity induced}(a)-(c) displays the interquartile range of disorder realizations.
	Figures \ref{fig:fidelity induced}(d)-(f) shows the expansion of the initial state in energy eigenstates at (d) weak disorder $W=0.5$, (e) strong disorder $W=5$ and (f) very strong disorder $W = 10$.
	As expected, the initial state is distributed over a wide range of eigenstates in the thermal phase similar to Fig.\ \ref{fig:fidelity}(d).
	As the disorder strength increases, the initial state has higher and higher overlap with eigenstates in an approximate tower of equidistant states.
	
\begin{figure}
\includegraphics{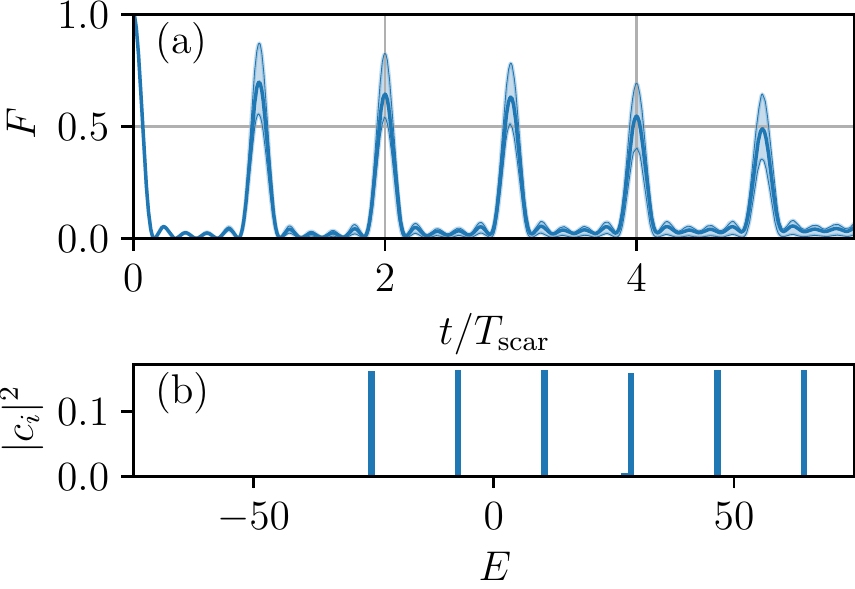}
\caption{(a) Average fidelity of the initial state Eq.\ \eqref{eq:psi induced 1} over $10^3$ disorder realizations with system size $L = 14$ and parameters $\lambda = \Delta = 1$, $J = 5$ and $W = 10$. The shaded area displays the interquartile range of the disorder realizations. The average fidelity displays persistent revivals with larger amplitude compared to Fig.\ \ref{fig:fidelity_mixed}. (b) Expansion of the initial state across energy eigenstates. The coefficients $|c_i|^2$ are sharply peaked around certain energies which are approximately equally spaced.}
\label{fig:MBL revival special}
\end{figure}

	Figure \ref{fig:fidelity induced} demonstrates that it is possible to observe revivals from generic linear combinations of the states $\{\ket{N_\mathrm{dw}, \bm D, n}|N_\mathrm{dw} = 0, 2, \ldots; n = 1, 2, \ldots\}$ at large disorder.
	However, the effects may be enhanced by choosing the initial state more carefully.
	The initial state in Eq.\ \eqref{eq:psi induced} is, in some sense, the worst case scenario.
	When all product states with disorder indices $\bm D$ are included in the sum, the initial state generally has significant overlap with all relevant energy eigenstates $\{\ket{E_{N_\mathrm{dw},\bm D,m}}|N_\mathrm{dw}=0,2,\ldots; m = 1, 2, \ldots\}$.
	This causes a large spread in the distribution of $|c_i|^2$ resulting in a faster decay of the average fidelity.
	If instead, we consider an initial state with exactly one product state from each symmetry sector, the spread of $|c_i|^2$ is smaller
	\begin{align}
	\begin{split}
		\ket{{\tilde \psi}_{\bm D_1}^\text{induced}} &= \frac{1}{\sqrt{\frac{L}{2} - 1}}\Big( \ket{\uparrow\uparrow\downarrow\downarrow\downarrow\downarrow\downarrow\ldots\downarrow}
		+ \ket{\uparrow\uparrow\downarrow\uparrow\downarrow\downarrow\downarrow\ldots\downarrow} \\
		& + \ket{\uparrow\uparrow\downarrow\uparrow\downarrow\uparrow\downarrow\ldots\downarrow}
		 + \ldots
		  + \ket{\uparrow\uparrow\downarrow\uparrow\downarrow\uparrow\ldots\downarrow\uparrow\downarrow\downarrow}
		  \Big).
	\end{split}
	\label{eq:psi induced 1}
	\end{align}
Figure \ref{fig:MBL revival special}(a) shows the average fidelity of this initial state over $10^3$ disorder realizations at strong disorder $W = 10$ and Fig.\ \ref{fig:MBL revival special}(b) displays the distribution of $|c_i|^2$ for a single realization.
	As expected, the distribution of $|c_i|^2$ is narrower and the revival amplitude larger compared to Fig.\ \ref{fig:fidelity induced}.

	The initial states Eq.\ \eqref{eq:psi induced} and \eqref{eq:psi induced 1} display revivals similar to the scar states.
	However, one may distinguish these initial states from the scar subspace by noting that the average fidelity in Fig.\ \ref{fig:fidelity induced} and \ref{fig:MBL revival special} decays to zero, while the amplitude in Fig.\ \ref{fig:fidelity}(c) and \ref{fig:fidelity_mixed} remain strictly larger than zero.
	The different dynamical behavior is caused by Eq.\ \eqref{eq:psi induced} and \eqref{eq:psi induced 1} being composed of eigenstates with approximately equal energy spacing while the scar states $\ket{\mathcal S_n}$ are exactly equally spaced in energy.

	\section{Conclusion}\label{sec:conclusion}
	Building on a known method to find parent Hamiltonians, we proposed a way to determine Hamiltonians hosting a tower of QMBS.
	Starting from the model by Iadecola and Schecter, we used this method to identify all local $1$- and $2$-body Hamiltonians of the scar tower $\ket{\mathcal S_n}$.
	Among these Hamiltonians, we found operators facilitating the implementation of local disorder while preserving the scar states.
	When introducing disorder, the mean level spacing statistics shifts from the GOE to the Poisson distribution and the entanglement entropy goes from volume-law to area-law scaling with system size.
	We conclude the system transitions from the thermal phase to being partially localized.
	A theory describing the partially localized eigenstates was developed and verified numerically.
	In total, we determined a system hosting a tower of scar states with the remaining spectrum being either thermal or partially localized depending on the disorder strength.
	
	We studied the properties of scar states embedded in a localized spectrum and compared with the corresponding features in a thermal spectrum.
	In contrast to thermal systems, the bipartite entanglement entropy does not enable the identification of scar states in a localized background.
	The Schmidt gap distinguishes scar states from fully MBL eigenstates, but is incapable of distinguishing scar states from partial MBL eigenstates.
	The average fidelity, on the other hand, effectively identifies the scar subspace in both a thermal and partial MBL background.
		
	We investigated the effect of localization on initial states with support both inside and outside the scar subspace.
	For a thermal system, the fidelity displays persistent revivals with rapidly decreasing amplitude.
	In contrast, the revival amplitude decays slower for a partially localized system.
	Hence, partial localization stabilizes the persistent revivals of states initialized partly outside the scar subspace.
	
	Finally, we demonstrated how additional approximate towers of eigenstates emerge as disorder is introduced.
	When initializing the system as a superposition of these eigenstates, the average fidelity displays revivals with the same period as the scar states.
	While this effect does not rely on fine-tuning the initial state, the revivals are amplified by choosing the initial state appropriately.
	
	\begin{acknowledgments}
	This work has been supported by the Carlsberg Foundation under grant number CF20-0658 and by the Independent Research Fund Denmark under grant number 8049-00074B.
	\end{acknowledgments}

	\appendix
	\section{Proof that $\ket{\mathcal S_n}$ are eigenstates of all operators in Tab.\ \ref{tab:operators} with equal energy spacing}\label{apx:new_operators}
	In section \ref{subsec:generalized_models}, we found $L+4$ operators having the scar states as eigenstates equidistantly spaced in energy.
	Since this analysis was carried out for finite system sizes $L=8, 10, 12, 14$, the validity of this statement is not guaranteed for larger system sizes.
	In this appendix, we rigorously prove the scar states $\ket{\mathcal S_n}$ are equally spaced eigenstates of all operators in Tab.\ \ref{tab:operators}.
	Since the scar states are constructed iteratively by applying the operator $Q^\dagger$, we generally prove this statement using proof by induction.
	
	First, we consider the operator $\hat H_z = \sum_i \hat \sigma_i^z$.
	The lowest scar state $\ket{\mathcal S_0} = \ket{\downarrow\downarrow\ldots\downarrow}$ is trivially an eigenstate of $\hat H_z$.
	A straightforward calculation shows that $[\hat H_z, \hat Q^\dagger] = 2\hat Q^\dagger$ and by induction all other scar states are eigenstates because
	\begin{align}
		\begin{split}
		\hat H_z \ket{\mathcal S_{n+1}} &\propto \hat H_z \hat Q^\dagger \ket{S_n} \\
		&= \big(E_{z,n} \hat Q^\dagger + 2 \hat Q^\dagger \big)\ket{S_n} \\
		&= \big(E_{z,n} + 2 \big)\ket{S_{n+1}},
		\end{split}
	\end{align}
	where $\hat H_z \ket{\mathcal S_n} = E_{z,n}\ket{\mathcal S_n}$.
	The scar states are also equally spaced in energy $E_{n+1,z} - E_{n,z} = 2$.
	A similar argument holds for $\hat H_{zz}^\text{odd}$ since $[\hat H_{zz}^\text{odd}, \hat Q^\dagger] = -4 \hat Q^\dagger$ where the energy gap between scar states is $-4$.

	Next, we consider the operators $\hat D_i = \hat \sigma_i^z + \hat \sigma_{i+1}^z + \hat \sigma_i^z \hat \sigma_{i+1}^z$.
	Recall that $\hat D_i$ is related to the projection operators through $\hat D_i = 4 \hat P_i^\uparrow \hat P_{i+1}^\uparrow - \hat {\mathds 1}$ where $\hat P_i^\uparrow = (\hat {\mathds 1} + \hat \sigma_i^z)/2$ projects site $i$ onto spin-up.
	First note that $\hat D_i\ket{\mathcal S_0} = (4 \hat P_i^\uparrow \hat P_{i+1}^\uparrow - \hat {\mathds 1})\ket{\downarrow \downarrow \ldots \downarrow} = -\ket{\downarrow\downarrow\ldots \downarrow}$.
	A simple calculation shows that $\hat D_i$ commutes with $\hat Q^\dagger$ by noting that $\hat P^\uparrow_i \hat P^\downarrow_i = 0$
	\begin{align}
	\begin{split}
		[\hat D_i, \hat Q^\dagger] &= 4
		\sum_{j=1}^L (-1)^j \Big(\hat P_{j-1}^\downarrow [\hat P_i^\uparrow, \hat \sigma_j^+] \hat P_{j+1}^\downarrow \hat P_{i+1}^\uparrow \\
		&\quad + \hat P_i^\uparrow \hat P_{j-1}^\downarrow [\hat P_{i+1}^\uparrow, \hat \sigma_j^+] \hat P_{j+1}^\downarrow \Big) \\
		&= 4(-1)^i \Big(\hat P_{i-1}^\downarrow \hat \sigma_i^+ \hat P_{i+1}^\downarrow \hat P_{i+1}^\uparrow - \hat P_i^\uparrow \hat P_i^\downarrow \hat \sigma^+_{i+1} \hat P_{i+2}^\downarrow \Big) \\
		&= 0.
	\end{split}
	\end{align}
	Thus, for all scar states we have $\hat D_i \ket{\mathcal S_n} = -\ket{\mathcal S_n}$.
	Alternatively, one may note that $\ket{\mathcal S_n}$ by construction does not contain adjacent sites being spin-up.
	Therefore, $\hat P_i^\uparrow \hat P_{i+1}^\uparrow$ naturally annihilates the state.

	Next, we consider the operator $\hat H_{xz}^\text{alt}$. Before studying the action of $\hat H_{xz}^\text{alt}$ on the scar states, we prove by induction that the commutator $[\hat H_{xz}^\text{alt}, \hat Q^\dagger]$ annihilates $\ket{\mathcal S_n}$.
	The commutator is given by
	\begin{align}
	\begin{split}
		[\hat H_{xz}^\text{alt}, \hat Q^\dagger]
		=& \sum_{i=1}^L \Big[2\big(\hat P_i^\downarrow \hat \sigma_{i+1}^+ \hat \sigma_{i+2}^-
			- \hat \sigma_i^+ \hat \sigma_{i+1}^+  \hat P_{i+2}^\downarrow \big) \\
		&+ i \big(
			\hat P_i^\downarrow  \hat \sigma_{i+1}^+ \hat \sigma_{i+2}^y
			+ \hat \sigma_i^y \hat \sigma_{i+1}^+ \hat P_{i+2}^\downarrow \\
			&+ \hat \sigma_i^z \hat \sigma_{i+1}^y \hat \sigma_{i+2}^+ \hat P_{i+3}^\downarrow
			- \hat P_i^\downarrow \hat \sigma_{i+1}^+ \hat \sigma_{i+2}^y \hat \sigma_{i+3}^z
		\big)\Big],
	\end{split}
	\end{align}
	where $\hat P_i^\downarrow = (\hat {\mathds 1} - \hat \sigma_i^z)/2$ is the local projection onto spin-down.
	By direct calculation, one can show the lowest scar state is annihilated by this expression $[\hat H_{xz}^\text{alt}, \hat Q^\dagger]\ket{\mathcal S_0} = 0$.
	A lengthy, yet straightforward, calculation also shows the nested commutator vanishes $\big[[\hat H_{xz}^\text{alt}, \hat Q^\dagger], \hat Q^\dagger\big] = 0$.
	We now prove by induction that the commutator annihilates all scar states.
	Assume $[\hat H_{xz}^\text{alt}, \hat Q^\dagger] \ket{S_n} = 0$ and consider,
	\begin{align} \label{eq:commutator_scar}
	\begin{split}
		[\hat H_{xz}^\text{alt}, \hat Q^\dagger] \ket{S_{n+1}} &\propto [\hat H_{xz}^\text{alt}, \hat Q^\dagger] \hat Q^\dagger \ket{S_n} \\
		&= \Big(\hat Q^\dagger [\hat H_{xz}^\text{alt}, \hat Q^\dagger] + \big[[\hat H_{xz}^\text{alt}, \hat Q^\dagger], \hat Q^\dagger\big]\Big) \ket{S_n} \\
		&= 0.
	\end{split}
	\end{align}
	Having shown this intermediate result, we prove by induction that the operator $\hat H_{xz}^\text{alt}$ annihilates the scar states.
	First we show the operator $\hat H_{xz}^\text{alt}$ annihilates $\ket{\mathcal S_0}$
	\begin{align}
	\begin{split}
			\hat H_{xz}^\text{alt} \ket{S_0} &= \sum_{i=1}^{L} (-1)^i(\hat \sigma_i^x \hat \sigma_{i+1}^z + \hat \sigma_i^z \hat \sigma_{i+1}^x) \ket{\downarrow\downarrow\ldots \downarrow} \\
			&= \sum_{i=1}^{L} (-1)^{i+1}(\hat \sigma_i^x + \hat \sigma_{i+1}^x) \ket{\downarrow\downarrow\ldots \downarrow} \\
			&= 0,
	\end{split}
	\end{align}
	where the second term cancels the first after changing summation index $i + 1 \to i$.
	Next, we show by induction that the $n$-th scar state is annihilated by $\hat H_{xy}^\text{alt}$.
	Assume $\hat H_{xz}^\text{alt}$ annihilates $\ket{S_n}$ and consider
	\begin{align}\label{eq:H_xy}
	\begin{split}
		\hat H_{xz}^\text{alt} \ket{S_{n+1}} &\propto \hat H_{xz}^\text{alt} \hat Q^\dagger \ket{S_n} \\
		&= (\hat Q^\dagger \hat H_{xz}^\text{alt} + [\hat H_{xz}^\text{alt}, \hat Q^\dagger])\ket{S_n} \\
		&= 0.
	\end{split}
	\end{align}
	The first term vanishes by assumption and the second term is exactly what we considered in Eq.\ \eqref{eq:commutator_scar}.
	In total, we conclude $\hat H_{xy}^\text{alt}$ has $\ket{\mathcal S_n}$ as eigenstates equidistantly separated in energy (with zero energy spacing).

	Finally we consider the operator $\hat H_{yz}^\text{alt}$.
	One can prove this operator annihilates the scar states using similar arguments to above.
	The commutator is given by
	\begin{align}
	\begin{split}
			[\hat H_{yz}^\text{alt}, \hat Q^\dagger] =&
			i \sum_{i=1}^L \Big[
			2 \big(\hat P_i^\downarrow \hat \sigma_{i+1}^+ \hat \sigma_{i+2}^-
			+ \hat \sigma_i^+ \hat \sigma_{i+1}^+ \hat P_{i+2}^\downarrow \big) \\
			 & - \hat \sigma_i^x \hat \sigma_{i+1}^+ \hat P_{i+2}^\downarrow
			 - \hat P_i^\downarrow \hat \sigma_{i+1}^+ \hat \sigma_{i+2}^x \\
			 & + \hat P_i^\downarrow \hat \sigma_{i+1}^+ \hat \sigma_{i+2}^x \hat \sigma_{i+3}^z
			 - \hat \sigma_i^z \hat \sigma_{i+1}^x \hat \sigma_{i+2}^+ \hat P_{i+3}^\downarrow
			\Big].
	\end{split}
	\end{align}
	Using induction, one can prove the commutator annihilates all scar states $[\hat H_{yz}^\text{alt}, \hat Q^\dagger] \ket{\mathcal S_n} = 0$ and the operator annihilates the lowest scar state $\hat H_{yz}^\text{alt}\ket{\mathcal S_0} = 0$. Retracing the steps in Eq.\ \eqref{eq:H_xy}, we find that $\hat H_{yz}^\text{alt}$ annihilates all scar states.


\begin{thebibliography}{84}%
\makeatletter
\providecommand \@ifxundefined [1]{%
 \@ifx{#1\undefined}
}%
\providecommand \@ifnum [1]{%
 \ifnum #1\expandafter \@firstoftwo
 \else \expandafter \@secondoftwo
 \fi
}%
\providecommand \@ifx [1]{%
 \ifx #1\expandafter \@firstoftwo
 \else \expandafter \@secondoftwo
 \fi
}%
\providecommand \natexlab [1]{#1}%
\providecommand \enquote  [1]{``#1''}%
\providecommand \bibnamefont  [1]{#1}%
\providecommand \bibfnamefont [1]{#1}%
\providecommand \citenamefont [1]{#1}%
\providecommand \href@noop [0]{\@secondoftwo}%
\providecommand \href [0]{\begingroup \@sanitize@url \@href}%
\providecommand \@href[1]{\@@startlink{#1}\@@href}%
\providecommand \@@href[1]{\endgroup#1\@@endlink}%
\providecommand \@sanitize@url [0]{\catcode `\\12\catcode `\$12\catcode
  `\&12\catcode `\#12\catcode `\^12\catcode `\_12\catcode `\%12\relax}%
\providecommand \@@startlink[1]{}%
\providecommand \@@endlink[0]{}%
\providecommand \url  [0]{\begingroup\@sanitize@url \@url }%
\providecommand \@url [1]{\endgroup\@href {#1}{\urlprefix }}%
\providecommand \urlprefix  [0]{URL }%
\providecommand \Eprint [0]{\href }%
\providecommand \doibase [0]{https://doi.org/}%
\providecommand \selectlanguage [0]{\@gobble}%
\providecommand \bibinfo  [0]{\@secondoftwo}%
\providecommand \bibfield  [0]{\@secondoftwo}%
\providecommand \translation [1]{[#1]}%
\providecommand \BibitemOpen [0]{}%
\providecommand \bibitemStop [0]{}%
\providecommand \bibitemNoStop [0]{.\EOS\space}%
\providecommand \EOS [0]{\spacefactor3000\relax}%
\providecommand \BibitemShut  [1]{\csname bibitem#1\endcsname}%
\let\auto@bib@innerbib\@empty
\bibitem [{\citenamefont {Deutsch}(1991)}]{Deutsch1991}%
  \BibitemOpen
  \bibfield  {author} {\bibinfo {author} {\bibfnamefont {J.~M.}\ \bibnamefont
  {Deutsch}},\ }\bibfield  {title} {\bibinfo {title} {Quantum statistical
  mechanics in a closed system},\ }\href
  {https://doi.org/10.1103/PhysRevA.43.2046} {\bibfield  {journal} {\bibinfo
  {journal} {Phys. Rev. A}\ }\textbf {\bibinfo {volume} {43}},\ \bibinfo
  {pages} {2046} (\bibinfo {year} {1991})}\BibitemShut {NoStop}%
\bibitem [{\citenamefont {Srednicki}(1994)}]{Srednicki1994}%
  \BibitemOpen
  \bibfield  {author} {\bibinfo {author} {\bibfnamefont {M.}~\bibnamefont
  {Srednicki}},\ }\bibfield  {title} {\bibinfo {title} {Chaos and quantum
  thermalization},\ }\href {https://doi.org/10.1103/PhysRevE.50.888} {\bibfield
   {journal} {\bibinfo  {journal} {Phys. Rev. E}\ }\textbf {\bibinfo {volume}
  {50}},\ \bibinfo {pages} {888} (\bibinfo {year} {1994})}\BibitemShut
  {NoStop}%
\bibitem [{\citenamefont {Rigol}\ \emph {et~al.}(2008)\citenamefont {Rigol},
  \citenamefont {Dunjko},\ and\ \citenamefont {Olshanii}}]{Rigol2008}%
  \BibitemOpen
  \bibfield  {author} {\bibinfo {author} {\bibfnamefont {M.}~\bibnamefont
  {Rigol}}, \bibinfo {author} {\bibfnamefont {V.}~\bibnamefont {Dunjko}},\ and\
  \bibinfo {author} {\bibfnamefont {M.}~\bibnamefont {Olshanii}},\ }\bibfield
  {title} {\bibinfo {title} {Thermalization and its mechanism for generic
  isolated quantum systems},\ }\href {https://doi.org/10.1038/nature06838}
  {\bibfield  {journal} {\bibinfo  {journal} {Nature}\ }\textbf {\bibinfo
  {volume} {452}},\ \bibinfo {pages} {854} (\bibinfo {year}
  {2008})}\BibitemShut {NoStop}%
\bibitem [{\citenamefont {Rigol}(2009{\natexlab{a}})}]{Rigol2009Sep}%
  \BibitemOpen
  \bibfield  {author} {\bibinfo {author} {\bibfnamefont {M.}~\bibnamefont
  {Rigol}},\ }\bibfield  {title} {\bibinfo {title} {Breakdown of thermalization
  in finite one-dimensional systems},\ }\href
  {https://doi.org/10.1103/PhysRevLett.103.100403} {\bibfield  {journal}
  {\bibinfo  {journal} {Phys. Rev. Lett.}\ }\textbf {\bibinfo {volume} {103}},\
  \bibinfo {pages} {100403} (\bibinfo {year} {2009}{\natexlab{a}})}\BibitemShut
  {NoStop}%
\bibitem [{\citenamefont {Rigol}(2009{\natexlab{b}})}]{Rigol2009Nov}%
  \BibitemOpen
  \bibfield  {author} {\bibinfo {author} {\bibfnamefont {M.}~\bibnamefont
  {Rigol}},\ }\bibfield  {title} {\bibinfo {title} {Quantum quenches and
  thermalization in one-dimensional fermionic systems},\ }\href
  {https://doi.org/10.1103/PhysRevA.80.053607} {\bibfield  {journal} {\bibinfo
  {journal} {Phys. Rev. A}\ }\textbf {\bibinfo {volume} {80}},\ \bibinfo
  {pages} {053607} (\bibinfo {year} {2009}{\natexlab{b}})}\BibitemShut
  {NoStop}%
\bibitem [{\citenamefont {Santos}\ and\ \citenamefont
  {Rigol}(2010)}]{Rigol2010}%
  \BibitemOpen
  \bibfield  {author} {\bibinfo {author} {\bibfnamefont {L.~F.}\ \bibnamefont
  {Santos}}\ and\ \bibinfo {author} {\bibfnamefont {M.}~\bibnamefont {Rigol}},\
  }\bibfield  {title} {\bibinfo {title} {Onset of quantum chaos in
  one-dimensional bosonic and fermionic systems and its relation to
  thermalization},\ }\href {https://doi.org/10.1103/PhysRevE.81.036206}
  {\bibfield  {journal} {\bibinfo  {journal} {Phys. Rev. E}\ }\textbf {\bibinfo
  {volume} {81}},\ \bibinfo {pages} {036206} (\bibinfo {year}
  {2010})}\BibitemShut {NoStop}%
\bibitem [{\citenamefont {Sorg}\ \emph {et~al.}(2014)\citenamefont {Sorg},
  \citenamefont {Vidmar}, \citenamefont {Pollet},\ and\ \citenamefont
  {Heidrich-Meisner}}]{Heidrick-Meisner2014}%
  \BibitemOpen
  \bibfield  {author} {\bibinfo {author} {\bibfnamefont {S.}~\bibnamefont
  {Sorg}}, \bibinfo {author} {\bibfnamefont {L.}~\bibnamefont {Vidmar}},
  \bibinfo {author} {\bibfnamefont {L.}~\bibnamefont {Pollet}},\ and\ \bibinfo
  {author} {\bibfnamefont {F.}~\bibnamefont {Heidrich-Meisner}},\ }\bibfield
  {title} {\bibinfo {title} {Relaxation and thermalization in the
  one-dimensional {B}ose-{H}ubbard model: A case study for the interaction
  quantum quench from the atomic limit},\ }\href
  {https://doi.org/10.1103/PhysRevA.90.033606} {\bibfield  {journal} {\bibinfo
  {journal} {Phys. Rev. A}\ }\textbf {\bibinfo {volume} {90}},\ \bibinfo
  {pages} {033606} (\bibinfo {year} {2014})}\BibitemShut {NoStop}%
\bibitem [{\citenamefont {Neuenhahn}\ and\ \citenamefont
  {Marquardt}(2012)}]{Marquardt2012}%
  \BibitemOpen
  \bibfield  {author} {\bibinfo {author} {\bibfnamefont {C.}~\bibnamefont
  {Neuenhahn}}\ and\ \bibinfo {author} {\bibfnamefont {F.}~\bibnamefont
  {Marquardt}},\ }\bibfield  {title} {\bibinfo {title} {Thermalization of
  interacting fermions and delocalization in {F}ock space},\ }\href
  {https://doi.org/10.1103/PhysRevE.85.060101} {\bibfield  {journal} {\bibinfo
  {journal} {Phys. Rev. E}\ }\textbf {\bibinfo {volume} {85}},\ \bibinfo
  {pages} {060101(R)} (\bibinfo {year} {2012})}\BibitemShut {NoStop}%
\bibitem [{\citenamefont {Steinigeweg}\ \emph {et~al.}(2014)\citenamefont
  {Steinigeweg}, \citenamefont {Khodja}, \citenamefont {Niemeyer},
  \citenamefont {Gogolin},\ and\ \citenamefont {Gemmer}}]{Gemmer2014}%
  \BibitemOpen
  \bibfield  {author} {\bibinfo {author} {\bibfnamefont {R.}~\bibnamefont
  {Steinigeweg}}, \bibinfo {author} {\bibfnamefont {A.}~\bibnamefont {Khodja}},
  \bibinfo {author} {\bibfnamefont {H.}~\bibnamefont {Niemeyer}}, \bibinfo
  {author} {\bibfnamefont {C.}~\bibnamefont {Gogolin}},\ and\ \bibinfo {author}
  {\bibfnamefont {J.}~\bibnamefont {Gemmer}},\ }\bibfield  {title} {\bibinfo
  {title} {Pushing the limits of the eigenstate thermalization hypothesis
  towards mesoscopic quantum systems},\ }\href
  {https://doi.org/10.1103/PhysRevLett.112.130403} {\bibfield  {journal}
  {\bibinfo  {journal} {Phys. Rev. Lett.}\ }\textbf {\bibinfo {volume} {112}},\
  \bibinfo {pages} {130403} (\bibinfo {year} {2014})}\BibitemShut {NoStop}%
\bibitem [{\citenamefont {Fratus}\ and\ \citenamefont
  {Srednicki}(2015)}]{Srednicki2015}%
  \BibitemOpen
  \bibfield  {author} {\bibinfo {author} {\bibfnamefont {K.~R.}\ \bibnamefont
  {Fratus}}\ and\ \bibinfo {author} {\bibfnamefont {M.}~\bibnamefont
  {Srednicki}},\ }\bibfield  {title} {\bibinfo {title} {Eigenstate
  thermalization in systems with spontaneously broken symmetry},\ }\href
  {https://doi.org/10.1103/PhysRevE.92.040103} {\bibfield  {journal} {\bibinfo
  {journal} {Phys. Rev. E}\ }\textbf {\bibinfo {volume} {92}},\ \bibinfo
  {pages} {040103(R)} (\bibinfo {year} {2015})}\BibitemShut {NoStop}%
\bibitem [{\citenamefont {Steinigeweg}\ \emph {et~al.}(2013)\citenamefont
  {Steinigeweg}, \citenamefont {Herbrych},\ and\ \citenamefont
  {Prelov\ifmmode~\check{s}\else \v{s}\fi{}ek}}]{Prelovsek2013}%
  \BibitemOpen
  \bibfield  {author} {\bibinfo {author} {\bibfnamefont {R.}~\bibnamefont
  {Steinigeweg}}, \bibinfo {author} {\bibfnamefont {J.}~\bibnamefont
  {Herbrych}},\ and\ \bibinfo {author} {\bibfnamefont {P.}~\bibnamefont
  {Prelov\ifmmode~\check{s}\else \v{s}\fi{}ek}},\ }\bibfield  {title} {\bibinfo
  {title} {Eigenstate thermalization within isolated spin-chain systems},\
  }\href {https://doi.org/10.1103/PhysRevE.87.012118} {\bibfield  {journal}
  {\bibinfo  {journal} {Phys. Rev. E}\ }\textbf {\bibinfo {volume} {87}},\
  \bibinfo {pages} {012118} (\bibinfo {year} {2013})}\BibitemShut {NoStop}%
\bibitem [{\citenamefont {Kim}\ \emph {et~al.}(2014)\citenamefont {Kim},
  \citenamefont {Ikeda},\ and\ \citenamefont {Huse}}]{Huse2014NovE}%
  \BibitemOpen
  \bibfield  {author} {\bibinfo {author} {\bibfnamefont {H.}~\bibnamefont
  {Kim}}, \bibinfo {author} {\bibfnamefont {T.~N.}\ \bibnamefont {Ikeda}},\
  and\ \bibinfo {author} {\bibfnamefont {D.~A.}\ \bibnamefont {Huse}},\
  }\bibfield  {title} {\bibinfo {title} {Testing whether all eigenstates obey
  the eigenstate thermalization hypothesis},\ }\href
  {https://doi.org/10.1103/PhysRevE.90.052105} {\bibfield  {journal} {\bibinfo
  {journal} {Phys. Rev. E}\ }\textbf {\bibinfo {volume} {90}},\ \bibinfo
  {pages} {052105} (\bibinfo {year} {2014})}\BibitemShut {NoStop}%
\bibitem [{\citenamefont {Mondaini}\ \emph {et~al.}(2016)\citenamefont
  {Mondaini}, \citenamefont {Fratus}, \citenamefont {Srednicki},\ and\
  \citenamefont {Rigol}}]{Rigol2016}%
  \BibitemOpen
  \bibfield  {author} {\bibinfo {author} {\bibfnamefont {R.}~\bibnamefont
  {Mondaini}}, \bibinfo {author} {\bibfnamefont {K.~R.}\ \bibnamefont
  {Fratus}}, \bibinfo {author} {\bibfnamefont {M.}~\bibnamefont {Srednicki}},\
  and\ \bibinfo {author} {\bibfnamefont {M.}~\bibnamefont {Rigol}},\ }\bibfield
   {title} {\bibinfo {title} {Eigenstate thermalization in the two-dimensional
  transverse field {I}sing model},\ }\href
  {https://doi.org/10.1103/PhysRevE.93.032104} {\bibfield  {journal} {\bibinfo
  {journal} {Phys. Rev. E}\ }\textbf {\bibinfo {volume} {93}},\ \bibinfo
  {pages} {032104} (\bibinfo {year} {2016})}\BibitemShut {NoStop}%
\bibitem [{\citenamefont {Basko}\ \emph {et~al.}(2006)\citenamefont {Basko},
  \citenamefont {Aleiner},\ and\ \citenamefont {Altshuler}}]{Basko2006}%
  \BibitemOpen
  \bibfield  {author} {\bibinfo {author} {\bibfnamefont {D.}~\bibnamefont
  {Basko}}, \bibinfo {author} {\bibfnamefont {I.}~\bibnamefont {Aleiner}},\
  and\ \bibinfo {author} {\bibfnamefont {B.}~\bibnamefont {Altshuler}},\
  }\bibfield  {title} {\bibinfo {title} {Metal–insulator transition in a
  weakly interacting many-electron system with localized single-particle
  states},\ }\href {https://doi.org/https://doi.org/10.1016/j.aop.2005.11.014}
  {\bibfield  {journal} {\bibinfo  {journal} {Annals of Physics}\ }\textbf
  {\bibinfo {volume} {321}},\ \bibinfo {pages} {1126} (\bibinfo {year}
  {2006})}\BibitemShut {NoStop}%
\bibitem [{\citenamefont {Gornyi}\ \emph {et~al.}(2005)\citenamefont {Gornyi},
  \citenamefont {Mirlin},\ and\ \citenamefont {Polyakov}}]{Polyakov2005}%
  \BibitemOpen
  \bibfield  {author} {\bibinfo {author} {\bibfnamefont {I.~V.}\ \bibnamefont
  {Gornyi}}, \bibinfo {author} {\bibfnamefont {A.~D.}\ \bibnamefont {Mirlin}},\
  and\ \bibinfo {author} {\bibfnamefont {D.~G.}\ \bibnamefont {Polyakov}},\
  }\bibfield  {title} {\bibinfo {title} {Interacting electrons in disordered
  wires: Anderson localization and low-{$T$} transport},\ }\href
  {https://doi.org/10.1103/PhysRevLett.95.206603} {\bibfield  {journal}
  {\bibinfo  {journal} {Phys. Rev. Lett.}\ }\textbf {\bibinfo {volume} {95}},\
  \bibinfo {pages} {206603} (\bibinfo {year} {2005})}\BibitemShut {NoStop}%
\bibitem [{\citenamefont {Oganesyan}\ and\ \citenamefont
  {Huse}(2007)}]{Huse2007}%
  \BibitemOpen
  \bibfield  {author} {\bibinfo {author} {\bibfnamefont {V.}~\bibnamefont
  {Oganesyan}}\ and\ \bibinfo {author} {\bibfnamefont {D.~A.}\ \bibnamefont
  {Huse}},\ }\bibfield  {title} {\bibinfo {title} {Localization of interacting
  fermions at high temperature},\ }\href
  {https://doi.org/10.1103/PhysRevB.75.155111} {\bibfield  {journal} {\bibinfo
  {journal} {Phys. Rev. B}\ }\textbf {\bibinfo {volume} {75}},\ \bibinfo
  {pages} {155111} (\bibinfo {year} {2007})}\BibitemShut {NoStop}%
\bibitem [{\citenamefont {Pal}\ and\ \citenamefont {Huse}(2010)}]{Huse2010}%
  \BibitemOpen
  \bibfield  {author} {\bibinfo {author} {\bibfnamefont {A.}~\bibnamefont
  {Pal}}\ and\ \bibinfo {author} {\bibfnamefont {D.~A.}\ \bibnamefont {Huse}},\
  }\bibfield  {title} {\bibinfo {title} {Many-body localization phase
  transition},\ }\href {https://doi.org/10.1103/PhysRevB.82.174411} {\bibfield
  {journal} {\bibinfo  {journal} {Phys. Rev. B}\ }\textbf {\bibinfo {volume}
  {82}},\ \bibinfo {pages} {174411} (\bibinfo {year} {2010})}\BibitemShut
  {NoStop}%
\bibitem [{\citenamefont {\ifmmode \check{Z}\else
  \v{Z}\fi{}nidari\ifmmode~\check{c}\else \v{c}\fi{}}\ \emph
  {et~al.}(2008)\citenamefont {\ifmmode \check{Z}\else
  \v{Z}\fi{}nidari\ifmmode~\check{c}\else \v{c}\fi{}}, \citenamefont {Prosen},\
  and\ \citenamefont {Prelov\ifmmode~\check{s}\else
  \v{s}\fi{}ek}}]{Znidaric2008}%
  \BibitemOpen
  \bibfield  {author} {\bibinfo {author} {\bibfnamefont {M.}~\bibnamefont
  {\ifmmode \check{Z}\else \v{Z}\fi{}nidari\ifmmode~\check{c}\else
  \v{c}\fi{}}}, \bibinfo {author} {\bibfnamefont {T.~c.~v.}\ \bibnamefont
  {Prosen}},\ and\ \bibinfo {author} {\bibfnamefont {P.}~\bibnamefont
  {Prelov\ifmmode~\check{s}\else \v{s}\fi{}ek}},\ }\bibfield  {title} {\bibinfo
  {title} {Many-body localization in the {H}eisenberg {$XXZ$} magnet in a
  random field},\ }\href {https://doi.org/10.1103/PhysRevB.77.064426}
  {\bibfield  {journal} {\bibinfo  {journal} {Phys. Rev. B}\ }\textbf {\bibinfo
  {volume} {77}},\ \bibinfo {pages} {064426} (\bibinfo {year}
  {2008})}\BibitemShut {NoStop}%
\bibitem [{\citenamefont {Iyer}\ \emph {et~al.}(2013)\citenamefont {Iyer},
  \citenamefont {Oganesyan}, \citenamefont {Refael},\ and\ \citenamefont
  {Huse}}]{Iyer2013}%
  \BibitemOpen
  \bibfield  {author} {\bibinfo {author} {\bibfnamefont {S.}~\bibnamefont
  {Iyer}}, \bibinfo {author} {\bibfnamefont {V.}~\bibnamefont {Oganesyan}},
  \bibinfo {author} {\bibfnamefont {G.}~\bibnamefont {Refael}},\ and\ \bibinfo
  {author} {\bibfnamefont {D.~A.}\ \bibnamefont {Huse}},\ }\bibfield  {title}
  {\bibinfo {title} {Many-body localization in a quasiperiodic system},\ }\href
  {https://doi.org/10.1103/PhysRevB.87.134202} {\bibfield  {journal} {\bibinfo
  {journal} {Phys. Rev. B}\ }\textbf {\bibinfo {volume} {87}},\ \bibinfo
  {pages} {134202} (\bibinfo {year} {2013})}\BibitemShut {NoStop}%
\bibitem [{\citenamefont {Setiawan}\ \emph {et~al.}(2017)\citenamefont
  {Setiawan}, \citenamefont {Deng},\ and\ \citenamefont
  {Pixley}}]{Setaiwan2017}%
  \BibitemOpen
  \bibfield  {author} {\bibinfo {author} {\bibfnamefont {F.}~\bibnamefont
  {Setiawan}}, \bibinfo {author} {\bibfnamefont {D.-L.}\ \bibnamefont {Deng}},\
  and\ \bibinfo {author} {\bibfnamefont {J.~H.}\ \bibnamefont {Pixley}},\
  }\bibfield  {title} {\bibinfo {title} {Transport properties across the
  many-body localization transition in quasiperiodic and random systems},\
  }\href {https://doi.org/10.1103/PhysRevB.96.104205} {\bibfield  {journal}
  {\bibinfo  {journal} {Phys. Rev. B}\ }\textbf {\bibinfo {volume} {96}},\
  \bibinfo {pages} {104205} (\bibinfo {year} {2017})}\BibitemShut {NoStop}%
\bibitem [{\citenamefont {Zhang}\ and\ \citenamefont {Yao}(2018)}]{Zhang2018}%
  \BibitemOpen
  \bibfield  {author} {\bibinfo {author} {\bibfnamefont {S.-X.}\ \bibnamefont
  {Zhang}}\ and\ \bibinfo {author} {\bibfnamefont {H.}~\bibnamefont {Yao}},\
  }\bibfield  {title} {\bibinfo {title} {Universal properties of many-body
  localization transitions in quasiperiodic systems},\ }\href
  {https://doi.org/10.1103/PhysRevLett.121.206601} {\bibfield  {journal}
  {\bibinfo  {journal} {Phys. Rev. Lett.}\ }\textbf {\bibinfo {volume} {121}},\
  \bibinfo {pages} {206601} (\bibinfo {year} {2018})}\BibitemShut {NoStop}%
\bibitem [{\citenamefont {Singh}\ \emph {et~al.}(2021)\citenamefont {Singh},
  \citenamefont {Ware}, \citenamefont {Vasseur},\ and\ \citenamefont
  {Gopalakrishnan}}]{Singh2021}%
  \BibitemOpen
  \bibfield  {author} {\bibinfo {author} {\bibfnamefont {H.}~\bibnamefont
  {Singh}}, \bibinfo {author} {\bibfnamefont {B.}~\bibnamefont {Ware}},
  \bibinfo {author} {\bibfnamefont {R.}~\bibnamefont {Vasseur}},\ and\ \bibinfo
  {author} {\bibfnamefont {S.}~\bibnamefont {Gopalakrishnan}},\ }\bibfield
  {title} {\bibinfo {title} {Local integrals of motion and the quasiperiodic
  many-body localization transition},\ }\href
  {https://doi.org/10.1103/PhysRevB.103.L220201} {\bibfield  {journal}
  {\bibinfo  {journal} {Phys. Rev. B}\ }\textbf {\bibinfo {volume} {103}},\
  \bibinfo {pages} {L220201} (\bibinfo {year} {2021})}\BibitemShut {NoStop}%
\bibitem [{\citenamefont {Sierant}\ \emph {et~al.}(2017)\citenamefont
  {Sierant}, \citenamefont {Delande},\ and\ \citenamefont
  {Zakrzewski}}]{Sierant2017}%
  \BibitemOpen
  \bibfield  {author} {\bibinfo {author} {\bibfnamefont {P.}~\bibnamefont
  {Sierant}}, \bibinfo {author} {\bibfnamefont {D.}~\bibnamefont {Delande}},\
  and\ \bibinfo {author} {\bibfnamefont {J.}~\bibnamefont {Zakrzewski}},\
  }\bibfield  {title} {\bibinfo {title} {Many-body localization due to random
  interactions},\ }\href {https://doi.org/10.1103/PhysRevA.95.021601}
  {\bibfield  {journal} {\bibinfo  {journal} {Phys. Rev. A}\ }\textbf {\bibinfo
  {volume} {95}},\ \bibinfo {pages} {021601(R)} (\bibinfo {year}
  {2017})}\BibitemShut {NoStop}%
\bibitem [{\citenamefont {Kj\"all}\ \emph {et~al.}(2014)\citenamefont
  {Kj\"all}, \citenamefont {Bardarson},\ and\ \citenamefont
  {Pollmann}}]{Pollmann2014}%
  \BibitemOpen
  \bibfield  {author} {\bibinfo {author} {\bibfnamefont {J.~A.}\ \bibnamefont
  {Kj\"all}}, \bibinfo {author} {\bibfnamefont {J.~H.}\ \bibnamefont
  {Bardarson}},\ and\ \bibinfo {author} {\bibfnamefont {F.}~\bibnamefont
  {Pollmann}},\ }\bibfield  {title} {\bibinfo {title} {Many-body localization
  in a disordered quantum {I}sing chain},\ }\href
  {https://doi.org/10.1103/PhysRevLett.113.107204} {\bibfield  {journal}
  {\bibinfo  {journal} {Phys. Rev. Lett.}\ }\textbf {\bibinfo {volume} {113}},\
  \bibinfo {pages} {107204} (\bibinfo {year} {2014})}\BibitemShut {NoStop}%
\bibitem [{\citenamefont {Vasseur}\ \emph {et~al.}(2016)\citenamefont
  {Vasseur}, \citenamefont {Friedman}, \citenamefont {Parameswaran},\ and\
  \citenamefont {Potter}}]{Potter2016}%
  \BibitemOpen
  \bibfield  {author} {\bibinfo {author} {\bibfnamefont {R.}~\bibnamefont
  {Vasseur}}, \bibinfo {author} {\bibfnamefont {A.~J.}\ \bibnamefont
  {Friedman}}, \bibinfo {author} {\bibfnamefont {S.~A.}\ \bibnamefont
  {Parameswaran}},\ and\ \bibinfo {author} {\bibfnamefont {A.~C.}\ \bibnamefont
  {Potter}},\ }\bibfield  {title} {\bibinfo {title} {Particle-hole symmetry,
  many-body localization, and topological edge modes},\ }\href
  {https://doi.org/10.1103/PhysRevB.93.134207} {\bibfield  {journal} {\bibinfo
  {journal} {Phys. Rev. B}\ }\textbf {\bibinfo {volume} {93}},\ \bibinfo
  {pages} {134207} (\bibinfo {year} {2016})}\BibitemShut {NoStop}%
\bibitem [{\citenamefont {Schulz}\ \emph {et~al.}(2019)\citenamefont {Schulz},
  \citenamefont {Hooley}, \citenamefont {Moessner},\ and\ \citenamefont
  {Pollmann}}]{Schulz2019}%
  \BibitemOpen
  \bibfield  {author} {\bibinfo {author} {\bibfnamefont {M.}~\bibnamefont
  {Schulz}}, \bibinfo {author} {\bibfnamefont {C.~A.}\ \bibnamefont {Hooley}},
  \bibinfo {author} {\bibfnamefont {R.}~\bibnamefont {Moessner}},\ and\
  \bibinfo {author} {\bibfnamefont {F.}~\bibnamefont {Pollmann}},\ }\bibfield
  {title} {\bibinfo {title} {{S}tark many-body localization},\ }\href
  {https://doi.org/10.1103/PhysRevLett.122.040606} {\bibfield  {journal}
  {\bibinfo  {journal} {Phys. Rev. Lett.}\ }\textbf {\bibinfo {volume} {122}},\
  \bibinfo {pages} {040606} (\bibinfo {year} {2019})}\BibitemShut {NoStop}%
\bibitem [{\citenamefont {van Nieuwenburg}\ \emph {et~al.}(2019)\citenamefont
  {van Nieuwenburg}, \citenamefont {Baum},\ and\ \citenamefont
  {Refael}}]{Evert2019}%
  \BibitemOpen
  \bibfield  {author} {\bibinfo {author} {\bibfnamefont {E.}~\bibnamefont {van
  Nieuwenburg}}, \bibinfo {author} {\bibfnamefont {Y.}~\bibnamefont {Baum}},\
  and\ \bibinfo {author} {\bibfnamefont {G.}~\bibnamefont {Refael}},\
  }\bibfield  {title} {\bibinfo {title} {From {B}loch oscillations to many-body
  localization in clean interacting systems},\ }\href
  {https://doi.org/10.1073/pnas.1819316116} {\bibfield  {journal} {\bibinfo
  {journal} {Proceedings of the National Academy of Sciences}\ }\textbf
  {\bibinfo {volume} {116}},\ \bibinfo {pages} {9269} (\bibinfo {year}
  {2019})}\BibitemShut {NoStop}%
\bibitem [{\citenamefont {Zhang}\ \emph {et~al.}(2021)\citenamefont {Zhang},
  \citenamefont {Ke}, \citenamefont {Liu},\ and\ \citenamefont
  {Lee}}]{Zhang2021}%
  \BibitemOpen
  \bibfield  {author} {\bibinfo {author} {\bibfnamefont {L.}~\bibnamefont
  {Zhang}}, \bibinfo {author} {\bibfnamefont {Y.}~\bibnamefont {Ke}}, \bibinfo
  {author} {\bibfnamefont {W.}~\bibnamefont {Liu}},\ and\ \bibinfo {author}
  {\bibfnamefont {C.}~\bibnamefont {Lee}},\ }\bibfield  {title} {\bibinfo
  {title} {Mobility edge of {S}tark many-body localization},\ }\href
  {https://doi.org/10.1103/PhysRevA.103.023323} {\bibfield  {journal} {\bibinfo
   {journal} {Phys. Rev. A}\ }\textbf {\bibinfo {volume} {103}},\ \bibinfo
  {pages} {023323} (\bibinfo {year} {2021})}\BibitemShut {NoStop}%
\bibitem [{\citenamefont {Bairey}\ \emph {et~al.}(2017)\citenamefont {Bairey},
  \citenamefont {Refael},\ and\ \citenamefont {Lindner}}]{Bairey2017}%
  \BibitemOpen
  \bibfield  {author} {\bibinfo {author} {\bibfnamefont {E.}~\bibnamefont
  {Bairey}}, \bibinfo {author} {\bibfnamefont {G.}~\bibnamefont {Refael}},\
  and\ \bibinfo {author} {\bibfnamefont {N.~H.}\ \bibnamefont {Lindner}},\
  }\bibfield  {title} {\bibinfo {title} {Driving induced many-body
  localization},\ }\href {https://doi.org/10.1103/PhysRevB.96.020201}
  {\bibfield  {journal} {\bibinfo  {journal} {Phys. Rev. B}\ }\textbf {\bibinfo
  {volume} {96}},\ \bibinfo {pages} {020201(R)} (\bibinfo {year}
  {2017})}\BibitemShut {NoStop}%
\bibitem [{\citenamefont {Choi}\ \emph {et~al.}(2018)\citenamefont {Choi},
  \citenamefont {Abanin},\ and\ \citenamefont {Lukin}}]{Choi2018}%
  \BibitemOpen
  \bibfield  {author} {\bibinfo {author} {\bibfnamefont {S.}~\bibnamefont
  {Choi}}, \bibinfo {author} {\bibfnamefont {D.~A.}\ \bibnamefont {Abanin}},\
  and\ \bibinfo {author} {\bibfnamefont {M.~D.}\ \bibnamefont {Lukin}},\
  }\bibfield  {title} {\bibinfo {title} {Dynamically induced many-body
  localization},\ }\href {https://doi.org/10.1103/PhysRevB.97.100301}
  {\bibfield  {journal} {\bibinfo  {journal} {Phys. Rev. B}\ }\textbf {\bibinfo
  {volume} {97}},\ \bibinfo {pages} {100301(R)} (\bibinfo {year}
  {2018})}\BibitemShut {NoStop}%
\bibitem [{\citenamefont {Bhakuni}\ \emph {et~al.}(2020)\citenamefont
  {Bhakuni}, \citenamefont {Nehra},\ and\ \citenamefont
  {Sharma}}]{Bhakuni2020}%
  \BibitemOpen
  \bibfield  {author} {\bibinfo {author} {\bibfnamefont {D.~S.}\ \bibnamefont
  {Bhakuni}}, \bibinfo {author} {\bibfnamefont {R.}~\bibnamefont {Nehra}},\
  and\ \bibinfo {author} {\bibfnamefont {A.}~\bibnamefont {Sharma}},\
  }\bibfield  {title} {\bibinfo {title} {Drive-induced many-body localization
  and coherent destruction of {S}tark many-body localization},\ }\href
  {https://doi.org/10.1103/PhysRevB.102.024201} {\bibfield  {journal} {\bibinfo
   {journal} {Phys. Rev. B}\ }\textbf {\bibinfo {volume} {102}},\ \bibinfo
  {pages} {024201} (\bibinfo {year} {2020})}\BibitemShut {NoStop}%
\bibitem [{\citenamefont {Yousefjani}\ \emph {et~al.}(2023)\citenamefont
  {Yousefjani}, \citenamefont {Bose},\ and\ \citenamefont
  {Bayat}}]{Yousefjani2023}%
  \BibitemOpen
  \bibfield  {author} {\bibinfo {author} {\bibfnamefont {R.}~\bibnamefont
  {Yousefjani}}, \bibinfo {author} {\bibfnamefont {S.}~\bibnamefont {Bose}},\
  and\ \bibinfo {author} {\bibfnamefont {A.}~\bibnamefont {Bayat}},\ }\bibfield
   {title} {\bibinfo {title} {Floquet-induced localization in long-range
  many-body systems},\ }\href
  {https://doi.org/10.1103/PhysRevResearch.5.013094} {\bibfield  {journal}
  {\bibinfo  {journal} {Phys. Rev. Res.}\ }\textbf {\bibinfo {volume} {5}},\
  \bibinfo {pages} {013094} (\bibinfo {year} {2023})}\BibitemShut {NoStop}%
\bibitem [{\citenamefont {Serbyn}\ \emph {et~al.}(2013)\citenamefont {Serbyn},
  \citenamefont {Papi\ifmmode~\acute{c}\else \'{c}\fi{}},\ and\ \citenamefont
  {Abanin}}]{Abanin2013}%
  \BibitemOpen
  \bibfield  {author} {\bibinfo {author} {\bibfnamefont {M.}~\bibnamefont
  {Serbyn}}, \bibinfo {author} {\bibfnamefont {Z.}~\bibnamefont
  {Papi\ifmmode~\acute{c}\else \'{c}\fi{}}},\ and\ \bibinfo {author}
  {\bibfnamefont {D.~A.}\ \bibnamefont {Abanin}},\ }\bibfield  {title}
  {\bibinfo {title} {Local conservation laws and the structure of the many-body
  localized states},\ }\href {https://doi.org/10.1103/PhysRevLett.111.127201}
  {\bibfield  {journal} {\bibinfo  {journal} {Phys. Rev. Lett.}\ }\textbf
  {\bibinfo {volume} {111}},\ \bibinfo {pages} {127201} (\bibinfo {year}
  {2013})}\BibitemShut {NoStop}%
\bibitem [{\citenamefont {Huse}\ \emph {et~al.}(2014)\citenamefont {Huse},
  \citenamefont {Nandkishore},\ and\ \citenamefont {Oganesyan}}]{Huse2014NovB}%
  \BibitemOpen
  \bibfield  {author} {\bibinfo {author} {\bibfnamefont {D.~A.}\ \bibnamefont
  {Huse}}, \bibinfo {author} {\bibfnamefont {R.}~\bibnamefont {Nandkishore}},\
  and\ \bibinfo {author} {\bibfnamefont {V.}~\bibnamefont {Oganesyan}},\
  }\bibfield  {title} {\bibinfo {title} {Phenomenology of fully
  many-body-localized systems},\ }\href
  {https://doi.org/10.1103/PhysRevB.90.174202} {\bibfield  {journal} {\bibinfo
  {journal} {Phys. Rev. B}\ }\textbf {\bibinfo {volume} {90}},\ \bibinfo
  {pages} {174202} (\bibinfo {year} {2014})}\BibitemShut {NoStop}%
\bibitem [{\citenamefont {Schreiber}\ \emph {et~al.}(2015)\citenamefont
  {Schreiber}, \citenamefont {Hodgman}, \citenamefont {Bordia}, \citenamefont
  {Lüschen}, \citenamefont {Fischer}, \citenamefont {Vosk}, \citenamefont
  {Altman}, \citenamefont {Schneider},\ and\ \citenamefont
  {Bloch}}]{Schreiber2015}%
  \BibitemOpen
  \bibfield  {author} {\bibinfo {author} {\bibfnamefont {M.}~\bibnamefont
  {Schreiber}}, \bibinfo {author} {\bibfnamefont {S.~S.}\ \bibnamefont
  {Hodgman}}, \bibinfo {author} {\bibfnamefont {P.}~\bibnamefont {Bordia}},
  \bibinfo {author} {\bibfnamefont {H.~P.}\ \bibnamefont {Lüschen}}, \bibinfo
  {author} {\bibfnamefont {M.~H.}\ \bibnamefont {Fischer}}, \bibinfo {author}
  {\bibfnamefont {R.}~\bibnamefont {Vosk}}, \bibinfo {author} {\bibfnamefont
  {E.}~\bibnamefont {Altman}}, \bibinfo {author} {\bibfnamefont
  {U.}~\bibnamefont {Schneider}},\ and\ \bibinfo {author} {\bibfnamefont
  {I.}~\bibnamefont {Bloch}},\ }\bibfield  {title} {\bibinfo {title}
  {Observation of many-body localization of interacting fermions in a
  quasirandom optical lattice},\ }\href
  {https://doi.org/10.1126/science.aaa7432} {\bibfield  {journal} {\bibinfo
  {journal} {Science}\ }\textbf {\bibinfo {volume} {349}},\ \bibinfo {pages}
  {842} (\bibinfo {year} {2015})}\BibitemShut {NoStop}%
\bibitem [{\citenamefont {yoon Choi}\ \emph {et~al.}(2016)\citenamefont {yoon
  Choi}, \citenamefont {Hild}, \citenamefont {Zeiher}, \citenamefont {Schauß},
  \citenamefont {Rubio-Abadal}, \citenamefont {Yefsah}, \citenamefont
  {Khemani}, \citenamefont {Huse}, \citenamefont {Bloch},\ and\ \citenamefont
  {Gross}}]{Choi2016}%
  \BibitemOpen
  \bibfield  {author} {\bibinfo {author} {\bibfnamefont {J.}~\bibnamefont {yoon
  Choi}}, \bibinfo {author} {\bibfnamefont {S.}~\bibnamefont {Hild}}, \bibinfo
  {author} {\bibfnamefont {J.}~\bibnamefont {Zeiher}}, \bibinfo {author}
  {\bibfnamefont {P.}~\bibnamefont {Schauß}}, \bibinfo {author} {\bibfnamefont
  {A.}~\bibnamefont {Rubio-Abadal}}, \bibinfo {author} {\bibfnamefont
  {T.}~\bibnamefont {Yefsah}}, \bibinfo {author} {\bibfnamefont
  {V.}~\bibnamefont {Khemani}}, \bibinfo {author} {\bibfnamefont {D.~A.}\
  \bibnamefont {Huse}}, \bibinfo {author} {\bibfnamefont {I.}~\bibnamefont
  {Bloch}},\ and\ \bibinfo {author} {\bibfnamefont {C.}~\bibnamefont {Gross}},\
  }\bibfield  {title} {\bibinfo {title} {Exploring the many-body localization
  transition in two dimensions},\ }\href
  {https://doi.org/10.1126/science.aaf8834} {\bibfield  {journal} {\bibinfo
  {journal} {Science}\ }\textbf {\bibinfo {volume} {352}},\ \bibinfo {pages}
  {1547} (\bibinfo {year} {2016})}\BibitemShut {NoStop}%
\bibitem [{\citenamefont {Smith}\ \emph {et~al.}(2016)\citenamefont {Smith},
  \citenamefont {Lee}, \citenamefont {Richerme}, \citenamefont {Neyenhuis},
  \citenamefont {Hess}, \citenamefont {Hauke}, \citenamefont {Heyl},
  \citenamefont {Huse},\ and\ \citenamefont {Monroe}}]{Smidt2016}%
  \BibitemOpen
  \bibfield  {author} {\bibinfo {author} {\bibfnamefont {J.}~\bibnamefont
  {Smith}}, \bibinfo {author} {\bibfnamefont {A.}~\bibnamefont {Lee}}, \bibinfo
  {author} {\bibfnamefont {P.}~\bibnamefont {Richerme}}, \bibinfo {author}
  {\bibfnamefont {B.}~\bibnamefont {Neyenhuis}}, \bibinfo {author}
  {\bibfnamefont {P.~W.}\ \bibnamefont {Hess}}, \bibinfo {author}
  {\bibfnamefont {P.}~\bibnamefont {Hauke}}, \bibinfo {author} {\bibfnamefont
  {M.}~\bibnamefont {Heyl}}, \bibinfo {author} {\bibfnamefont {D.~A.}\
  \bibnamefont {Huse}},\ and\ \bibinfo {author} {\bibfnamefont
  {C.}~\bibnamefont {Monroe}},\ }\bibfield  {title} {\bibinfo {title}
  {Many-body localization in a quantum simulator with programmable random
  disorder},\ }\href {https://doi.org/10.1038/nphys3783} {\bibfield  {journal}
  {\bibinfo  {journal} {Nature Physics}\ }\textbf {\bibinfo {volume} {12}},\
  \bibinfo {pages} {907} (\bibinfo {year} {2016})}\BibitemShut {NoStop}%
\bibitem [{\citenamefont {Xu}\ \emph {et~al.}(2018)\citenamefont {Xu},
  \citenamefont {Chen}, \citenamefont {Zeng}, \citenamefont {Zhang},
  \citenamefont {Song}, \citenamefont {Liu}, \citenamefont {Guo}, \citenamefont
  {Zhang}, \citenamefont {Xu}, \citenamefont {Deng}, \citenamefont {Huang},
  \citenamefont {Wang}, \citenamefont {Zhu}, \citenamefont {Zheng},\ and\
  \citenamefont {Fan}}]{Xu2018}%
  \BibitemOpen
  \bibfield  {author} {\bibinfo {author} {\bibfnamefont {K.}~\bibnamefont
  {Xu}}, \bibinfo {author} {\bibfnamefont {J.-J.}\ \bibnamefont {Chen}},
  \bibinfo {author} {\bibfnamefont {Y.}~\bibnamefont {Zeng}}, \bibinfo {author}
  {\bibfnamefont {Y.-R.}\ \bibnamefont {Zhang}}, \bibinfo {author}
  {\bibfnamefont {C.}~\bibnamefont {Song}}, \bibinfo {author} {\bibfnamefont
  {W.}~\bibnamefont {Liu}}, \bibinfo {author} {\bibfnamefont {Q.}~\bibnamefont
  {Guo}}, \bibinfo {author} {\bibfnamefont {P.}~\bibnamefont {Zhang}}, \bibinfo
  {author} {\bibfnamefont {D.}~\bibnamefont {Xu}}, \bibinfo {author}
  {\bibfnamefont {H.}~\bibnamefont {Deng}}, \bibinfo {author} {\bibfnamefont
  {K.}~\bibnamefont {Huang}}, \bibinfo {author} {\bibfnamefont
  {H.}~\bibnamefont {Wang}}, \bibinfo {author} {\bibfnamefont {X.}~\bibnamefont
  {Zhu}}, \bibinfo {author} {\bibfnamefont {D.}~\bibnamefont {Zheng}},\ and\
  \bibinfo {author} {\bibfnamefont {H.}~\bibnamefont {Fan}},\ }\bibfield
  {title} {\bibinfo {title} {Emulating many-body localization with a
  superconducting quantum processor},\ }\href
  {https://doi.org/10.1103/PhysRevLett.120.050507} {\bibfield  {journal}
  {\bibinfo  {journal} {Phys. Rev. Lett.}\ }\textbf {\bibinfo {volume} {120}},\
  \bibinfo {pages} {050507} (\bibinfo {year} {2018})}\BibitemShut {NoStop}%
\bibitem [{\citenamefont {Imbrie}(2016)}]{Imbrie2016}%
  \BibitemOpen
  \bibfield  {author} {\bibinfo {author} {\bibfnamefont {J.~Z.}\ \bibnamefont
  {Imbrie}},\ }\bibfield  {title} {\bibinfo {title} {Diagonalization and
  many-body localization for a disordered quantum spin chain},\ }\href
  {https://doi.org/10.1103/PhysRevLett.117.027201} {\bibfield  {journal}
  {\bibinfo  {journal} {Phys. Rev. Lett.}\ }\textbf {\bibinfo {volume} {117}},\
  \bibinfo {pages} {027201} (\bibinfo {year} {2016})}\BibitemShut {NoStop}%
\bibitem [{\citenamefont {\ifmmode~\check{S}\else \v{S}\fi{}untajs}\ \emph
  {et~al.}(2020{\natexlab{a}})\citenamefont {\ifmmode~\check{S}\else
  \v{S}\fi{}untajs}, \citenamefont {Bon\ifmmode~\check{c}\else \v{c}\fi{}a},
  \citenamefont {Prosen},\ and\ \citenamefont {Vidmar}}]{Suntajs2020}%
  \BibitemOpen
  \bibfield  {author} {\bibinfo {author} {\bibfnamefont {J.}~\bibnamefont
  {\ifmmode~\check{S}\else \v{S}\fi{}untajs}}, \bibinfo {author} {\bibfnamefont
  {J.}~\bibnamefont {Bon\ifmmode~\check{c}\else \v{c}\fi{}a}}, \bibinfo
  {author} {\bibfnamefont {T.~c.~v.}\ \bibnamefont {Prosen}},\ and\ \bibinfo
  {author} {\bibfnamefont {L.}~\bibnamefont {Vidmar}},\ }\bibfield  {title}
  {\bibinfo {title} {Ergodicity breaking transition in finite disordered spin
  chains},\ }\href {https://doi.org/10.1103/PhysRevB.102.064207} {\bibfield
  {journal} {\bibinfo  {journal} {Phys. Rev. B}\ }\textbf {\bibinfo {volume}
  {102}},\ \bibinfo {pages} {064207} (\bibinfo {year}
  {2020}{\natexlab{a}})}\BibitemShut {NoStop}%
\bibitem [{\citenamefont {Luitz}\ and\ \citenamefont {Lev}(2020)}]{Luitz2020}%
  \BibitemOpen
  \bibfield  {author} {\bibinfo {author} {\bibfnamefont {D.~J.}\ \bibnamefont
  {Luitz}}\ and\ \bibinfo {author} {\bibfnamefont {Y.~B.}\ \bibnamefont
  {Lev}},\ }\bibfield  {title} {\bibinfo {title} {Absence of slow particle
  transport in the many-body localized phase},\ }\href
  {https://doi.org/10.1103/PhysRevB.102.100202} {\bibfield  {journal} {\bibinfo
   {journal} {Phys. Rev. B}\ }\textbf {\bibinfo {volume} {102}},\ \bibinfo
  {pages} {100202(R)} (\bibinfo {year} {2020})}\BibitemShut {NoStop}%
\bibitem [{\citenamefont {\ifmmode~\check{S}\else \v{S}\fi{}untajs}\ \emph
  {et~al.}(2020{\natexlab{b}})\citenamefont {\ifmmode~\check{S}\else
  \v{S}\fi{}untajs}, \citenamefont {Bon\ifmmode~\check{c}\else \v{c}\fi{}a},
  \citenamefont {Prosen},\ and\ \citenamefont {Vidmar}}]{Suntajs2020Dec}%
  \BibitemOpen
  \bibfield  {author} {\bibinfo {author} {\bibfnamefont {J.}~\bibnamefont
  {\ifmmode~\check{S}\else \v{S}\fi{}untajs}}, \bibinfo {author} {\bibfnamefont
  {J.}~\bibnamefont {Bon\ifmmode~\check{c}\else \v{c}\fi{}a}}, \bibinfo
  {author} {\bibfnamefont {T.~c.~v.}\ \bibnamefont {Prosen}},\ and\ \bibinfo
  {author} {\bibfnamefont {L.}~\bibnamefont {Vidmar}},\ }\bibfield  {title}
  {\bibinfo {title} {Quantum chaos challenges many-body localization},\ }\href
  {https://doi.org/10.1103/PhysRevE.102.062144} {\bibfield  {journal} {\bibinfo
   {journal} {Phys. Rev. E}\ }\textbf {\bibinfo {volume} {102}},\ \bibinfo
  {pages} {062144} (\bibinfo {year} {2020}{\natexlab{b}})}\BibitemShut
  {NoStop}%
\bibitem [{\citenamefont {Kiefer-Emmanouilidis}\ \emph
  {et~al.}(2021)\citenamefont {Kiefer-Emmanouilidis}, \citenamefont {Unanyan},
  \citenamefont {Fleischhauer},\ and\ \citenamefont
  {Sirker}}]{Kiefer-Emmanouilidis2021}%
  \BibitemOpen
  \bibfield  {author} {\bibinfo {author} {\bibfnamefont {M.}~\bibnamefont
  {Kiefer-Emmanouilidis}}, \bibinfo {author} {\bibfnamefont {R.}~\bibnamefont
  {Unanyan}}, \bibinfo {author} {\bibfnamefont {M.}~\bibnamefont
  {Fleischhauer}},\ and\ \bibinfo {author} {\bibfnamefont {J.}~\bibnamefont
  {Sirker}},\ }\bibfield  {title} {\bibinfo {title} {Slow delocalization of
  particles in many-body localized phases},\ }\href
  {https://doi.org/10.1103/PhysRevB.103.024203} {\bibfield  {journal} {\bibinfo
   {journal} {Phys. Rev. B}\ }\textbf {\bibinfo {volume} {103}},\ \bibinfo
  {pages} {024203} (\bibinfo {year} {2021})}\BibitemShut {NoStop}%
\bibitem [{\citenamefont {Abanin}\ \emph {et~al.}(2021)\citenamefont {Abanin},
  \citenamefont {Bardarson}, \citenamefont {{De Tomasi}}, \citenamefont
  {Gopalakrishnan}, \citenamefont {Khemani}, \citenamefont {Parameswaran},
  \citenamefont {Pollmann}, \citenamefont {Potter}, \citenamefont {Serbyn},\
  and\ \citenamefont {Vasseur}}]{Abanin2021}%
  \BibitemOpen
  \bibfield  {author} {\bibinfo {author} {\bibfnamefont {D.}~\bibnamefont
  {Abanin}}, \bibinfo {author} {\bibfnamefont {J.}~\bibnamefont {Bardarson}},
  \bibinfo {author} {\bibfnamefont {G.}~\bibnamefont {{De Tomasi}}}, \bibinfo
  {author} {\bibfnamefont {S.}~\bibnamefont {Gopalakrishnan}}, \bibinfo
  {author} {\bibfnamefont {V.}~\bibnamefont {Khemani}}, \bibinfo {author}
  {\bibfnamefont {S.}~\bibnamefont {Parameswaran}}, \bibinfo {author}
  {\bibfnamefont {F.}~\bibnamefont {Pollmann}}, \bibinfo {author}
  {\bibfnamefont {A.}~\bibnamefont {Potter}}, \bibinfo {author} {\bibfnamefont
  {M.}~\bibnamefont {Serbyn}},\ and\ \bibinfo {author} {\bibfnamefont
  {R.}~\bibnamefont {Vasseur}},\ }\bibfield  {title} {\bibinfo {title}
  {Distinguishing localization from chaos: Challenges in finite-size systems},\
  }\href {https://doi.org/https://doi.org/10.1016/j.aop.2021.168415} {\bibfield
   {journal} {\bibinfo  {journal} {Annals of Physics}\ }\textbf {\bibinfo
  {volume} {427}},\ \bibinfo {pages} {168415} (\bibinfo {year}
  {2021})}\BibitemShut {NoStop}%
\bibitem [{\citenamefont {Moudgalya}\ \emph
  {et~al.}(2018{\natexlab{a}})\citenamefont {Moudgalya}, \citenamefont
  {Rachel}, \citenamefont {Bernevig},\ and\ \citenamefont
  {Regnault}}]{Moudgalya_I}%
  \BibitemOpen
  \bibfield  {author} {\bibinfo {author} {\bibfnamefont {S.}~\bibnamefont
  {Moudgalya}}, \bibinfo {author} {\bibfnamefont {S.}~\bibnamefont {Rachel}},
  \bibinfo {author} {\bibfnamefont {B.~A.}\ \bibnamefont {Bernevig}},\ and\
  \bibinfo {author} {\bibfnamefont {N.}~\bibnamefont {Regnault}},\ }\bibfield
  {title} {\bibinfo {title} {Exact excited states of nonintegrable models},\
  }\href {https://doi.org/10.1103/PhysRevB.98.235155} {\bibfield  {journal}
  {\bibinfo  {journal} {Phys. Rev. B}\ }\textbf {\bibinfo {volume} {98}},\
  \bibinfo {pages} {235155} (\bibinfo {year} {2018}{\natexlab{a}})}\BibitemShut
  {NoStop}%
\bibitem [{\citenamefont {Moudgalya}\ \emph
  {et~al.}(2018{\natexlab{b}})\citenamefont {Moudgalya}, \citenamefont
  {Regnault},\ and\ \citenamefont {Bernevig}}]{Moudgalya_II}%
  \BibitemOpen
  \bibfield  {author} {\bibinfo {author} {\bibfnamefont {S.}~\bibnamefont
  {Moudgalya}}, \bibinfo {author} {\bibfnamefont {N.}~\bibnamefont
  {Regnault}},\ and\ \bibinfo {author} {\bibfnamefont {B.~A.}\ \bibnamefont
  {Bernevig}},\ }\bibfield  {title} {\bibinfo {title} {Entanglement of exact
  excited states of {A}ffleck-{K}ennedy-{L}ieb-{T}asaki models: Exact results,
  many-body scars, and violation of the strong eigenstate thermalization
  hypothesis},\ }\href {https://doi.org/10.1103/PhysRevB.98.235156} {\bibfield
  {journal} {\bibinfo  {journal} {Phys. Rev. B}\ }\textbf {\bibinfo {volume}
  {98}},\ \bibinfo {pages} {235156} (\bibinfo {year}
  {2018}{\natexlab{b}})}\BibitemShut {NoStop}%
\bibitem [{\citenamefont {Bernien}\ \emph {et~al.}(2017)\citenamefont
  {Bernien}, \citenamefont {Schwartz}, \citenamefont {Keesling}, \citenamefont
  {Levine}, \citenamefont {Omran}, \citenamefont {Pichler}, \citenamefont
  {Choi}, \citenamefont {Zibrov}, \citenamefont {Endres}, \citenamefont
  {Greiner}, \citenamefont {Vuleti{\'{c}}},\ and\ \citenamefont
  {Lukin}}]{Bernien2017}%
  \BibitemOpen
  \bibfield  {author} {\bibinfo {author} {\bibfnamefont {H.}~\bibnamefont
  {Bernien}}, \bibinfo {author} {\bibfnamefont {S.}~\bibnamefont {Schwartz}},
  \bibinfo {author} {\bibfnamefont {A.}~\bibnamefont {Keesling}}, \bibinfo
  {author} {\bibfnamefont {H.}~\bibnamefont {Levine}}, \bibinfo {author}
  {\bibfnamefont {A.}~\bibnamefont {Omran}}, \bibinfo {author} {\bibfnamefont
  {H.}~\bibnamefont {Pichler}}, \bibinfo {author} {\bibfnamefont
  {S.}~\bibnamefont {Choi}}, \bibinfo {author} {\bibfnamefont {A.~S.}\
  \bibnamefont {Zibrov}}, \bibinfo {author} {\bibfnamefont {M.}~\bibnamefont
  {Endres}}, \bibinfo {author} {\bibfnamefont {M.}~\bibnamefont {Greiner}},
  \bibinfo {author} {\bibfnamefont {V.}~\bibnamefont {Vuleti{\'{c}}}},\ and\
  \bibinfo {author} {\bibfnamefont {M.~D.}\ \bibnamefont {Lukin}},\ }\bibfield
  {title} {\bibinfo {title} {Probing many-body dynamics on a 51-atom quantum
  simulator},\ }\href {https://doi.org/10.1038/nature24622} {\bibfield
  {journal} {\bibinfo  {journal} {Nature}\ }\textbf {\bibinfo {volume} {551}},\
  \bibinfo {pages} {579} (\bibinfo {year} {2017})}\BibitemShut {NoStop}%
\bibitem [{\citenamefont {Turner}\ \emph
  {et~al.}(2018{\natexlab{a}})\citenamefont {Turner}, \citenamefont
  {Michailidis}, \citenamefont {Abanin}, \citenamefont {Serbyn},\ and\
  \citenamefont {Papi{\'{c}}}}]{Papic2018Jul}%
  \BibitemOpen
  \bibfield  {author} {\bibinfo {author} {\bibfnamefont {C.~J.}\ \bibnamefont
  {Turner}}, \bibinfo {author} {\bibfnamefont {A.~A.}\ \bibnamefont
  {Michailidis}}, \bibinfo {author} {\bibfnamefont {D.~A.}\ \bibnamefont
  {Abanin}}, \bibinfo {author} {\bibfnamefont {M.}~\bibnamefont {Serbyn}},\
  and\ \bibinfo {author} {\bibfnamefont {Z.}~\bibnamefont {Papi{\'{c}}}},\
  }\bibfield  {title} {\bibinfo {title} {Weak ergodicity breaking from quantum
  many-body scars},\ }\href {https://doi.org/10.1038/s41567-018-0137-5}
  {\bibfield  {journal} {\bibinfo  {journal} {Nature Physics}\ }\textbf
  {\bibinfo {volume} {14}},\ \bibinfo {pages} {745} (\bibinfo {year}
  {2018}{\natexlab{a}})}\BibitemShut {NoStop}%
\bibitem [{\citenamefont {Turner}\ \emph
  {et~al.}(2018{\natexlab{b}})\citenamefont {Turner}, \citenamefont
  {Michailidis}, \citenamefont {Abanin}, \citenamefont {Serbyn},\ and\
  \citenamefont {Papi\ifmmode~\acute{c}\else \'{c}\fi{}}}]{Papic2018Oct}%
  \BibitemOpen
  \bibfield  {author} {\bibinfo {author} {\bibfnamefont {C.~J.}\ \bibnamefont
  {Turner}}, \bibinfo {author} {\bibfnamefont {A.~A.}\ \bibnamefont
  {Michailidis}}, \bibinfo {author} {\bibfnamefont {D.~A.}\ \bibnamefont
  {Abanin}}, \bibinfo {author} {\bibfnamefont {M.}~\bibnamefont {Serbyn}},\
  and\ \bibinfo {author} {\bibfnamefont {Z.}~\bibnamefont
  {Papi\ifmmode~\acute{c}\else \'{c}\fi{}}},\ }\bibfield  {title} {\bibinfo
  {title} {Quantum scarred eigenstates in a {R}ydberg atom chain: Entanglement,
  breakdown of thermalization, and stability to perturbations},\ }\href
  {https://doi.org/10.1103/PhysRevB.98.155134} {\bibfield  {journal} {\bibinfo
  {journal} {Phys. Rev. B}\ }\textbf {\bibinfo {volume} {98}},\ \bibinfo
  {pages} {155134} (\bibinfo {year} {2018}{\natexlab{b}})}\BibitemShut
  {NoStop}%
\bibitem [{\citenamefont {Lin}\ and\ \citenamefont
  {Motrunich}(2019)}]{Olexei2019}%
  \BibitemOpen
  \bibfield  {author} {\bibinfo {author} {\bibfnamefont {C.-J.}\ \bibnamefont
  {Lin}}\ and\ \bibinfo {author} {\bibfnamefont {O.~I.}\ \bibnamefont
  {Motrunich}},\ }\bibfield  {title} {\bibinfo {title} {Exact quantum many-body
  scar states in the {R}ydberg-blockaded atom chain},\ }\href
  {https://doi.org/10.1103/PhysRevLett.122.173401} {\bibfield  {journal}
  {\bibinfo  {journal} {Phys. Rev. Lett.}\ }\textbf {\bibinfo {volume} {122}},\
  \bibinfo {pages} {173401} (\bibinfo {year} {2019})}\BibitemShut {NoStop}%
\bibitem [{\citenamefont {Iadecola}\ \emph {et~al.}(2019)\citenamefont
  {Iadecola}, \citenamefont {Schecter},\ and\ \citenamefont
  {Xu}}]{Iadecola2019}%
  \BibitemOpen
  \bibfield  {author} {\bibinfo {author} {\bibfnamefont {T.}~\bibnamefont
  {Iadecola}}, \bibinfo {author} {\bibfnamefont {M.}~\bibnamefont {Schecter}},\
  and\ \bibinfo {author} {\bibfnamefont {S.}~\bibnamefont {Xu}},\ }\bibfield
  {title} {\bibinfo {title} {Quantum many-body scars from magnon
  condensation},\ }\href {https://doi.org/10.1103/PhysRevB.100.184312}
  {\bibfield  {journal} {\bibinfo  {journal} {Phys. Rev. B}\ }\textbf {\bibinfo
  {volume} {100}},\ \bibinfo {pages} {184312} (\bibinfo {year}
  {2019})}\BibitemShut {NoStop}%
\bibitem [{\citenamefont {Schecter}\ and\ \citenamefont
  {Iadecola}(2019)}]{Schecter2019}%
  \BibitemOpen
  \bibfield  {author} {\bibinfo {author} {\bibfnamefont {M.}~\bibnamefont
  {Schecter}}\ and\ \bibinfo {author} {\bibfnamefont {T.}~\bibnamefont
  {Iadecola}},\ }\bibfield  {title} {\bibinfo {title} {Weak ergodicity breaking
  and quantum many-body scars in spin-1 {$XY$} magnets},\ }\href
  {https://doi.org/10.1103/PhysRevLett.123.147201} {\bibfield  {journal}
  {\bibinfo  {journal} {Phys. Rev. Lett.}\ }\textbf {\bibinfo {volume} {123}},\
  \bibinfo {pages} {147201} (\bibinfo {year} {2019})}\BibitemShut {NoStop}%
\bibitem [{\citenamefont {Iadecola}\ and\ \citenamefont
  {Schecter}(2020)}]{Iadecola_Schecter}%
  \BibitemOpen
  \bibfield  {author} {\bibinfo {author} {\bibfnamefont {T.}~\bibnamefont
  {Iadecola}}\ and\ \bibinfo {author} {\bibfnamefont {M.}~\bibnamefont
  {Schecter}},\ }\bibfield  {title} {\bibinfo {title} {Quantum many-body scar
  states with emergent kinetic constraints and finite-entanglement revivals},\
  }\href {https://doi.org/10.1103/PhysRevB.101.024306} {\bibfield  {journal}
  {\bibinfo  {journal} {Phys. Rev. B}\ }\textbf {\bibinfo {volume} {101}},\
  \bibinfo {pages} {024306} (\bibinfo {year} {2020})}\BibitemShut {NoStop}%
\bibitem [{\citenamefont {Mark}\ and\ \citenamefont
  {Motrunich}(2020)}]{MarkDaniel2020}%
  \BibitemOpen
  \bibfield  {author} {\bibinfo {author} {\bibfnamefont {D.~K.}\ \bibnamefont
  {Mark}}\ and\ \bibinfo {author} {\bibfnamefont {O.~I.}\ \bibnamefont
  {Motrunich}},\ }\bibfield  {title} {\bibinfo {title}
  {$\ensuremath{\eta}$-pairing states as true scars in an extended {H}ubbard
  model},\ }\href {https://doi.org/10.1103/PhysRevB.102.075132} {\bibfield
  {journal} {\bibinfo  {journal} {Phys. Rev. B}\ }\textbf {\bibinfo {volume}
  {102}},\ \bibinfo {pages} {075132} (\bibinfo {year} {2020})}\BibitemShut
  {NoStop}%
\bibitem [{\citenamefont {Moudgalya}\ \emph
  {et~al.}(2020{\natexlab{a}})\citenamefont {Moudgalya}, \citenamefont
  {O'Brien}, \citenamefont {Bernevig}, \citenamefont {Fendley},\ and\
  \citenamefont {Regnault}}]{MoudgalyaSanjay2020}%
  \BibitemOpen
  \bibfield  {author} {\bibinfo {author} {\bibfnamefont {S.}~\bibnamefont
  {Moudgalya}}, \bibinfo {author} {\bibfnamefont {E.}~\bibnamefont {O'Brien}},
  \bibinfo {author} {\bibfnamefont {B.~A.}\ \bibnamefont {Bernevig}}, \bibinfo
  {author} {\bibfnamefont {P.}~\bibnamefont {Fendley}},\ and\ \bibinfo {author}
  {\bibfnamefont {N.}~\bibnamefont {Regnault}},\ }\bibfield  {title} {\bibinfo
  {title} {Large classes of quantum scarred {H}amiltonians from matrix product
  states},\ }\href {https://doi.org/10.1103/PhysRevB.102.085120} {\bibfield
  {journal} {\bibinfo  {journal} {Phys. Rev. B}\ }\textbf {\bibinfo {volume}
  {102}},\ \bibinfo {pages} {085120} (\bibinfo {year}
  {2020}{\natexlab{a}})}\BibitemShut {NoStop}%
\bibitem [{\citenamefont {Shibata}\ \emph {et~al.}(2020)\citenamefont
  {Shibata}, \citenamefont {Yoshioka},\ and\ \citenamefont
  {Katsura}}]{Shibata2020}%
  \BibitemOpen
  \bibfield  {author} {\bibinfo {author} {\bibfnamefont {N.}~\bibnamefont
  {Shibata}}, \bibinfo {author} {\bibfnamefont {N.}~\bibnamefont {Yoshioka}},\
  and\ \bibinfo {author} {\bibfnamefont {H.}~\bibnamefont {Katsura}},\
  }\bibfield  {title} {\bibinfo {title} {{O}nsager's scars in disordered spin
  chains},\ }\href {https://doi.org/10.1103/PhysRevLett.124.180604} {\bibfield
  {journal} {\bibinfo  {journal} {Phys. Rev. Lett.}\ }\textbf {\bibinfo
  {volume} {124}},\ \bibinfo {pages} {180604} (\bibinfo {year}
  {2020})}\BibitemShut {NoStop}%
\bibitem [{\citenamefont {Mark}\ \emph {et~al.}(2020)\citenamefont {Mark},
  \citenamefont {Lin},\ and\ \citenamefont {Motrunich}}]{Mark2020}%
  \BibitemOpen
  \bibfield  {author} {\bibinfo {author} {\bibfnamefont {D.~K.}\ \bibnamefont
  {Mark}}, \bibinfo {author} {\bibfnamefont {C.-J.}\ \bibnamefont {Lin}},\ and\
  \bibinfo {author} {\bibfnamefont {O.~I.}\ \bibnamefont {Motrunich}},\
  }\bibfield  {title} {\bibinfo {title} {Unified structure for exact towers of
  scar states in the {A}ffleck-{K}ennedy-{L}ieb-{T}asaki and other models},\
  }\href {https://doi.org/10.1103/PhysRevB.101.195131} {\bibfield  {journal}
  {\bibinfo  {journal} {Phys. Rev. B}\ }\textbf {\bibinfo {volume} {101}},\
  \bibinfo {pages} {195131} (\bibinfo {year} {2020})}\BibitemShut {NoStop}%
\bibitem [{\citenamefont {Moudgalya}\ \emph
  {et~al.}(2020{\natexlab{b}})\citenamefont {Moudgalya}, \citenamefont
  {Regnault},\ and\ \citenamefont {Bernevig}}]{Moudgalya2020}%
  \BibitemOpen
  \bibfield  {author} {\bibinfo {author} {\bibfnamefont {S.}~\bibnamefont
  {Moudgalya}}, \bibinfo {author} {\bibfnamefont {N.}~\bibnamefont
  {Regnault}},\ and\ \bibinfo {author} {\bibfnamefont {B.~A.}\ \bibnamefont
  {Bernevig}},\ }\bibfield  {title} {\bibinfo {title}
  {$\ensuremath{\eta}$-pairing in {H}ubbard models: From spectrum generating
  algebras to quantum many-body scars},\ }\href
  {https://doi.org/10.1103/PhysRevB.102.085140} {\bibfield  {journal} {\bibinfo
   {journal} {Phys. Rev. B}\ }\textbf {\bibinfo {volume} {102}},\ \bibinfo
  {pages} {085140} (\bibinfo {year} {2020}{\natexlab{b}})}\BibitemShut
  {NoStop}%
\bibitem [{\citenamefont {Shiraishi}\ and\ \citenamefont
  {Mori}(2017)}]{Shiraishi2017}%
  \BibitemOpen
  \bibfield  {author} {\bibinfo {author} {\bibfnamefont {N.}~\bibnamefont
  {Shiraishi}}\ and\ \bibinfo {author} {\bibfnamefont {T.}~\bibnamefont
  {Mori}},\ }\bibfield  {title} {\bibinfo {title} {Systematic construction of
  counterexamples to the eigenstate thermalization hypothesis},\ }\href
  {https://doi.org/10.1103/PhysRevLett.119.030601} {\bibfield  {journal}
  {\bibinfo  {journal} {Phys. Rev. Lett.}\ }\textbf {\bibinfo {volume} {119}},\
  \bibinfo {pages} {030601} (\bibinfo {year} {2017})}\BibitemShut {NoStop}%
\bibitem [{\citenamefont {Ren}\ \emph {et~al.}(2021)\citenamefont {Ren},
  \citenamefont {Liang},\ and\ \citenamefont {Fang}}]{Ren2021}%
  \BibitemOpen
  \bibfield  {author} {\bibinfo {author} {\bibfnamefont {J.}~\bibnamefont
  {Ren}}, \bibinfo {author} {\bibfnamefont {C.}~\bibnamefont {Liang}},\ and\
  \bibinfo {author} {\bibfnamefont {C.}~\bibnamefont {Fang}},\ }\bibfield
  {title} {\bibinfo {title} {Quasisymmetry groups and many-body scar
  dynamics},\ }\href {https://doi.org/10.1103/PhysRevLett.126.120604}
  {\bibfield  {journal} {\bibinfo  {journal} {Phys. Rev. Lett.}\ }\textbf
  {\bibinfo {volume} {126}},\ \bibinfo {pages} {120604} (\bibinfo {year}
  {2021})}\BibitemShut {NoStop}%
\bibitem [{\citenamefont {Ren}\ \emph {et~al.}(2022)\citenamefont {Ren},
  \citenamefont {Liang},\ and\ \citenamefont {Fang}}]{Ren2022}%
  \BibitemOpen
  \bibfield  {author} {\bibinfo {author} {\bibfnamefont {J.}~\bibnamefont
  {Ren}}, \bibinfo {author} {\bibfnamefont {C.}~\bibnamefont {Liang}},\ and\
  \bibinfo {author} {\bibfnamefont {C.}~\bibnamefont {Fang}},\ }\bibfield
  {title} {\bibinfo {title} {Deformed symmetry structures and quantum many-body
  scar subspaces},\ }\href {https://doi.org/10.1103/PhysRevResearch.4.013155}
  {\bibfield  {journal} {\bibinfo  {journal} {Phys. Rev. Res.}\ }\textbf
  {\bibinfo {volume} {4}},\ \bibinfo {pages} {013155} (\bibinfo {year}
  {2022})}\BibitemShut {NoStop}%
\bibitem [{\citenamefont {Wildeboer}\ \emph {et~al.}(2022)\citenamefont
  {Wildeboer}, \citenamefont {Langlett}, \citenamefont {Yang}, \citenamefont
  {Gorshkov}, \citenamefont {Iadecola},\ and\ \citenamefont
  {Xu}}]{Wildeboer2022}%
  \BibitemOpen
  \bibfield  {author} {\bibinfo {author} {\bibfnamefont {J.}~\bibnamefont
  {Wildeboer}}, \bibinfo {author} {\bibfnamefont {C.~M.}\ \bibnamefont
  {Langlett}}, \bibinfo {author} {\bibfnamefont {Z.-C.}\ \bibnamefont {Yang}},
  \bibinfo {author} {\bibfnamefont {A.~V.}\ \bibnamefont {Gorshkov}}, \bibinfo
  {author} {\bibfnamefont {T.}~\bibnamefont {Iadecola}},\ and\ \bibinfo
  {author} {\bibfnamefont {S.}~\bibnamefont {Xu}},\ }\bibfield  {title}
  {\bibinfo {title} {Quantum many-body scars from {E}instein-{P}odolsky-{R}osen
  states in bilayer systems},\ }\href
  {https://doi.org/10.1103/PhysRevB.106.205142} {\bibfield  {journal} {\bibinfo
   {journal} {Phys. Rev. B}\ }\textbf {\bibinfo {volume} {106}},\ \bibinfo
  {pages} {205142} (\bibinfo {year} {2022})}\BibitemShut {NoStop}%
\bibitem [{\citenamefont {Chen}\ \emph {et~al.}(2022)\citenamefont {Chen},
  \citenamefont {Burdick}, \citenamefont {Yao}, \citenamefont {Orth},\ and\
  \citenamefont {Iadecola}}]{Chen2022}%
  \BibitemOpen
  \bibfield  {author} {\bibinfo {author} {\bibfnamefont {I.-C.}\ \bibnamefont
  {Chen}}, \bibinfo {author} {\bibfnamefont {B.}~\bibnamefont {Burdick}},
  \bibinfo {author} {\bibfnamefont {Y.}~\bibnamefont {Yao}}, \bibinfo {author}
  {\bibfnamefont {P.~P.}\ \bibnamefont {Orth}},\ and\ \bibinfo {author}
  {\bibfnamefont {T.}~\bibnamefont {Iadecola}},\ }\bibfield  {title} {\bibinfo
  {title} {Error-mitigated simulation of quantum many-body scars on quantum
  computers with pulse-level control},\ }\href
  {https://doi.org/10.1103/PhysRevResearch.4.043027} {\bibfield  {journal}
  {\bibinfo  {journal} {Phys. Rev. Res.}\ }\textbf {\bibinfo {volume} {4}},\
  \bibinfo {pages} {043027} (\bibinfo {year} {2022})}\BibitemShut {NoStop}%
\bibitem [{\citenamefont {Bluvstein}\ \emph {et~al.}(2021)\citenamefont
  {Bluvstein}, \citenamefont {Omran}, \citenamefont {Levine}, \citenamefont
  {Keesling}, \citenamefont {Semeghini}, \citenamefont {Ebadi}, \citenamefont
  {Wang}, \citenamefont {Michailidis}, \citenamefont {Maskara}, \citenamefont
  {Ho}, \citenamefont {Choi}, \citenamefont {Serbyn}, \citenamefont {Greiner},
  \citenamefont {Vuletić},\ and\ \citenamefont {Lukin}}]{Bluvstein2021}%
  \BibitemOpen
  \bibfield  {author} {\bibinfo {author} {\bibfnamefont {D.}~\bibnamefont
  {Bluvstein}}, \bibinfo {author} {\bibfnamefont {A.}~\bibnamefont {Omran}},
  \bibinfo {author} {\bibfnamefont {H.}~\bibnamefont {Levine}}, \bibinfo
  {author} {\bibfnamefont {A.}~\bibnamefont {Keesling}}, \bibinfo {author}
  {\bibfnamefont {G.}~\bibnamefont {Semeghini}}, \bibinfo {author}
  {\bibfnamefont {S.}~\bibnamefont {Ebadi}}, \bibinfo {author} {\bibfnamefont
  {T.~T.}\ \bibnamefont {Wang}}, \bibinfo {author} {\bibfnamefont {A.~A.}\
  \bibnamefont {Michailidis}}, \bibinfo {author} {\bibfnamefont
  {N.}~\bibnamefont {Maskara}}, \bibinfo {author} {\bibfnamefont {W.~W.}\
  \bibnamefont {Ho}}, \bibinfo {author} {\bibfnamefont {S.}~\bibnamefont
  {Choi}}, \bibinfo {author} {\bibfnamefont {M.}~\bibnamefont {Serbyn}},
  \bibinfo {author} {\bibfnamefont {M.}~\bibnamefont {Greiner}}, \bibinfo
  {author} {\bibfnamefont {V.}~\bibnamefont {Vuletić}},\ and\ \bibinfo
  {author} {\bibfnamefont {M.~D.}\ \bibnamefont {Lukin}},\ }\bibfield  {title}
  {\bibinfo {title} {Controlling quantum many-body dynamics in driven {R}ydberg
  atom arrays},\ }\href {https://doi.org/10.1126/science.abg2530} {\bibfield
  {journal} {\bibinfo  {journal} {Science}\ }\textbf {\bibinfo {volume}
  {371}},\ \bibinfo {pages} {1355} (\bibinfo {year} {2021})}\BibitemShut
  {NoStop}%
\bibitem [{\citenamefont {Zhang}\ \emph {et~al.}(2023)\citenamefont {Zhang},
  \citenamefont {Dong}, \citenamefont {Gao}, \citenamefont {Zhao},
  \citenamefont {Hao}, \citenamefont {Desaules}, \citenamefont {Guo},
  \citenamefont {Chen}, \citenamefont {Deng}, \citenamefont {Liu},
  \citenamefont {Ren}, \citenamefont {Yao}, \citenamefont {Zhang},
  \citenamefont {Xu}, \citenamefont {Wang}, \citenamefont {Jin}, \citenamefont
  {Zhu}, \citenamefont {Zhang}, \citenamefont {Li}, \citenamefont {Song},
  \citenamefont {Wang}, \citenamefont {Liu}, \citenamefont {Papić},
  \citenamefont {Ying}, \citenamefont {Wang},\ and\ \citenamefont
  {Lai}}]{ZhangPengfei2021}%
  \BibitemOpen
  \bibfield  {author} {\bibinfo {author} {\bibfnamefont {P.}~\bibnamefont
  {Zhang}}, \bibinfo {author} {\bibfnamefont {H.}~\bibnamefont {Dong}},
  \bibinfo {author} {\bibfnamefont {Y.}~\bibnamefont {Gao}}, \bibinfo {author}
  {\bibfnamefont {L.}~\bibnamefont {Zhao}}, \bibinfo {author} {\bibfnamefont
  {J.}~\bibnamefont {Hao}}, \bibinfo {author} {\bibfnamefont {J.-Y.}\
  \bibnamefont {Desaules}}, \bibinfo {author} {\bibfnamefont {Q.}~\bibnamefont
  {Guo}}, \bibinfo {author} {\bibfnamefont {J.}~\bibnamefont {Chen}}, \bibinfo
  {author} {\bibfnamefont {J.}~\bibnamefont {Deng}}, \bibinfo {author}
  {\bibfnamefont {B.}~\bibnamefont {Liu}}, \bibinfo {author} {\bibfnamefont
  {W.}~\bibnamefont {Ren}}, \bibinfo {author} {\bibfnamefont {Y.}~\bibnamefont
  {Yao}}, \bibinfo {author} {\bibfnamefont {X.}~\bibnamefont {Zhang}}, \bibinfo
  {author} {\bibfnamefont {S.}~\bibnamefont {Xu}}, \bibinfo {author}
  {\bibfnamefont {K.}~\bibnamefont {Wang}}, \bibinfo {author} {\bibfnamefont
  {F.}~\bibnamefont {Jin}}, \bibinfo {author} {\bibfnamefont {X.}~\bibnamefont
  {Zhu}}, \bibinfo {author} {\bibfnamefont {B.}~\bibnamefont {Zhang}}, \bibinfo
  {author} {\bibfnamefont {H.}~\bibnamefont {Li}}, \bibinfo {author}
  {\bibfnamefont {C.}~\bibnamefont {Song}}, \bibinfo {author} {\bibfnamefont
  {Z.}~\bibnamefont {Wang}}, \bibinfo {author} {\bibfnamefont {F.}~\bibnamefont
  {Liu}}, \bibinfo {author} {\bibfnamefont {Z.}~\bibnamefont {Papić}},
  \bibinfo {author} {\bibfnamefont {L.}~\bibnamefont {Ying}}, \bibinfo {author}
  {\bibfnamefont {H.}~\bibnamefont {Wang}},\ and\ \bibinfo {author}
  {\bibfnamefont {Y.-C.}\ \bibnamefont {Lai}},\ }\bibfield  {title} {\bibinfo
  {title} {Many-body {H}ilbert space scarring on a superconducting processor},\
  }\href {https://doi.org/10.1038/s41567-022-01784-9} {\bibfield  {journal}
  {\bibinfo  {journal} {Nature Physics}\ }\textbf {\bibinfo {volume} {19}},\
  \bibinfo {pages} {120} (\bibinfo {year} {2023})}\BibitemShut {NoStop}%
\bibitem [{\citenamefont {Srivatsa}\ \emph {et~al.}(2020)\citenamefont
  {Srivatsa}, \citenamefont {Moessner},\ and\ \citenamefont
  {Nielsen}}]{Anne2020}%
  \BibitemOpen
  \bibfield  {author} {\bibinfo {author} {\bibfnamefont {N.~S.}\ \bibnamefont
  {Srivatsa}}, \bibinfo {author} {\bibfnamefont {R.}~\bibnamefont {Moessner}},\
  and\ \bibinfo {author} {\bibfnamefont {A.~E.~B.}\ \bibnamefont {Nielsen}},\
  }\bibfield  {title} {\bibinfo {title} {Many-body delocalization via emergent
  symmetry},\ }\href {https://doi.org/10.1103/PhysRevLett.125.240401}
  {\bibfield  {journal} {\bibinfo  {journal} {Phys. Rev. Lett.}\ }\textbf
  {\bibinfo {volume} {125}},\ \bibinfo {pages} {240401} (\bibinfo {year}
  {2020})}\BibitemShut {NoStop}%
\bibitem [{\citenamefont {Iversen}\ \emph {et~al.}(2022)\citenamefont
  {Iversen}, \citenamefont {Srivatsa},\ and\ \citenamefont
  {Nielsen}}]{Iversen2022}%
  \BibitemOpen
  \bibfield  {author} {\bibinfo {author} {\bibfnamefont {M.}~\bibnamefont
  {Iversen}}, \bibinfo {author} {\bibfnamefont {N.~S.}\ \bibnamefont
  {Srivatsa}},\ and\ \bibinfo {author} {\bibfnamefont {A.~E.~B.}\ \bibnamefont
  {Nielsen}},\ }\bibfield  {title} {\bibinfo {title} {Escaping many-body
  localization in an exact eigenstate},\ }\href
  {https://doi.org/10.1103/PhysRevB.106.214201} {\bibfield  {journal} {\bibinfo
   {journal} {Phys. Rev. B}\ }\textbf {\bibinfo {volume} {106}},\ \bibinfo
  {pages} {214201} (\bibinfo {year} {2022})}\BibitemShut {NoStop}%
\bibitem [{\citenamefont {Srivatsa}\ \emph {et~al.}(2022)\citenamefont
  {Srivatsa}, \citenamefont {Yarloo}, \citenamefont {Moessner},\ and\
  \citenamefont {Nielsen}}]{Srivatsa2022}%
  \BibitemOpen
  \bibfield  {author} {\bibinfo {author} {\bibfnamefont {N.~S.}\ \bibnamefont
  {Srivatsa}}, \bibinfo {author} {\bibfnamefont {H.}~\bibnamefont {Yarloo}},
  \bibinfo {author} {\bibfnamefont {R.}~\bibnamefont {Moessner}},\ and\
  \bibinfo {author} {\bibfnamefont {A.~E.~B.}\ \bibnamefont {Nielsen}},\ }\href
  {https://doi.org/10.48550/ARXIV.2208.01054} {\bibinfo {title} {Mobility edges
  through inverted quantum many-body scarring}} (\bibinfo {year} {2022}),\
  \Eprint {https://arxiv.org/abs/2208.01054} {arXiv:2208.01054} \BibitemShut
  {NoStop}%
\bibitem [{\citenamefont {Nandkishore}\ and\ \citenamefont
  {Huse}(2015)}]{nandkishore2015}%
  \BibitemOpen
  \bibfield  {author} {\bibinfo {author} {\bibfnamefont {R.}~\bibnamefont
  {Nandkishore}}\ and\ \bibinfo {author} {\bibfnamefont {D.~A.}\ \bibnamefont
  {Huse}},\ }\bibfield  {title} {\bibinfo {title} {Many-body localization and
  thermalization in quantum statistical mechanics},\ }\href@noop {} {\bibfield
  {journal} {\bibinfo  {journal} {Annu. Rev. Condens. Matter Phys.}\ }\textbf
  {\bibinfo {volume} {6}},\ \bibinfo {pages} {15} (\bibinfo {year}
  {2015})}\BibitemShut {NoStop}%
\bibitem [{\citenamefont {Dooley}(2021)}]{Dooley2021}%
  \BibitemOpen
  \bibfield  {author} {\bibinfo {author} {\bibfnamefont {S.}~\bibnamefont
  {Dooley}},\ }\bibfield  {title} {\bibinfo {title} {Robust quantum sensing in
  strongly interacting systems with many-body scars},\ }\href
  {https://doi.org/10.1103/PRXQuantum.2.020330} {\bibfield  {journal} {\bibinfo
   {journal} {PRX Quantum}\ }\textbf {\bibinfo {volume} {2}},\ \bibinfo {pages}
  {020330} (\bibinfo {year} {2021})}\BibitemShut {NoStop}%
\bibitem [{\citenamefont {Dooley}\ \emph {et~al.}(2023)\citenamefont {Dooley},
  \citenamefont {Pappalardi},\ and\ \citenamefont {Goold}}]{Dooley2023}%
  \BibitemOpen
  \bibfield  {author} {\bibinfo {author} {\bibfnamefont {S.}~\bibnamefont
  {Dooley}}, \bibinfo {author} {\bibfnamefont {S.}~\bibnamefont {Pappalardi}},\
  and\ \bibinfo {author} {\bibfnamefont {J.}~\bibnamefont {Goold}},\ }\bibfield
   {title} {\bibinfo {title} {Entanglement enhanced metrology with quantum
  many-body scars},\ }\href {https://doi.org/10.1103/PhysRevB.107.035123}
  {\bibfield  {journal} {\bibinfo  {journal} {Phys. Rev. B}\ }\textbf {\bibinfo
  {volume} {107}},\ \bibinfo {pages} {035123} (\bibinfo {year}
  {2023})}\BibitemShut {NoStop}%
\bibitem [{\citenamefont {Chertkov}\ and\ \citenamefont
  {Clark}(2018)}]{Chertkov}%
  \BibitemOpen
  \bibfield  {author} {\bibinfo {author} {\bibfnamefont {E.}~\bibnamefont
  {Chertkov}}\ and\ \bibinfo {author} {\bibfnamefont {B.~K.}\ \bibnamefont
  {Clark}},\ }\bibfield  {title} {\bibinfo {title} {Computational inverse
  method for constructing spaces of quantum models from wave functions},\
  }\href {https://doi.org/10.1103/PhysRevX.8.031029} {\bibfield  {journal}
  {\bibinfo  {journal} {Phys. Rev. X}\ }\textbf {\bibinfo {volume} {8}},\
  \bibinfo {pages} {031029} (\bibinfo {year} {2018})}\BibitemShut {NoStop}%
\bibitem [{\citenamefont {Greiter}\ \emph {et~al.}(2018)\citenamefont
  {Greiter}, \citenamefont {Schnells},\ and\ \citenamefont
  {Thomale}}]{Greiter}%
  \BibitemOpen
  \bibfield  {author} {\bibinfo {author} {\bibfnamefont {M.}~\bibnamefont
  {Greiter}}, \bibinfo {author} {\bibfnamefont {V.}~\bibnamefont {Schnells}},\
  and\ \bibinfo {author} {\bibfnamefont {R.}~\bibnamefont {Thomale}},\
  }\bibfield  {title} {\bibinfo {title} {Method to identify parent
  {H}amiltonians for trial states},\ }\href
  {https://doi.org/10.1103/PhysRevB.98.081113} {\bibfield  {journal} {\bibinfo
  {journal} {Phys. Rev. B}\ }\textbf {\bibinfo {volume} {98}},\ \bibinfo
  {pages} {081113(R)} (\bibinfo {year} {2018})}\BibitemShut {NoStop}%
\bibitem [{\citenamefont {Qu}\ \emph {et~al.}(2016)\citenamefont {Qu},
  \citenamefont {Sun},\ and\ \citenamefont {Wright}}]{QU}%
  \BibitemOpen
  \bibfield  {author} {\bibinfo {author} {\bibfnamefont {Q.}~\bibnamefont
  {Qu}}, \bibinfo {author} {\bibfnamefont {J.}~\bibnamefont {Sun}},\ and\
  \bibinfo {author} {\bibfnamefont {J.}~\bibnamefont {Wright}},\ }\bibfield
  {title} {\bibinfo {title} {Finding a sparse vector in a subspace: Linear
  sparsity using alternating directions},\ }\href
  {https://doi.org/10.1109/TIT.2016.2601599} {\bibfield  {journal} {\bibinfo
  {journal} {IEEE Transactions on Information Theory}\ }\textbf {\bibinfo
  {volume} {62}},\ \bibinfo {pages} {5855} (\bibinfo {year}
  {2016})}\BibitemShut {NoStop}%
\bibitem [{\citenamefont {Qi}\ and\ \citenamefont {Ranard}(2019)}]{Qi2019}%
  \BibitemOpen
  \bibfield  {author} {\bibinfo {author} {\bibfnamefont {X.-L.}\ \bibnamefont
  {Qi}}\ and\ \bibinfo {author} {\bibfnamefont {D.}~\bibnamefont {Ranard}},\
  }\bibfield  {title} {\bibinfo {title} {Determining a local {H}amiltonian from
  a single eigenstate},\ }\href {https://doi.org/10.22331/q-2019-07-08-159}
  {\bibfield  {journal} {\bibinfo  {journal} {{Quantum}}\ }\textbf {\bibinfo
  {volume} {3}},\ \bibinfo {pages} {159} (\bibinfo {year} {2019})}\BibitemShut
  {NoStop}%
\bibitem [{\citenamefont {D'Alessio}\ \emph {et~al.}(2016)\citenamefont
  {D'Alessio}, \citenamefont {Kafri}, \citenamefont {Polkovnikov},\ and\
  \citenamefont {Rigol}}]{Luca}%
  \BibitemOpen
  \bibfield  {author} {\bibinfo {author} {\bibfnamefont {L.}~\bibnamefont
  {D'Alessio}}, \bibinfo {author} {\bibfnamefont {Y.}~\bibnamefont {Kafri}},
  \bibinfo {author} {\bibfnamefont {A.}~\bibnamefont {Polkovnikov}},\ and\
  \bibinfo {author} {\bibfnamefont {M.}~\bibnamefont {Rigol}},\ }\bibfield
  {title} {\bibinfo {title} {From quantum chaos and eigenstate thermalization
  to statistical mechanics and thermodynamics},\ }\href
  {https://doi.org/10.1080/00018732.2016.1198134} {\bibfield  {journal}
  {\bibinfo  {journal} {Advances in Physics}\ }\textbf {\bibinfo {volume}
  {65}},\ \bibinfo {pages} {239} (\bibinfo {year} {2016})}\BibitemShut
  {NoStop}%
\bibitem [{\citenamefont {Guhr}\ \emph {et~al.}(1998)\citenamefont {Guhr},
  \citenamefont {Müller–Groeling},\ and\ \citenamefont
  {Weidenmüller}}]{Guhr}%
  \BibitemOpen
  \bibfield  {author} {\bibinfo {author} {\bibfnamefont {T.}~\bibnamefont
  {Guhr}}, \bibinfo {author} {\bibfnamefont {A.}~\bibnamefont
  {Müller–Groeling}},\ and\ \bibinfo {author} {\bibfnamefont {H.~A.}\
  \bibnamefont {Weidenmüller}},\ }\bibfield  {title} {\bibinfo {title}
  {Random-matrix theories in quantum physics: Common concepts},\ }\href
  {https://doi.org/https://doi.org/10.1016/S0370-1573(97)00088-4} {\bibfield
  {journal} {\bibinfo  {journal} {Physics Reports}\ }\textbf {\bibinfo {volume}
  {299}},\ \bibinfo {pages} {189} (\bibinfo {year} {1998})}\BibitemShut
  {NoStop}%
\bibitem [{\citenamefont {Abul-Magd}\ and\ \citenamefont
  {Abul-Magd}(2014)}]{Abdulmagd}%
  \BibitemOpen
  \bibfield  {author} {\bibinfo {author} {\bibfnamefont {A.~A.}\ \bibnamefont
  {Abul-Magd}}\ and\ \bibinfo {author} {\bibfnamefont {A.~Y.}\ \bibnamefont
  {Abul-Magd}},\ }\bibfield  {title} {\bibinfo {title} {Unfolding of the
  spectrum for chaotic and mixed systems},\ }\href
  {https://doi.org/https://doi.org/10.1016/j.physa.2013.11.012} {\bibfield
  {journal} {\bibinfo  {journal} {Physica A: Statistical Mechanics and its
  Applications}\ }\textbf {\bibinfo {volume} {396}},\ \bibinfo {pages} {185}
  (\bibinfo {year} {2014})}\BibitemShut {NoStop}%
\bibitem [{\citenamefont {Atas}\ \emph {et~al.}(2013)\citenamefont {Atas},
  \citenamefont {Bogomolny}, \citenamefont {Giraud},\ and\ \citenamefont
  {Roux}}]{Atas}%
  \BibitemOpen
  \bibfield  {author} {\bibinfo {author} {\bibfnamefont {Y.~Y.}\ \bibnamefont
  {Atas}}, \bibinfo {author} {\bibfnamefont {E.}~\bibnamefont {Bogomolny}},
  \bibinfo {author} {\bibfnamefont {O.}~\bibnamefont {Giraud}},\ and\ \bibinfo
  {author} {\bibfnamefont {G.}~\bibnamefont {Roux}},\ }\bibfield  {title}
  {\bibinfo {title} {Distribution of the ratio of consecutive level spacings in
  random matrix ensembles},\ }\href
  {https://doi.org/10.1103/PhysRevLett.110.084101} {\bibfield  {journal}
  {\bibinfo  {journal} {Phys. Rev. Lett.}\ }\textbf {\bibinfo {volume} {110}},\
  \bibinfo {pages} {084101} (\bibinfo {year} {2013})}\BibitemShut {NoStop}%
\bibitem [{\citenamefont {Page}(1993)}]{Page1993}%
  \BibitemOpen
  \bibfield  {author} {\bibinfo {author} {\bibfnamefont {D.~N.}\ \bibnamefont
  {Page}},\ }\bibfield  {title} {\bibinfo {title} {Average entropy of a
  subsystem},\ }\href {https://doi.org/10.1103/PhysRevLett.71.1291} {\bibfield
  {journal} {\bibinfo  {journal} {Phys. Rev. Lett.}\ }\textbf {\bibinfo
  {volume} {71}},\ \bibinfo {pages} {1291} (\bibinfo {year}
  {1993})}\BibitemShut {NoStop}%
\bibitem [{\citenamefont {Bauer}\ and\ \citenamefont
  {Nayak}(2013)}]{Bauer2013}%
  \BibitemOpen
  \bibfield  {author} {\bibinfo {author} {\bibfnamefont {B.}~\bibnamefont
  {Bauer}}\ and\ \bibinfo {author} {\bibfnamefont {C.}~\bibnamefont {Nayak}},\
  }\bibfield  {title} {\bibinfo {title} {Area laws in a many-body localized
  state and its implications for topological order},\ }\href
  {https://doi.org/10.1088/1742-5468/2013/09/p09005} {\bibfield  {journal}
  {\bibinfo  {journal} {Journal of Statistical Mechanics: Theory and
  Experiment}\ }\textbf {\bibinfo {volume} {2013}},\ \bibinfo {pages} {P09005}
  (\bibinfo {year} {2013})}\BibitemShut {NoStop}%
\bibitem [{\citenamefont {Gray}\ \emph {et~al.}(2018)\citenamefont {Gray},
  \citenamefont {Bose},\ and\ \citenamefont {Bayat}}]{Gray2018}%
  \BibitemOpen
  \bibfield  {author} {\bibinfo {author} {\bibfnamefont {J.}~\bibnamefont
  {Gray}}, \bibinfo {author} {\bibfnamefont {S.}~\bibnamefont {Bose}},\ and\
  \bibinfo {author} {\bibfnamefont {A.}~\bibnamefont {Bayat}},\ }\bibfield
  {title} {\bibinfo {title} {Many-body localization transition: {S}chmidt gap,
  entanglement length, and scaling},\ }\href
  {https://doi.org/10.1103/PhysRevB.97.201105} {\bibfield  {journal} {\bibinfo
  {journal} {Phys. Rev. B}\ }\textbf {\bibinfo {volume} {97}},\ \bibinfo
  {pages} {201105(R)} (\bibinfo {year} {2018})}\BibitemShut {NoStop}%
\bibitem [{\citenamefont {Santos}\ \emph
  {et~al.}(2012{\natexlab{a}})\citenamefont {Santos}, \citenamefont
  {Borgonovi},\ and\ \citenamefont {Izrailev}}]{Santos2012a}%
  \BibitemOpen
  \bibfield  {author} {\bibinfo {author} {\bibfnamefont {L.~F.}\ \bibnamefont
  {Santos}}, \bibinfo {author} {\bibfnamefont {F.}~\bibnamefont {Borgonovi}},\
  and\ \bibinfo {author} {\bibfnamefont {F.~M.}\ \bibnamefont {Izrailev}},\
  }\bibfield  {title} {\bibinfo {title} {Chaos and statistical relaxation in
  quantum systems of interacting particles},\ }\href
  {https://doi.org/10.1103/PhysRevLett.108.094102} {\bibfield  {journal}
  {\bibinfo  {journal} {Phys. Rev. Lett.}\ }\textbf {\bibinfo {volume} {108}},\
  \bibinfo {pages} {094102} (\bibinfo {year} {2012}{\natexlab{a}})}\BibitemShut
  {NoStop}%
\bibitem [{\citenamefont {Santos}\ \emph
  {et~al.}(2012{\natexlab{b}})\citenamefont {Santos}, \citenamefont
  {Borgonovi},\ and\ \citenamefont {Izrailev}}]{Santos2012b}%
  \BibitemOpen
  \bibfield  {author} {\bibinfo {author} {\bibfnamefont {L.~F.}\ \bibnamefont
  {Santos}}, \bibinfo {author} {\bibfnamefont {F.}~\bibnamefont {Borgonovi}},\
  and\ \bibinfo {author} {\bibfnamefont {F.~M.}\ \bibnamefont {Izrailev}},\
  }\bibfield  {title} {\bibinfo {title} {Onset of chaos and relaxation in
  isolated systems of interacting spins: Energy shell approach},\ }\href
  {https://doi.org/10.1103/PhysRevE.85.036209} {\bibfield  {journal} {\bibinfo
  {journal} {Phys. Rev. E}\ }\textbf {\bibinfo {volume} {85}},\ \bibinfo
  {pages} {036209} (\bibinfo {year} {2012}{\natexlab{b}})}\BibitemShut
  {NoStop}%
\end{thebibliography}
\end{document}